%% file: Main_preprint.tex
\documentclass[reprint,amsmath,amssymb,longbibliography]{revtex4-2}
\usepackage{graphicx}% Include figure files
\usepackage{dcolumn}% Align table columns on decimal point
\usepackage{bm}% bold math
\usepackage{lineno}
\usepackage{physics}
\usepackage{bbold}
\usepackage{placeins}
\usepackage{multirow, makecell}
\usepackage{xcolor}
\usepackage{ifthen}

\newcommand{\natcommphys}{0}% set to 1 for natcommphys

\input commands.tex

\begin{document}
%TC:ignore
%\linenumbers
%\title[]{Bayesian inferencing and deterministic anisotropy for molecular geometry retrieval in gas-phase diffraction experiments}%
%\title[]{Addressing the gas-phase diffraction inverse problem using Bayesian inference and deterministic anisotropy to retrieve the molecular structure probability distribution $|\Psi(\mathbf{R})|^2$}%
\title[]{Applying Bayesian Inference and deterministic anisotropy to retrieve the molecular structure $|\Psi(\boldsymbol{R})|^2$ distribution from gas-phase diffraction experiments}

\author{Kareem Hegazy$^{1,2,\dagger}$} 
\author{Varun Makhija$^{3}$}
\author{Phil Bucksbaum$^{1,2,4}$}
\author{Jeff Corbett$^5$}
\author{James Cryan$^2$}
%\author{Markus G{\"u}hr$^{2,6}$}
\author{Nick Hartmann$^6$}
\author{Markus Ilchen$^{2,7,8}$}
\author{Keith Jobe$^5$}
\author{Renkai Li$^{9}$}
\author{Igor Makasyuk$^5$}
\author{Xiaozhe Shen$^5$}
%\author{Theodore Vecchione$^5$}
\author{Xijie Wang$^5$}
\author{Stephen Weathersby$^5$}
\author{Jie Yang$^{10}$}
\author{Ryan Coffee$^{2,6,\dagger}$}
 
 \affiliation{$^1$Department of Physics, Stanford University, Stanford, California 94305, USA}
 \affiliation{$^2$Stanford PULSE Institute, SLAC National Accelerator Laboratory, 2575 Sand Hill Road, Menlo Park, California 94025, USA}
 \affiliation{$^3$Department of Chemistry and Physics, University of Mary Washington, Fredericksburg, Virginia 22401, USA}
 \affiliation{$^4$Department of Applied Physics, Stanford University, Stanford, California 94305, USA}
 \affiliation{\mbox{$^5$SLAC National Accelerator Laboratory, 2575 Sand Hill Road, Menlo Park, California 94025, USA}}
 %\affiliation{$^6$Institut f{\"u}r Physik and Astronomie, Universit{\"a}t Potsdam, Potsdam 14476, Germany}
 \affiliation{\mbox{$^6$Linac Coherent Light Source, SLAC National Accelerator Laboratory, Menlo Park, California 94025, USA}}
 \affiliation{$^7$European XFEL, Holzkoppel 4, 22869 Schenefeld, Germany}
 \affiliation{$^8$Universit{\"a}t Kassel, Heinrich-Plett-Str. 40, 34132 Kassel, Germany}
 \affiliation{\mbox{$^{9}$Department of Engineering Physics, Tsinghua University, Beijing 100084, China}}
 \affiliation{\mbox{$^{10}$Department of Chemistry, Tsinghua University, Beijing 100084, China}}
 \affiliation{$^\dagger$ Corresponding Author Emails: KHegazy@stanford.edu, Coffee@slac.stanford.edu}
%TC:endignore

%TC:ignore
\date{\today}% It is always \today, today,
             %  but any date may be explicitly specified

\begin{abstract}
\input sections/01_abstract
\end{abstract}

\maketitle
%TC:endignore

%\keywords{Suggested keywords}%Use showkeys class option if keyword
                              %display desired

%\begin{refsection}
\section{Introduction}
\input sections/02_introduction

\ifthenelse{\equal{\natcommphys}{0}}{
    %TC:ignore
    \section{Methods}
    \input sections/03_methods
    %TC:endignore

    \section{Results}
        \input sections/04_results

    \section{Discussion}
    \input sections/05_discussion_intro
        \subsection{The Role of Anisotropy}
        \input sections/06_discussion_derivation

        \subsection{Bayesian Inference and the MHA}
        \input sections/07_discussion_mcmc
        
        \subsection{\label{sc:excited_states}Outlook and potential Extension to Excited State Dynamics}
        \input sections/08_discussion_excited_state

    %TC:ignore
    \section{Conclusion}
    \input sections/09_conclusion
    %TC:endignore
}

\ifthenelse{\equal{\natcommphys}{1}}{
    \section{Results}

\input sections/04_results
        \subsection{Effects of Bayesian Inference}
        \input sections/07.5_discussion_mcmc

    \section{Discussion}
        \input sections/05_discussion_intro
        %\subsection{The Role of Anisotropy}

\input sections/06_discussion_derivation
        %\subsection{Bayesian \edit{Inference} and the MHA}

\input sections/07_discussion_mcmc
        %\subsection{\label{sc:excited_states}Outlook and potential Extension to Excited State Dynamics}

\input sections/08_discussion_excited_state
    %TC:ignore
    \section{Conclusion}

\input sections/09_conclusion
    \section{Methods}
    \input sections/03_methods
    %TC:endignore
}

%TC:ignore
\input sections/10_availability

\begin{acknowledgments}
\input sections/11_acknowledgements
\end{acknowledgments}

%\section{Effects of too many degrees of freedom\label{ap:dof}}
%\input sections/appendix_dof.tex

%\section{Comparison of \dns and \popt \label{ap:mcmc_wtf}}
%\input sections/mcmcAppendix

%\printbibliography[section=1, heading=references]
%\nocite{*}
\bibliography{alignment}% Produces the bibliography via BibTeX.
%\bibliography{aipsamp}% Produces the bibliography via BibTeX.
%TC:endignore
%\end{refsection}

%%%%%%%%%%%%%%%%%%%%%%%%%%%%%%%%%%%%
%%%%%  Supplemental Materials  %%%%%
%%%%%%%%%%%%%%%%%%%%%%%%%%%%%%%%%%%%

%TC:ignore
\pagebreak
\newpage
\widetext
\begin{center}
\textbf{\large Supplemental Information}
\end{center}
%TC:endignore
%%%%%%%%%% Merge with supplemental materials %%%%%%%%%%
%%%%%%%%%% Prefix a "S" to all equations, figures, tables and reset the counter %%%%%%%%%%
\setcounter{equation}{0}
\setcounter{figure}{0}
\setcounter{table}{0}
\setcounter{page}{1}
\setcounter{section}{0}
\makeatletter
\renewcommand{\theequation}{S\arabic{equation}}
\renewcommand{\thefigure}{S\arabic{figure}}
\renewcommand{\thetable}{S\arabic{table}}
\renewcommand{\thesection}{Supplementary Note \arabic{section}}
\renewcommand{\bibnumfmt}[1]{[S#1]}
\renewcommand{\citenumfont}[1]{#1}
%%%%%%%%%% Prefix a "S" to all equations, figures, tables and reset the counter %%%%%%%%%%
%\input sections/supplemental
%\begin{refsection}
%TC:ignore
\section{Calculating the axis distribution moments\label{ap:ADM_calc}}
\input sections/ap_ADM_calculations

\section{Anisotropy Derivation\label{ap:anisotropy_derivation}}
\input sections/ap_anisotropy_derivation

\section{\label{ap:fitting}Fitting for \blm{} and $C_{lmk}(q)$, and common mistakes}
\input sections/ap_fitting_legendres_ADMs

\section{\label{ap:mcmc}Using Bayesian Inference and the Metropolis-Hastings Algorithm}
\input sections/ap_mcmc

\section{\label{ap:fit_error_prop}Calculating error bars for \blm{} and $C_{lmk}(q)$ coefficients}
\input sections/ap_fitting_error_propagation

\section{\label{ap:mode_search}Searching for the optimal $\boldsymbol{\Theta}$ parameters}
\input sections/ap_mode_search

\section{\label{ap:delta_dist}Results of the Delta distribution posterior}
\input sections/ap_delta_distribution

\section{Fitting for the $\mathcal{I}$ coefficient\label{ap:fitting_I}}
\input sections/ap_fittingIcoeff

\section{Initial Rotational and Vibrational Thermal Distribution of N$_2$O\label{ap:n2o_thermal}}
\input sections/ap_n2o_thermal_dist
%TC:endignore

%%%%%%%%%%%%%%%%%%%%%%%%%%%%%%%%
%%%%%  Reviewer Responses  %%%%%
%%%%%%%%%%%%%%%%%%%%%%%%%%%%%%%%

\iffalse
%TC:ignore
\pagebreak
\begin{equation*}
    
\end{equation*}
\newpage
\widetext
\begin{center}
\textbf{\Large Referee 1 Response}
\end{center}
\setcounter{page}{1}

\input reviewer_responses/reviewer1_response

\pagebreak
\newpage
\widetext
\begin{center}
\textbf{\Large Referee 2 Response}
\end{center}
\setcounter{page}{1}

\input reviewer_responses/reviewer2_response
%TC:endignore
\fi

\end{document}

%% file: commands.tex
\newcommand{\angs}{\mbox{\AA}}

\newcommand{\threeJ}[6]{ 
	\begin{pmatrix}
   		#1 & #2 & #3 \\
   		#4 & #5 & #6 
	\end{pmatrix}
}
\newcommand{\textapprox}{\raisebox{0.5ex}{\texttildelow}}

\newcommand{\iang}{\AA$^{-1}$}

\newcommand{\mmprth}{P \left(\boldsymbol{R} \right| \left. \boldsymbol{\Theta}, C\right)}
\newcommand{\prth}{$\mmprth$}
\newcommand{\mmprtht}{P \left(\boldsymbol{R}, t \right| \left. \boldsymbol{\Theta}, C\right)}
\newcommand{\prtht}{$\mmprtht$}
\newcommand{\mmprthopt}{P (\boldsymbol{R} | \mmthopt, C )}
\newcommand{\prthopt}{$\mmprthopt$}
\newcommand{\mmprthoptt}{P (\boldsymbol{R}, t | \mmthopt, C )}
\newcommand{\prthoptt}{$\mmprthoptt$}

\newcommand{\mmptheta}{P \left(\boldsymbol{\Theta} | C\right)}
\newcommand{\ptheta}{$\mmptheta$}
\newcommand{\mmpthetag}{P^{(\mathcal{N})} \left(\boldsymbol{\Theta} | C\right)}
\newcommand{\pthetag}{$\mmpthetag$}
\newcommand{\mmpthetad}{P^{(\delta)} \left(\boldsymbol{\Theta} | C\right)}
\newcommand{\pthetad}{$\mmpthetad$}

\newcommand{\mmprthd}{P^{(\delta)} \left(\boldsymbol{R} \right| \left. \boldsymbol{\Theta}, C\right)}
\newcommand{\prthd}{$\mmprthd$}
\newcommand{\mmprthg}{P^{(\mathcal{N})} \left(\boldsymbol{R} \right| \left. \boldsymbol{\Theta}, C\right)}
\newcommand{\prthg}{$\mmprthg$}

\newcommand{\mmprthoptg}{P^{(\mathcal{N})} (\boldsymbol{R} | \mmthopt, C )}
\newcommand{\prthoptg}{$\mmprthoptg$}

\newcommand{\mmadmnt}[3]{\mathcal{A}^{#1}_{#2#3}}
\newcommand{\mmadm}[3]{\mmadmnt{#1}{#2}{#3}(t)}
\newcommand{\adm}{$\mmadm{l}{m}{k}$}
\newcommand{\mmblm}{B^m_{l}(q, t)}
\newcommand{\blm}{$\mmblm$}

\newcommand{\mmthd}{\theta^{(\text{d})}}
\newcommand{\thd}{$\mmthd$}
\newcommand{\mmthlf}{\theta^{(\text{lf})}_{q}}
\newcommand{\mmthlfu}[1]{\theta^{(\text{lf})}_{q#1}}
\newcommand{\thlf}{$\mmthlf$}
\newcommand{\mmphlf}{\phi^{(\text{lf})}_q}
\newcommand{\mmphlfu}[1]{\phi^{(\text{lf})}_{q#1}}
\newcommand{\phlf}{$\mmphlf$}
\newcommand{\mmthopt}{\boldsymbol{\Theta}^*}
\newcommand{\thopt}{$\mmthopt$}

\newcommand{\mmmaono}{\ev**{\angle \text{ONO}}}
\newcommand{\maono}{$\mmmaono$}
\newcommand{\mmmrnoa}{\ev**{\text{NO}^{(1)}}}
\newcommand{\mrnoa}{$\mmmrnoa$}
\newcommand{\mmmrnob}{\ev**{\text{NO}^{(2)}}}
\newcommand{\mrnob}{$\mmmrnob$}

\newcommand{\mmsrnoa}{\sigma\left(\text{NO}^{(1)}\right)}
\newcommand{\srnoa}{$\mmsrnoa$}
\newcommand{\mmsrnob}{\sigma\left(\text{NO}^{(2)}\right)}
\newcommand{\srnob}{$\mmsrnob$}
\newcommand{\mmsaono}{\sigma\left(\angle \text{ONO}\right)}
\newcommand{\saono}{$\mmsaono$}
\newcommand{\sth}{$\sigma^{\Theta}$}

\RequirePackage{color}\definecolor{RED}{rgb}{1,0,0}\definecolor{BLUE}{rgb}{0,0,1}\definecolor{GREEN}{rgb}{0,1,0} %DIF

%% file: sections/01_abstract.tex
Currently, our general approach to retrieving molecular structures from ultrafast gas-phase diffraction heavily relies on complex \textit{ab initio} electronic or vibrational excited state simulations to make conclusive interpretations.
Without such simulations, inverting this measurement for the structural probability distribution is typically intractable.
This creates a so-called inverse problem.
In this work, we develop a broadly applicable method that addresses this inverse problem by approximating the molecular frame structure $|\Psi(\boldsymbol{R}, t)|^2$ distribution independent of these complex simulations.
%\edit{We address two problems, the so-called inverse problem of gas-phase diffraction, and the curse of dimensionality.
%With Bayesian Inference and data, we effectively invert the diffraction patterns for unique molecular structure distributions.
%Applying Markov-Chain Monte Carlos to this statistical reinterpretation we address the curse of dimensionality by drastically narrowing the search space of likely structures and sampling structures based on their agreement with the data.}
%\edit{This method does not require complex electronic or vibronic excited state molecular dynamic simulations and can identify the unique molecular structure.}
We retrieve the vibronic ground state $|\Psi(\boldsymbol{R})|^2$ for both simulated stretched NO$_2$ and measured N$_2$O.
From measured N$_2$O, we observe 40~m\AA{} coordinate-space resolution from 3.75~\iang{} reciprocal space range and poor signal-to-noise, a 50X improvement over traditional Fourier transform methods.
In simulated NO$_2$, typical to high signal-to-noise levels predict 100--1000X resolution improvements, down to 0.1~m\AA.
By directly measuring the width of $|\Psi(\boldsymbol{R})|^2$, we open ultrafast gas-phase diffraction capabilities to measurements beyond current analysis approaches.
%since this width is generally only accessible through simulation.
%Our method also leverages deterministic ensemble anisotropy; this provides an explicit dependence on the molecular frame angles.
%SNR over q and well suited for LCLS II?
%This method's ability to retrieve the unique molecular structure with high resolution, and without complex simulations, 
This method has the potential to effectively turn gas-phase ultrafast diffraction into a discovery-oriented technique to probe systems that are prohibitively difficult to simulate.

%% file: sections/02_introduction.tex
Ultrafast molecular gas-phase diffraction, from either x-rays \cite{Stankus.xrayDiff.2020, Minitti.Fits.2015} or electrons \cite{Zewail.UED.1997, Schafer.UED.1992, Ewbank.UED.1993, shen.UED_machine.2019}, is a vital tool for retrieving time-dependent molecular structures.
In elastic molecular gas-phase diffraction experiments, x-rays or electrons scatter off of electrons and nuclei, with differing proportionality.
Each pairwise atomic distance creates a pattern of scattered x-rays or electrons as a function of their transverse momentum $q$.
The measured diffraction pattern is the sum of all such contributions, this is orientationally averaged over the lab frame ensemble distribution.
We lose pairwise directional information and thus the ability to explicitly distinguish individual atomic distances.
Consequently, directly inverting diffraction patterns for the molecular structure is generally intractable, this is a so-called inverse problem.
Typically, we avoid this inverse problem and retrieve both the molecular structures and the molecular frame orientations by simulating the forward excited state process.
These are generally time-dependent \textit{ab initio} electronic and vibrational excited state simulations that explore a large parameter space (rovibration, structure, and electronic state) with trajectory bifurcations due to effects like conical intersections \cite{Ben-Nun.AIMS.2000, Siegbahn.CASSCF.1981, Mai.SHARC.2018, Meyer.MCTDH.1990}.
We refer to such simulations as complex simulations, that are typically validated through comparisons with measured diffraction patterns or pair-distribution functions (PDFs -- a weighted histogram of pairwise distances).
Consequently, ultrafast gas-phase diffraction is generally limited by the ability to perform these complex simulations.
We aim to expand diffraction measurements for high-resolution reconstructions of molecular structure probability distribution $|\Psi(\boldsymbol{R}, t)|^2$ without relying on complex molecular dynamics simulations by effectively solving this inverse problem with a statistical interpretation.

%, where $\boldsymbol{R}$ is the molecular frame nuclear coordinates,
\begin{figure}[!htbp] 
\begin{center}
\includegraphics[scale=0.75]{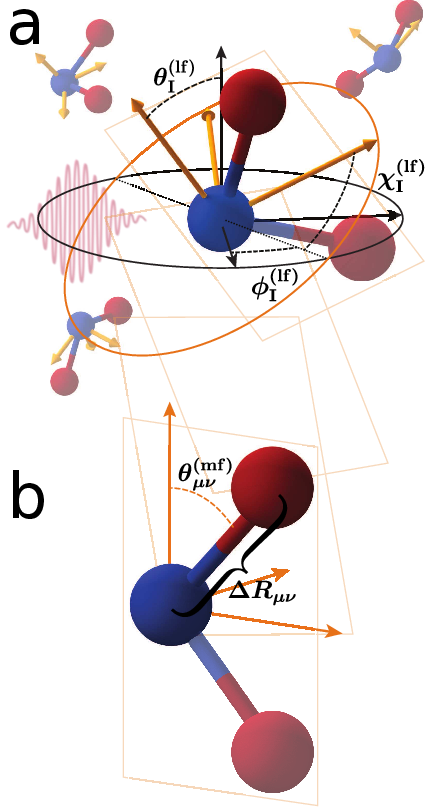}
\caption{\label{fig:rotations_LFMF}\textbf{Correspondence between the lab and molecular frame} Our analysis considers each pairwise distance independently and we define the origin of both the lab and molecular frames by one of the pairwise vectors. For the highlighted NO bond, the nitrogen atom (blue) defines the origin. The lab frame (panel a) is defined by the laser polarization $(\hat{\mathbf{z}})$ and propagation direction $(\hat{\mathbf{y}})$. The molecular frame (panel b) is defined by the molecule's rovibronic ground state principal moments of inertia, where the molecular A, B, and C axes define $\hat{\mathbf{z}}^{(\text{mf})}$, $\hat{\mathbf{y}}^{(\text{mf})}$, and $\hat{\mathbf{x}}^{(\text{mf})}$. Here the NO is described by $\Delta \boldsymbol{R}_{\mu\nu}$, $\theta_{\mu\nu}^{(mf)}$, and $\phi_{\mu\nu}^{(mf)}$ which correspond to its distance, polar angle, and azimuthal angle respectively. One accesses the molecular frame by rotating the lab frame by the Euler angles $\theta^{(\text{lf})}_\text{I}$, $\phi^{(\text{lf})}_\text{I}$, and $\chi^{(\text{lf})}_\text{I}$.
}
\end{center}
\end{figure}

\begin{figure}[!htbp] 
    \centering
    \includegraphics[scale=0.33]{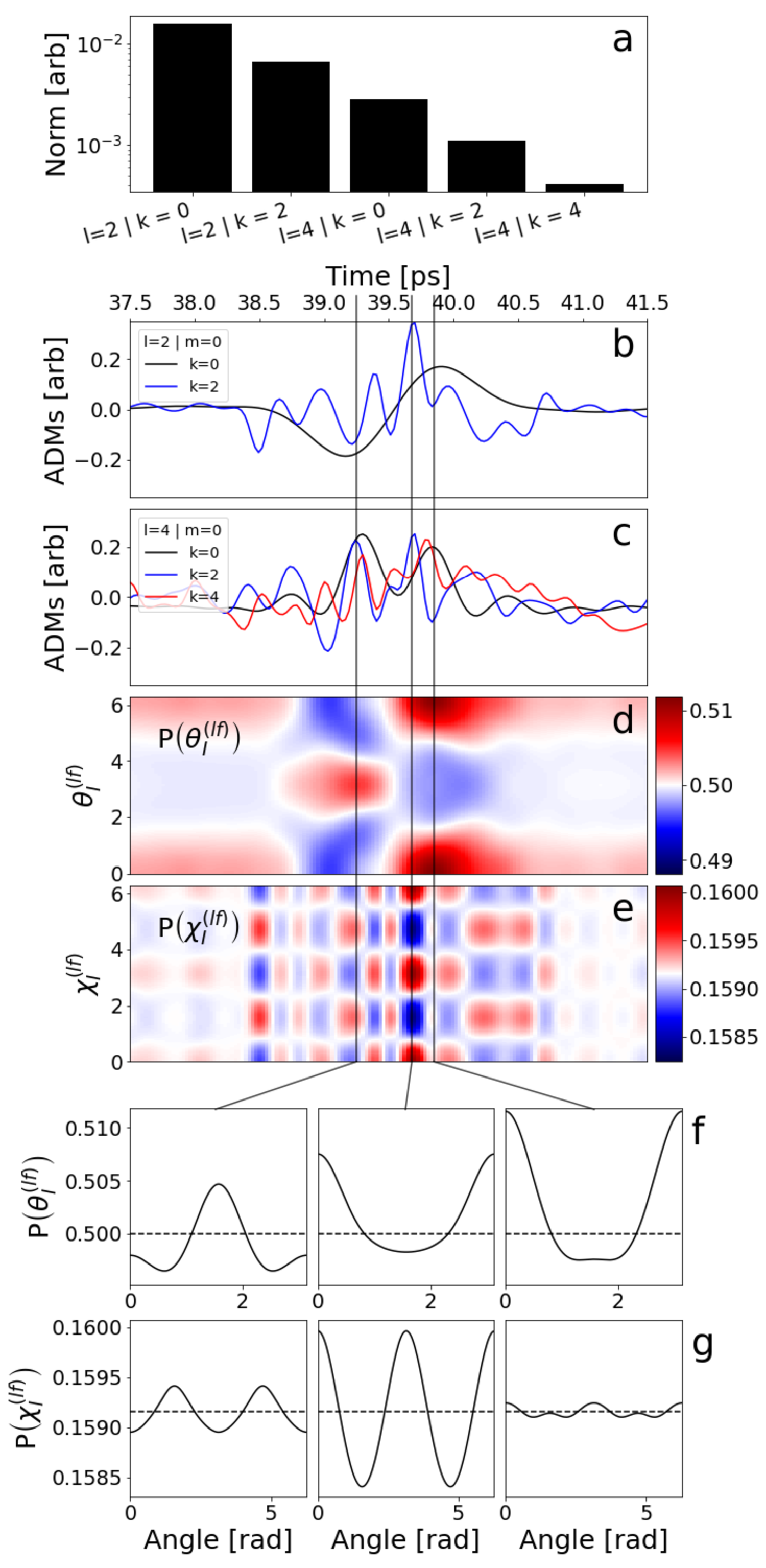}
    \caption{\textbf{Axis distribution moments and ensemble anisotropy} The Axis distribution moments (ADMs) encapsulate the ensemble anisotropy which provides various constraints on the molecular frame as a function of time. The ADMs are parameterized by the three angular momentum quantum numbers $l$, $m$, and $k$ which correspond to the total angular momentum, the projection along the lab frame $(\hat{\mathbf{z}})$ axis, and projection along the molecular frame $(\hat{\mathbf{z}})$ axis respectively. Panel a shows the square norms of the ADMs. Panel b and c show these normalized ADMs, highlighting their time dependence. Panels d and e show the time-dependent ensemble anisotropy probability distribution for $\theta^{(\text{lf})}_\text{I}$ and $\chi^{(\text{lf})}_\text{I}$, respectively. Panels f and g show illustrative line-outs of these Euler angle distributions for $\theta^{(\text{lf})}_\text{I}$ and $\chi^{(\text{lf})}_\text{I}$, respectively, with isotropy indicated by the dashed lines.}
    \label{fig:alignment}
\end{figure}

A variety of studies sought to reduce reliance on complex simulations, but are either limited in the systems they address or quickly run into the curse of dimensionality.
Fourier transforming the time dependence exposes dissociative and vibronic signals \cite{Ware.TRXS.2018, Ware.TRXS.2019, Ware.TRXS.2020} but it is insensitive to classes of isomerizations.
Methods employing ensemble anisotropy have garnered much interest \cite{Saldin.aligned_diffraction.2010, Starodub.aligned_diffraction.2010, Ho.aligned_symTop_diffraction.2009, Elser.correlations_aligned.2011, Poon.fiber_diffraction.2013, Reckenthaeler.observe_anisotropy.2009, Pabst.3d_alignment.2010, Yang.align.2014, Hensley.mFrame.2012, Wilkin.MF2D_retrieval.2022} yet they struggle to get sub-Angstrom resolution and the full 3d structure for generic molecular structures.
%When pattern matching against ensembles of simulations \cite{Minitti.Fits.2015} and 
%Pattern matching against many individual isomerizations determined by simulation \cite{Stankus.Fits.2019} or a grid search \cite{Natan.PDF_inversion.2021}, quickly becomes intractable due to the curse of dimensionality or may still heavily relies on complex molecular dynamics simulations.
Optimization methods, while capable of exposing large-scale motion, are susceptible to local minima~\cite{Yang.align.2014}.
% Check martin's paper
%Our method is unique in that it retrieves the labeled pairwise distances and angles under conventional experimental without over-burdensome molecular dynamic simulations, only the known vibronic ground state structure.
%conditions with and without deterministic ensemble anisotropy and requiring tractable rotational simulations.
Pattern matching measured data against sampled isomers~\cite{Stankus.Fits.2019, Natan.PDF_inversion.2021, Natan.high_order_anisotropy.2021} becomes intractable for moderately large molecules due to the curse of dimensionality.
For example, a molecule with $N_{\text{atoms}}$ atoms has $3N_{\text{atoms}}-6$ degrees of freedom.
To independently sample each degree of freedom 10 times would require $10^{3N_\text{atoms}-6}$ structures, becoming intractable for molecules with 7 or more atoms.
Simulations reduce the structure-space of isomers to select, but this trade-off requires previous knowledge~\cite{Stankus.Fits.2019} that potentially imparts biases.

We employ insights from molecular ensemble anisotropy methods, applied statistics, and machine learning principles to address the inverse problem and the curse of dimensionality to approximate the molecular structure probability density $|\Psi(\boldsymbol{R}, t)|^2$.
It is important to note that instead of sampling individual molecular structures and comparing single structures to the measured data, we are sampling entire $|\Psi(\boldsymbol{R}, t)|^2$ probability distributions.
We access the molecular frame by decomposing measured data onto anisotropic components.
Then, we iteratively approximate $|\Psi(\boldsymbol{R}, t)|^2$ with a statistical approach uniquely suited for high repetition-rate diffraction facilities.
We observe that resolution strongly improves with signal-to-noise much faster than increasing the $q$ range beyond moderate values.
%while recovering labeled pairwise distances and molecular frame angles.
%Our method eliminates the PDF and improves resolution over the PDF by \textapprox100X.
Unlike the PDF approach, this method retrieves the molecular distances and angles required to define a unique molecular structure.

In our method, we recover the molecular frame through time-dependent ensemble anisotropy~\cite{makhija.orientation.2016, Marceau.mFrame.2017, gregory.mfpad.2021, Sandor.anisotropy_MF.2018, Mikosch.anisotropy_MF.2013,Peter.mFrame_SFI.2019, Peter.mFrame_SFI_OCS.2018}.
One rotates into the molecular frame with the lab frame Euler angles $\theta^{(\text{lf})}_\text{I}$ (polar), $\phi^{(\text{lf})}_\text{I}$ (azimuthal), and $\chi^{(\text{lf})}_\text{I}$ (Fig.~\ref{fig:rotations_LFMF}).
An induced rotational wavepacket creates ensemble anisotropy given by $|\Psi(\theta^{(\text{lf})}, \phi^{(\text{lf})},t)|^2$.
Axis distribution moments (ADMs) \cite{stolow.calc_adm.2008,underwood.calc_adm.2000,gregory.mfpad.2021} are the coefficients in the Wigner D matrix expansion of $|\Psi(\theta^{(\text{lf})}, \phi^{(\text{lf})},t)|^2$
\begin{equation}
    \mathcal{A}^l_{mk}(t) = \frac{2l+1}{8\pi^2}\expectationvalue**{D^{l}_{mk}\left(\phi_{\text{I}}^{(\text{lf})}, \theta_{\text{I}}^{(\text{lf})}, \chi_{\text{I}}^{(\text{lf})}\right)}{\Psi(t)}.
    \label{eq:ADMs}
\end{equation}
These ADMs describe the ensemble of molecular frame orientations with respect to the lab frame.
When calculating the ADMs, the $l$, $m$, and $k$ are difference and sum of quantum numbers between rotational eigenstates, respectively for the total angular momentum, the projection onto the lab frame z-axis, and the projection onto the molecular frame z-axis.
%To illustrate this concept, we impulsively align the asymmetric top NO$_2$ and show ensemble's molecular frame (Fig.~\ref{fig:NO2molFrame}) orientation Euler angles $\theta_I$ (polar), $\phi_I$ (azimuthal), and $\chi_I$ (spin), with respect to the lab frame.
%We show this in Fig.~\ref{fig:alignment} with impulsive molecular alignment of the asymmetric top NO2 through the Euler angles $\theta$ (polar), $\phi$ (azimuthal), and $\chi$ (spin) that describe the molecular frame's orientation, shown in Fig.~\ref{fig:NO2molFrame}, with respect to the lab frame coordinates.
%We access the molecular frame by inducing time-dependent anisotropy by preparing a coherent molecular rotational wavepacket via impulsive alignment.
These ADMs transform the lab frame into the molecular frame by decomposing the measured lab frame anisotropy into $C_{lmk}(q)$ coefficients, which are dependent on molecular frame pairwise distances and angles ($\theta_{\mu\nu}^{(\text{mf})}$ and $\phi_{\mu\nu}^{(\text{mf})}$) shown in Fig.~\ref{fig:rotations_LFMF}b.
The PDF is not directly sensitive to these angles.
After impulsively aligning the molecular ensemble, Fig.~\ref{fig:alignment} illustrates how transient anisotropy (panels b and c) provides constraints on these Euler angles and consequently the molecular frame (panels d-g).
For example, at 39.25~ps the anisotropy provides simultaneous constraints on $\theta^{(\text{lf})}_\text{I}$ and $\chi^{(\text{lf})}_\text{I}$.
At 39.68~ps, $\chi^{(\text{lf})}_\text{I}$ (the molecular frame azimuthal plane) is highly constrained.
At 39.85~ps the ensemble is well localized in $\theta^{(\text{lf})}_\text{I}$, resolving measurements along the molecular frame $\hat{\mathbf{z}}$.
Here, P$\left(\phi^{(\text{lf})}_\text{I}\right)$ is uniform due to cylindrical symmetry imparted by a linearly polarized pulse.
%This framework encourages measurements over picosecond time intervals in regions where the ensemble exhibits high variations of anisotropy.

To effectively invert the molecular diffraction pattern and approximate $|\Psi(\boldsymbol{R},t)|^2$, we use Bayesian Inference.
Bayesian Inference describes a class of statistical inference techniques using Bayes's Theorem to update one's model based on observed data~\cite{box.bayesian_inference.2011}.
We first approximate $|\Psi(\boldsymbol{R},t)|^2$ as the probability distribution \prtht, which is parameterized by the molecular structure degrees of freedom $\boldsymbol{\Theta}$.
Using Bayesian Inference, we then relate \ptheta{} to the measured molecular diffraction pattern.
With this framework, we use Markov-chain Monte Carlo (MCMC) techniques to build \ptheta{} and tackle the curse of dimensionality by efficiently sampling structures most consistent with the measured $C_{lmk}(q)$.
This method is unbiased and naturally avoids regions in our sampling space that are inconsistent with the $C_{lmk}(q)$.
We retrieve \prtht{} with neither the PDF nor complex molecular dynamics simulations since we will analytically relate the molecular frame pairwise distances and angles to the $C_{lmk}(q)$.
Further intuition is provided in \ref{ap:mcmc} and Ref.~\cite{Hegazy.dissertation.2023}.

Instead of complex molecular dynamics simulations this method has fewer simulation requirements.
%the much more tractable simulation of the rovibronic vibronic ground state structure and limited \textit{a priori} knowledge of $|\Psi(\boldsymbol{R},t)|^2$.
In this method's simplest form, when probing structural dynamics it only requires the much more tractable simulation of the rovibronic ground state structure to define the molecular frame.
When measuring the equilibrium vibronic ground state, one does not require \textit{a priori} knowledge of the structure they wish to find.
This is because each sampled structure will define a new molecular frame.
When using anisotropy components, we require time-dependent rotational simulations for the ADMs.
This requires rotational constants and molecular polarizability, all of which can be measured or calculated from the rovibronic ground state structure.
When applying this method to excited states, we require the transition dipole, which is also measured or calculated from the rovibronic ground state structure.
As discussed later, depending on the desired accuracy, one must select a functional form for \prtht{} based on \textit{a priori} knowledge of the excitation or use normal distributions as a ``first-order" approximation.

In this manuscript, we validate these principles by retrieving $|\Psi(\boldsymbol{R})|^2$ for the vibronic ground states of both simulated NO$_2$ and measured N$_2$O rotational wavepackets.
Here NO$_2$, an asymmetric top, serves as a test case to show our method's broad capabilities and behavior under various experimental conditions.
Furthermore, we validated these capabilities with measured N$_2$O data from the ultrafast MeV electron diffraction facility at SLAC (UED).
We chose these molecules to specifically be amenable to conventional methods since triatomics do not suffer significantly from the curse of dimensionality.
In this lower dimensional realm, we benchmark and validate our method against conventional methods with intentions to later expand to larger molecules. 
In the following, all simulations and equations correspond to ultrafast electron diffraction experiments but are easily extended to x-ray diffraction.
%As discussed, it is important to measure our ensemble throughout the entire anisotropy signal to aggregate constraints on the molecular frame.

In this work, we rigorously and qualitatively describe this method in addition to quantitatively benchmarking both its advantages and shortcomings.
We provide intuition and mathematically describe how induced anisotropy accesses the molecular frame structural angles ($\theta^{(\text{mf})}_{\mu\nu}$ and $\phi^{(\text{mf})}_{\mu\nu}$) and how to retrieve this molecular frame structure using Bayesian Inference.
We evaluate this method on simulated and measured data, showing how \prth{} significantly improves upon the traditional Fourier limited PDF.
Firstly, \prth{} unambiguously defines a unique molecular mean structure without complex molecular dynamics simulations.
This is generally not possible from the PDF alone.
Secondly, we report pairwise distance resolutions of order 10~m\AA{} and down to 0.1~m\AA{} from measured and simulated data, respectively.
These resolutions are respectively a factor of 50 and 1000 times smaller than their corresponding PDF resolutions.
Thirdly, we investigate this method's behaviors and systematic errors as a function of experimental factors and analysis choices.
We find this procedure depends more strongly on signal-to-noise than it does by extending measured momentum transfer.
Fourthly, we demonstrate how this method expands ultrafast gas-phase diffraction experiments to quantitatively measure additional parameters, such as the width of $|\Psi(\boldsymbol{R}, t)|^2$.
Lastly, we describe how one can apply this method to excited state dynamics.
With these advancements, this method has the potential to expand ultrafast gas-phase diffraction into a more discovery-oriented technique, one that is free of complex excited state simulation limitations and is applicable to currently inaccessible molecular systems.% and should accommodating non Born-Oppenheimer dynamics.

%% file: sections/03_methods.tex
\begin{figure*}[!htbp]
    \centering
    \includegraphics[scale=0.38]{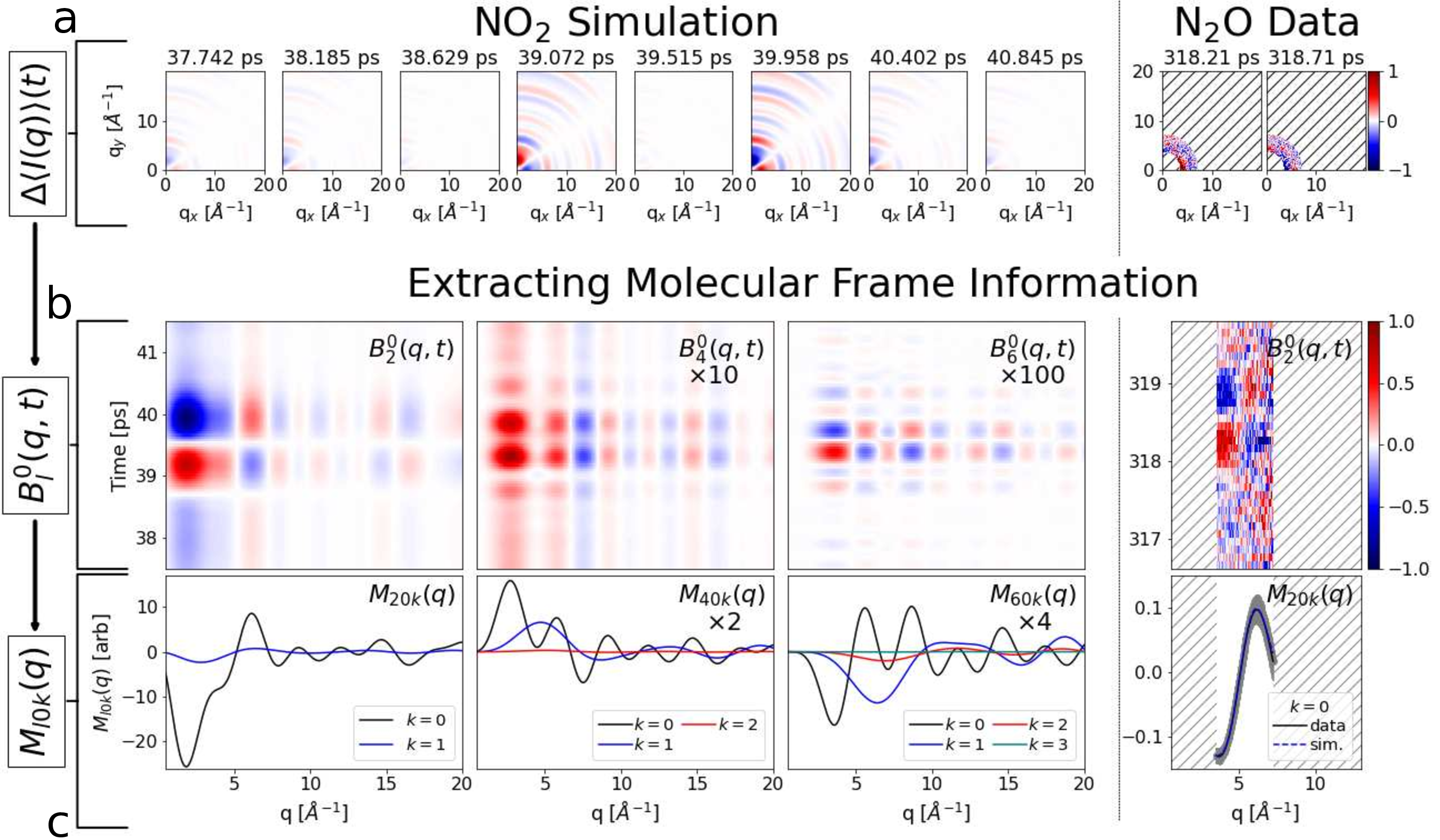}
    \caption{\textbf{Analysis to access the molecular frame signal} To access the molecular structure term, in the molecular frame, one must remove the lab frame anisotropy dependence and fit onto the ADMs. For the NO$_2$ simulation (left) and N$_2$O data (right), we illustrate the analysis steps. One first measures the difference diffraction pattern ($\Delta \langle I(\boldsymbol{q},t) \rangle$), given by Eq.~\ref{eq:diffRigAvg} (row a). Removing the detector angular dependence, one retrieves \blm{} of Eq.~\ref{eq:Blm} (row b). Removing the time-dependent ensemble anisotropy (ADMs) yields the molecular frame $M_{lmk}(q)$ coefficients Eq.~\ref{eq:Ccoeffs} (row c). All as described in the text. We note that in the N$_2$O data (right) we have limited visibility of data due to experimental limitations illustrated by the hashes.}
    \label{fig:analysis_steps}
\end{figure*}

Our method can be subdivided into three principal concepts.
Firstly, we use ensemble anisotropy, described by the ADMs, to access the molecular frame by projecting the data onto anisotropic components.
Secondly, we select a model, \prth, to approximate $|\Psi(\boldsymbol{R})|^2$ and develop our statistical approach to solve for $\boldsymbol{\Theta}$ using Bayesian Inference.
That is, through the statistical nature of our measurement we use Bayesian Inference to effectively invert the diffraction signal for $\boldsymbol{\Theta}$.
Lastly, we take our statistical description and use MCMC techniques to solve for \ptheta{} to retrieve the optimal $\boldsymbol{\Theta}$ parameters (\thopt).
The code used for this analysis~\cite{Hegazy.github.2022} can be run to reproduce the following results or adapted for other molecules.

\subsection{Extracting Molecular Frame Information}
We describe our analysis procedure for a system given an induced deterministic ensemble anisotropy under experimental conditions at the SLAC MeV ultrafast electron diffraction facility (UED) \cite{shen.UED_machine.2019}.
Our generic pump-probe setup is similar to most ultrafast diffraction setups, consisting of an 800~nm Ti:Sapphire pump laser and a 120~fs FWHM electron bunch probe.
For the simulated NO$_2$ results, we consider using a single 10~TW/cm$^2$ 800~nm pump pulse to impulsively induce a coherent rotational wave packet and probing it within a window of high anisotropy variation: [37.5, 41.5]~ps.
For the measured N$_2$O sample, a train of 8 identical 800~nm pulses (40~fs duration and $5\times 10^{12}$~W/cm$^2$ irradiance) separated by full quantum revivals induced such rotational wavepacket \cite{Cryan.N2_stacker.2009}.
We measured the first field free full quantum revival over a window of \textapprox3~ps.
We masked $q$ regions $[0,3.5]$~\iang{} and above 7.25~\iang{} due to ellipticity in the imaging of the diffraction pattern and poor signal-to-noise, respectively.
Linearly polarized pump pulses induce azimuthal symmetry, which sets $m=0$ in Eq.~\ref{eq:ADMs} (P$(\phi_{\text{I}}^{(\text{lf})}, t)=1/2\pi$), while the Raman excitation of the wavepacket requires $l$ being even in Eq.~\ref{eq:ADMs}.

We define anisotropy in two equivalent ways and quantify it through the ADMs.
%, with $\theta$, $\phi$, and $\chi$ representing the polar, azimuthal, and spin Euler angles, respectively, with respect to the laser polarization (quantization axis) and the diffraction probe path $(\hat{\mathbf{x}})$.
Firstly, anisotropy is defined by a non-zero projection of the measured diffraction pattern onto any $Y_l^m$ with even $l>0$ for a given $\Delta q$ range.
Secondly, anisotropy exists when there is a non-zero \adm{} for $l>0$.
%We are only concerned with even $l$ since a measured diffraction pattern is even with respect to the direction of each pairwise distance.
To calculate the ADMs, one must know the rotational (A, B, C) and ideally the centrifugal distortion (D) constants, as well as the differential polarizability, which can be calculated from the known ground state structure or measured from Raman spectroscopy.
For N$_2$O, we used the measured rotational constants~\cite{Toth.n2o_rovib_spect.1991, Bohlin.n2o_rovib_spect.2012} to model the rotational wavepacket for the fitted ensemble temperature and laser intensity described in \ref{ap:ADM_calc}.
We note other methodologies to calculate the ADMs~\cite{stolow.calc_adm.2008, Hockett.calc_adm.2015, underwood.calc_adm.2000}.
\ref{ap:ADM_calc} describes both our calculation of the ADMs and our search for the best-fit ADMs.

We access the molecular pairwise distances and angles in the molecular frame.
Using the ADMs and the Independent Atom Approximation, we relate measured lab frame anisotropy in diffraction patterns, $\langle I(\boldsymbol{q},t) \rangle$, to the molecular structure
\begin{widetext}
    \begin{equation}
        \begin{aligned}
            \langle I(\boldsymbol{q},t) \rangle ={}&\mathcal{I} \bigg( \sum_\mu |f_\mu(q)|^2 + \sum_{\mu,\nu : \mu \neq \nu} \text{Re} \bigg\{ f_\mu(q) f^*_\nu(q)  \sum_l 4 \pi i^l \\
            &\times \sum_{m, k} (-1)^{k} \underbrace{Y^{m}_l \left( \theta^{(\text{lf})}_q, \phi^{(\text{lf})}_q \right)}_{\text{\small Lab Frame}} \ev{\underbrace{D^l_{m k} \left( \phi_{\text{I}}^{(\text{lf})}, \theta_{\text{I}}^{(\text{lf})}, \chi_{\text{I}}^{(\text{lf})} \right)}_{\text{\small Ensemble Anisotropy}} \underbrace{j_l(q \Delta R_{\mu\nu}) Y_l^{-k} \left( \theta_{\mu\nu}^{(\text{mf})}, \phi_{\mu\nu}^{(\text{mf})}\right)}_{\text{\small Molecular Frame Structure}}}{\Psi(t)} \bigg\} \bigg)
        \end{aligned} \label{eq:ImolGen}
    \end{equation}
\end{widetext}
In Eq.~\ref{eq:ImolGen}, derived in \ref{ap:anisotropy_derivation}, $f_\mu(q)$ is the scattering amplitude of the $\mu^{\text{th}}$ atom, $j_l(qr)$ are the spherical Bessel functions of the first kind, $\mathcal{I}$ is the diffraction beam intensity, and the momentum transfer vector is given by $\boldsymbol{q}=[q,\theta_q^{(\text{lf})}, \phi_q^{(\text{lf})}]$.
The difference vector $\Delta\boldsymbol{R}_{\mu\nu}=\boldsymbol{R}_\mu - \boldsymbol{R}_\nu = [\Delta R_{\mu\nu}, \theta^{(\text{mf})}_{\mu\nu}, \phi^{(\text{mf})}_{\mu\nu}]$ is the molecular frame pairwise distance and angles between the $\mu^{\text{th}}$ and $\nu^{\text{th}}$ atoms, illustrated in Fig.~\ref{fig:rotations_LFMF}b.
Equation~\ref{eq:ImolGen} shows how the ensemble anisotropy connects the lab frame to the molecular frame structure.
Directly accessing the molecular frame pairwise angles $(\theta^{(\text{mf})}_{\mu\nu}, \phi^{(\text{mf})}_{\mu\nu})$ requires anisotropy and is otherwise inaccessible through the PDF and isotropic contributions alone.
This is evident by isolating the isotropic component ($l=0$, $m=0$, $k=0$) which sets $Y_0^{0} \left( \theta_{\mu\nu}^{(\text{mf})}, \phi_{\mu\nu}^{(\text{mf})}\right) = 1/(2\sqrt{\pi})$.

For our method, we describe optimal representations of the lab and molecular frames used in Eq.~\ref{eq:ImolGen}.
The molecular frame is defined by the molecule's principal moments of inertia before laser excitation with the $\hat{\mathbf{z}}^{(\text{mf})}$, $\hat{\mathbf{x}}^{(\text{mf})}$, and $\hat{\mathbf{y}}^{(\text{mf})}$ corresponding to the principle moments of inertia in decreasing order: A, B, and C respectively.
This necessitates knowledge of the rovibronic ground state structure when one is measuring an excited rovibronic structure.
When looking at the $\Delta\boldsymbol{R}_{\mu\nu}$ contribution, we isolate the $\mu^{\text{th}}$ and $\nu^{\text{th}}$ atoms while ignoring other atoms and translate the atom pair such that $\boldsymbol{R}_\nu$ defines the origin.
This is highlighted in Fig.~\ref{fig:rotations_LFMF}b where the nitrogen is translated to the origin.
This translation allows us to define the pairwise angles and derive Eq.~\ref{eq:ImolGen}.
Since we are concerned with a difference in locations $\Delta\boldsymbol{R}_{\mu\nu}$, Eq.~\ref{eq:ImolGen} is invariant under such molecular frame translations.
In the lab frame, the laser polarization defines $\hat{\mathbf{z}}^{(\text{lf})}$ and the propagation direction of the probe pulse defines $\hat{\mathbf{y}}^{(\text{lf})}$.
The measured signals in the lab frame, on a 2D detector, are defined by detector parameters $q=|\boldsymbol{q}|$ and the azimuthal angle \thd{} defined by $\hat{\mathbf{z}}^{(\text{lf})}$.
\ref{ap:anisotropy_derivation} describes how to rewrite $\boldsymbol{q}$ in terms of the detector coordinates.
For small angle scattering at UED $\mmthlf \approx \mmthd$ and $\mmphlf \approx 0$.

The primary difficulty of working with Eq.~\ref{eq:ImolGen} comes from the expectation value including both the ensemble anisotropy and molecular frame structure.
We want to separate the ensemble anisotropy into the ADMs.
This isolates the time-dependent molecular structure term that we would like to retrieve.
By doing this, we only require more tractable molecular rotation simulations with respect to the known rovibronic ground state structure in order to retrieve the time-dependent molecular structure.
Otherwise, as Eq.~\ref{eq:ImolGen} is written, it requires \textit{a priori} knowledge of exactly the unknown time-dependent structures for which we are solving.
In this work, we describe various ways to do this under common experimental conditions.

Focusing on the vibronic ground state of NO$_2$, we can separate the ADMs and molecular structure contribution in Eq.~\ref{eq:ImolGen} by applying a rigid rotor approximation:
\begin{widetext}
    \begin{equation}
        \begin{aligned}
           \langle I(\boldsymbol{q},t) \rangle_{\text{rigid}} ={}&\mathcal{I} \bigg( \sum_\mu |f_\mu(q)|^2 +  \sum_{\mu,\nu : \mu \neq \nu} \text{Re} \bigg\{ f_\mu(q) f^*_\nu(q)  \sum_l  \frac{32 \pi^3 i^l}{2l+1} \\
            &\times \sum_{m, k} (-1)^{k} \underbrace{Y^{m}_l \left( \theta^{(\text{lf})}_q, \phi^{(\text{lf})}_q \right)}_{\text{\small Lab Frame}} \ev{\underbrace{j_l(q\Delta R_{\mu\nu}) Y_l^{-k} \left( \theta_{\mu\nu}^{(\text{mf})}, \phi_{\mu\nu}^{(\text{mf})}\right)}_{\text{\small Molecular Frame Structure}}}{\Psi(0)} \underbrace{\left.\mmadm{l}{m}{k}\right|_{\text{rigid}}}_{\text{\small Anisotropy}} \bigg\} \bigg).
            \label{eq:diffRigAvg}
        \end{aligned}
    \end{equation}
\end{widetext}
Equation~\ref{eq:diffRigAvg} is the general form, which we adapt to our specific case by setting $m=0$ and replacing $\mmthlf \approx \mmthd$ and $\mmphlf \approx 0$.
The resulting lab frame measurements are shown in Fig.~\ref{fig:analysis_steps}a. 

\begin{figure*}[!htbp]
    \centering
    \includegraphics[scale=0.35]{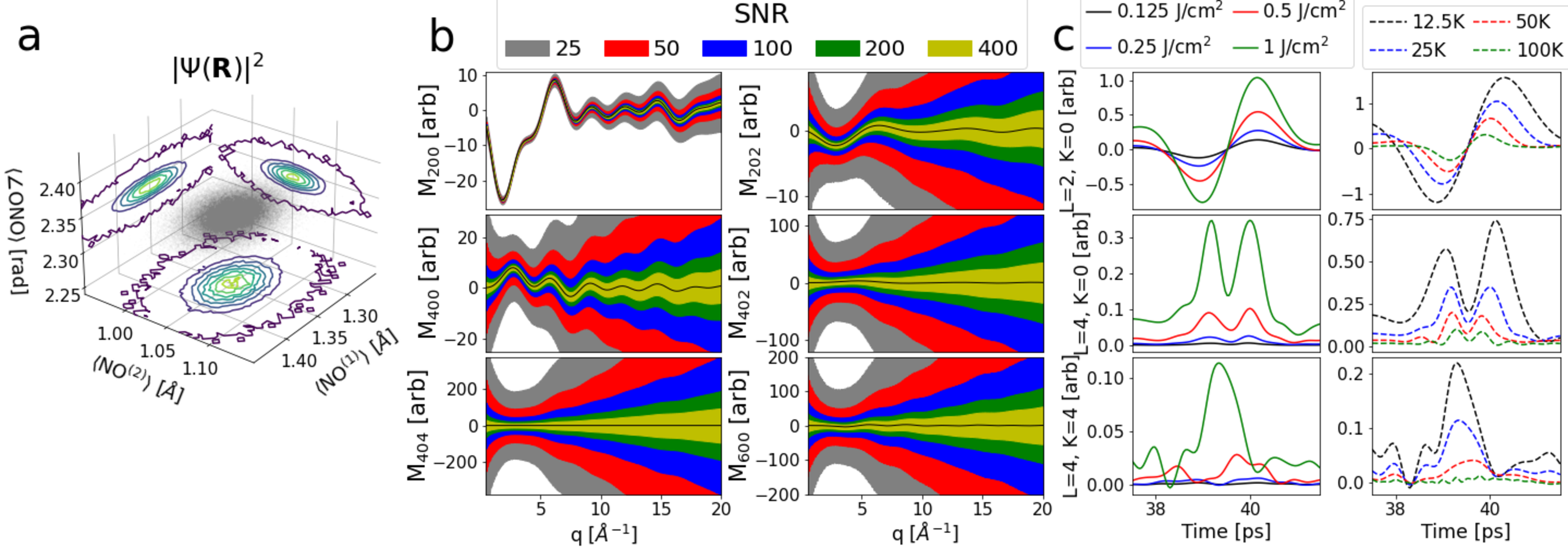}
    \caption{\textbf{Simulated NO$_2$ data at various experimental conditions} For simulated NO$_2$ we defined a $|\Psi(\boldsymbol{R})|^2$ distribution, from which we calculated the $C_{lmk}(q)$ under various experimental conditions. Panel a shows the simulated NO$_2$ distribution that we use to calculate the simulated NO$_2$ responses ($C_{lmk}(q)$ and $M_{lmk}(q)$).
    Panel b shows $M_{lmk}(q)$ for various signal-to-noise ratios (SNR) for the case of an ensemble temperature of 100~K and kick fluence of 1~J/cm$^2$. Panel c shows two ADM dependencies: pump strength (constant ensemble temperature of 25~K) on the left and temperature (constant pump fluence of 1~J/cm$^2$) on the right.}
    \label{fig:Mlmk_vs_error}
\end{figure*}

To retrieve \prth, we first isolate the molecular frame structure terms from Eq.~\ref{eq:diffRigAvg} with a series of fits.
%, but $m_1$ may be nonzero \rnc{in the general case of an asymmetric top.} since NO$_2$ is an asymmetric top.
%We show examples of the initial diffraction patterns in Fig.~\ref{fig:densityExt}(stage 1).
The first fit removes the initial diffraction beam intensity ($\mathcal{I}$), described \ref{ap:fitting_I}.
The second fit projects out the measured lab frame anisotropy $\left( Y^{m}_l \left( \theta^{(\text{lf})}_q, \phi^{(\text{lf})}_q \right)\right)$ from Eq.~\ref{eq:diffRigAvg} by fitting the angular dependence of the measured diffraction.
%The fit benefits from the orthogonality of the Legendre polynomials, since $m=0$:%, as shown in Fig.~\ref{fig:densityExt}(stage 2).
\begin{widetext}
    \begin{align}
        \mmblm ={}& \int_{0}^{\pi} \Big \langle I \Big ( \boldsymbol{q}(q,\mmthd),t \Big ) \Big \rangle_{\text{rigid}} Y_l^{m} \Big (\mmthlf \left(q,\mmthd\right), \mmphlf \left(q,\mmthd\right) \Big ) \sin \Big ( \mmthlf\left(q,\mmthd\right) \Big) d\mmthd \nonumber \\
        ={}& \mathcal{I} \sum_{\mu,\nu : \mu \neq \nu} \text{Re} \bigg\{ f_\mu(q) f^*_\nu(q) \frac{32 \pi^3 i^l}{2l+1}  (-1)^{k}  \ev{\underbrace{j_l(q\Delta R_{\mu\nu}) Y_l^{-k} \left( \theta_{\mu\nu}^{(\text{mf})}, \phi_{\mu\nu}^{(\text{mf})}\right)}_{\text{\small Molecular Frame Structure}}}{\Psi(0)} \underbrace{\left.\mathcal{A}^l_{m k}(t)\right|_{\text{rigid}}}_{\text{\small Anisotropy}} \bigg\}
        \label{eq:Blm}
    \end{align}
\end{widetext}
This yields the time $(t)$ and $q$ dependent \blm{} coefficients shown in Fig.~\ref{fig:analysis_steps}b.
The third fit isolates the molecular frame information by fitting out the time dependence of \blm{} with the simulated ADMs, \adm.
The resulting coefficients, $C_{lmk}(q)$, relate measured data to the molecular frame pairwise structure.
\begin{align}
    \begin{split}
        C_{lmk}(q)& ={} \mathcal{I} \sum_{\mu,\nu : \mu \neq \nu} \text{Re} \bigg\{ f_\mu(q) f^*_\nu(q) (-1)^{k} \frac{32 \pi^3i^l}{2l+1}  \\
        \times & \ev{\underbrace{j_l(q\Delta R_{\mu\nu}) Y_l^{-k} \left( \theta_{\mu\nu}^{(\text{mf})}, \phi_{\mu\nu}^{(\text{mf})}\right)}_{\text{\small Molecular Frame Structure}}}{\Psi(0)} \bigg\} 
    \end{split} \label{eq:Ccoeffs} \\
    M_{lmk}(q)& ={} \frac{C_{lmk}(q)}{\sum_\mu |f_\mu(q)|^2}.
    \label{eq:Mlmk}
\end{align}
Here, $M_{lmk}(q)$ are the modified $C_{lmk}(q)$ coefficients that compensate for the rapid $q^{-4}$ falloff in the electron scattering amplitudes.
Figure~\ref{fig:analysis_steps}c  shows the retrieved $M_{lmk}(q)$ for both the simulated and measured data.
For the N$_2$O data, the poor signal-to-noise precludes all contributions except $C_{200}(q)$.
%In the Fig.~\ref{fig:densityExt}(stage 3), we show the $M_{lmk}(q)$ molecular frame fit coefficients.
%The two described fits, for $B^l_m(q,t)$ and $C_{lmk}(t)$, will likely be the most important and difficult part of the analysis.
%In Supplementary Section \ref{ap:fitting} we discuss various fitting methods, best practices, and likely issues that may arise, potentially skewing one's results.
Depending on the data quality and degree of orthogonality in the ADMs, one may need to employ regularization to retrieve physical fit values.
Regularization adds a fitting cost to extraneous coefficients, thus minimizing the impact of non-orthogonal ADMs.
\ref{ap:fitting} provides a further discussion on fitting the ADMs and regularization.

The standard error of the mean $\sigma_{lmk}(q)$ for each $C_{lmk}(q)$ is calculated from a distribution of measured $C_{lmk}(q)$ coefficients.
For the N$_2$O data, \ref{ap:fit_error_prop} describes the data processing and retrieval of $\sigma_{lmk}(q)$.
For the NO$_2$ simulation, we add Poisson noise to the diffraction patterns and propagate that noise through the lab frame anisotropy and ADM fit (see Supplementary Section~\ref{ap:fit_error_prop}).

\subsection{Applying Bayesian Inference}
We approximate $|\Psi(\boldsymbol{R})|^2$ with the probability distribution \prth, which is parameterized by $\boldsymbol{\Theta}$ and conditioned on the observed $C_{lmk}(q)$ coefficients.
This requires one to choose a functional form of \prth{} dependent on the system's state and the desired degree of accuracy.
Depending on the desired accuracy and precision of the desired results, this requires varying degrees of \textit{a priori} knowledge.
For example, one may choose a multivariate delta function for a single molecule response, a normal distribution to model the ground vibrational states, or harmonic oscillator eigenfunctions to describe arbitrary individual vibrational states.
%The delta distribution is analogous to calculating the diffraction pattern from a single structure and comparing it to one's measurement.
%This is important for intermediate results and in some cases may be the only tractable solution.
%Here we explicitly write out \prth{} and the $\boldsymbol{\Theta}$ parameters
\begin{widetext}
\begin{align}
    \mmprth &\approx \left| \Psi \left(\boldsymbol{R} \right) \right|^2\\
    \mmprthd &= \delta\left(\boldsymbol{\Theta}^{(\delta)} - \boldsymbol{R}\right) \\
    \boldsymbol{\Theta}^{(\delta)} &= \left[ \mmmrnoa, \mmmrnob, \mmmaono \right] \\
    \mmprthg &= \frac{1}{\sqrt{2\pi}^{N_{\text{dof}}} \prod^{i<N_{\text{dof}}}_{i=0} \boldsymbol{\Theta}^{(\mathcal{N})}_{2i+1}}  \exp \Bigg \{ \frac{-1}{2}  \sum_{i=0}^{i<N_{\text{dof}}} \left( \frac{\boldsymbol{\Theta}^{(\mathcal{N})}_{2i} - \boldsymbol{R}_i}{\boldsymbol{\Theta}^{(\mathcal{N})}_{2i+1}} \right)^2 \Bigg \} \\
    \boldsymbol{\Theta}^{(\mathcal{N})} &= \left[ \mmmrnoa, \mmsrnoa, \mmmrnob, \mmsrnob, \mmmaono, \mmsaono  \right] \label{eq:post_gauss}
\end{align}
\end{widetext}
The $\boldsymbol{\Theta}$ parameters include the $3N_{\text{atom}}-6$ structural degrees of freedom $(N_{\text{dof}})$ needed to define a unique molecular structure, and the width parameters in the case of \prthg.
Here, $\boldsymbol{\Theta}$ has the minimal number of parameters needed to define \prth, and adding redundant parameters can significantly alter one's results.
%, as shown in Supplementary Section~\ref{ap:extra_dof}.

Having isolated the molecular frame structure terms $(C_{lmk}(q))$ and chosen \prth, we apply Bayesian Inference to address the diffraction inverse problem \cite{Hegazy.dissertation.2023, box.bayesian_inference.2011, foreman.mcmc.2013} by effectively inverting $C_{lmk}(q)$ to approximate $|\Psi\left(\boldsymbol{R}\right)|^2$.
With Bayes rule,
\begin{equation}
    \mmptheta = \frac{P\left(C | \boldsymbol{\Theta}\right) P\left( \boldsymbol{\Theta} \right)}{P(C)}
    \label{eq:Bayes_rule}
\end{equation}
we use the statistical nature of our measurement to analytically relate the desired $\boldsymbol{\Theta}$ parameters to the measured $C_{lmk}(q)$.
In Eq.~\ref{eq:Bayes_rule}, \ptheta{} is the posterior distribution we wish to build.
%Here, \ptheta{} is the posterior distribution we wish to build, $P\left(C | \boldsymbol{\Theta}\right)$ is the likelihood distribution which relates $\boldsymbol{\Theta}$ to the measured data, $P\left(\boldsymbol{\Theta}\right)$ is the prior, and $P(C)$ is the marginal likelihood.
The likelihood $P\left(C | \boldsymbol{\Theta}\right)$ relates the measured data to the $\boldsymbol{\Theta}$ parameters and is the probability of observing $C_{lmk}(q)$ given the parameters $\boldsymbol{\Theta}$
\begin{equation}
    \begin{split}
        &P\left(C | \boldsymbol{\Theta}\right) = \left[ \prod_{lmk,q} \frac{1}{\sigma_{lmk}(q)\sqrt{2\pi}}\right] \\
        &\times \exp{\frac{-1}{2} \sum_{lmk,q} \left(\frac{C_{lmk}(q) - C^{(\text{calc})}_{lmk}(q,\boldsymbol{\Theta})}{\sigma_{lmk}(q)} \right)^2}.
    \end{split}
    \label{eq:MHlikelihood}
\end{equation}
Here, $C^{(\text{calc})}_{lmk}(q,\boldsymbol{\Theta})$ are the calculated $C_{lmk}(q)$ coefficients, and $\sigma_{lmk}(q)$ are the standard errors of the means for $C_{lmk}(q)$.
The prior, $P\left(\boldsymbol{\Theta}\right)$ contains our \textit{a priori} knowledge of the system, and in this work is used to constrain $\boldsymbol{\Theta}$ to physicality (e.g., $\boldsymbol{\Theta} > 0 \text{ and } \angle\text{ONO} < \pi$).
This is because we do not assume any prior knowledge or simulations of the system.
Calculating the marginal likelihood $P(C)$ is generally, and in our case, intractable.
Further intuition regarding how the statistical nature of our measurement allows us to invert for $\boldsymbol{\Theta}$ is described in Ref.~\cite{Hegazy.dissertation.2023}. 

Given the functional forms of \ptheta, $P\left(C | \boldsymbol{\Theta}\right)$, and the presumed functional form of \prth, we now find the globally optimal $\boldsymbol{\Theta}$ parameters (\thopt) by building \ptheta{} and finding its mode.
%The \thopt{} parameters provide the best agreement between $C_{lmk}(q)$ and $C^\text{calc}_{lmk}(q)$, which is therefore the mode of the posterior and likelihood distributions.
%Generally, the mean of the marginalized \ptheta{} distributions and \thopt{} will disagree due to correlations between $\boldsymbol{\Theta}$ parameters within \ptheta.
To converge on the mode of \ptheta, one must use the correlations between the $\boldsymbol{\Theta}$ parameters by building \ptheta{} in the full $\boldsymbol{\Theta}$-space rather than sampling each parameter individually.
Consequently, we must next address the curse of dimensionality.

\ifthenelse{\equal{\natcommphys}{0}}{
    \begin{figure}[!htbp]
        \centering
        \includegraphics[scale=0.38]{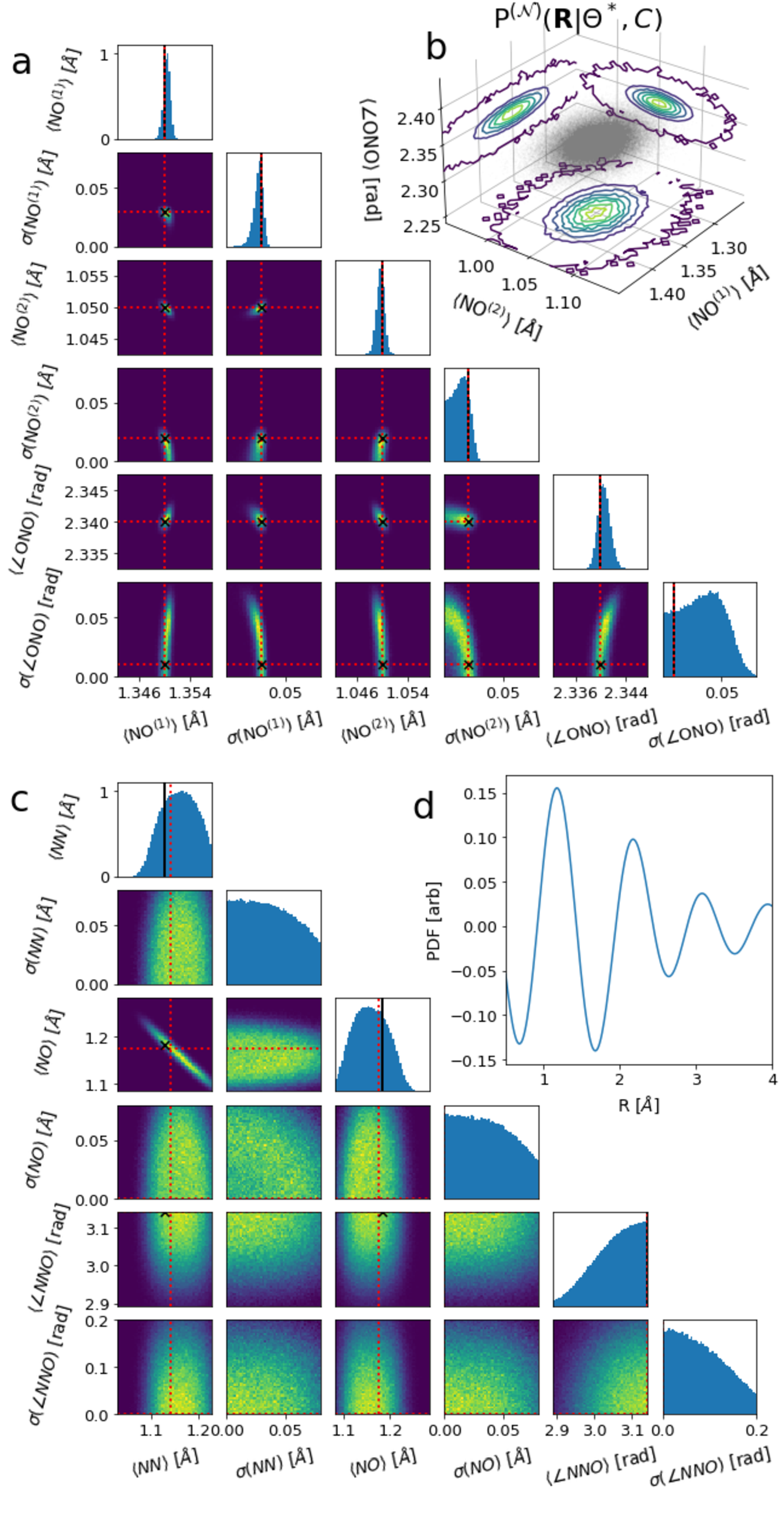}
        \caption{\textbf{Retrieving \pthetag, \prthg, and the molecular structure parameters} We successfully retrieve the multivariate posterior \pthetag{} for NO$_2$ and N$_2$O from which we find \thopt. The axes of panels a and b are the $\boldsymbol{\Theta}$ parameters: the mean and standard deviations of the pairwise distances and angles that define \prthg. Panel a shows the 1d and 2d projections of \pthetag{} distributions for the simulated NO$_2$ response. The recovered \prthoptg{} (panel b) is what we compare to the simulated $|\Psi(\boldsymbol{R})|^2$ in Fig.~\ref{fig:Mlmk_vs_error}a. The red dashed lines indicate the retrieved mode (\thopt), while the black ``x" and solid black lines indicate the ground truth, respectively. Panel c shows the 1d and 2d projections \pthetag{} distributions for N$_2$O data, though only using the $C_{200}(q)$ contribution. The black ``X" and solid black lines indicate previously measured values for N$_2$O~\cite{herzberg.structure_spectroscopy.1966, Teffo.n2o_structure.1989}. For comparison, panel d shows the simulated Pairwise Distribution Function (PDF) from the same $q$ range.}
        \label{fig:response_gauss}
    \end{figure}
    
    \begin{figure*}[!htbp]
        \centering
        \includegraphics[scale=0.38]{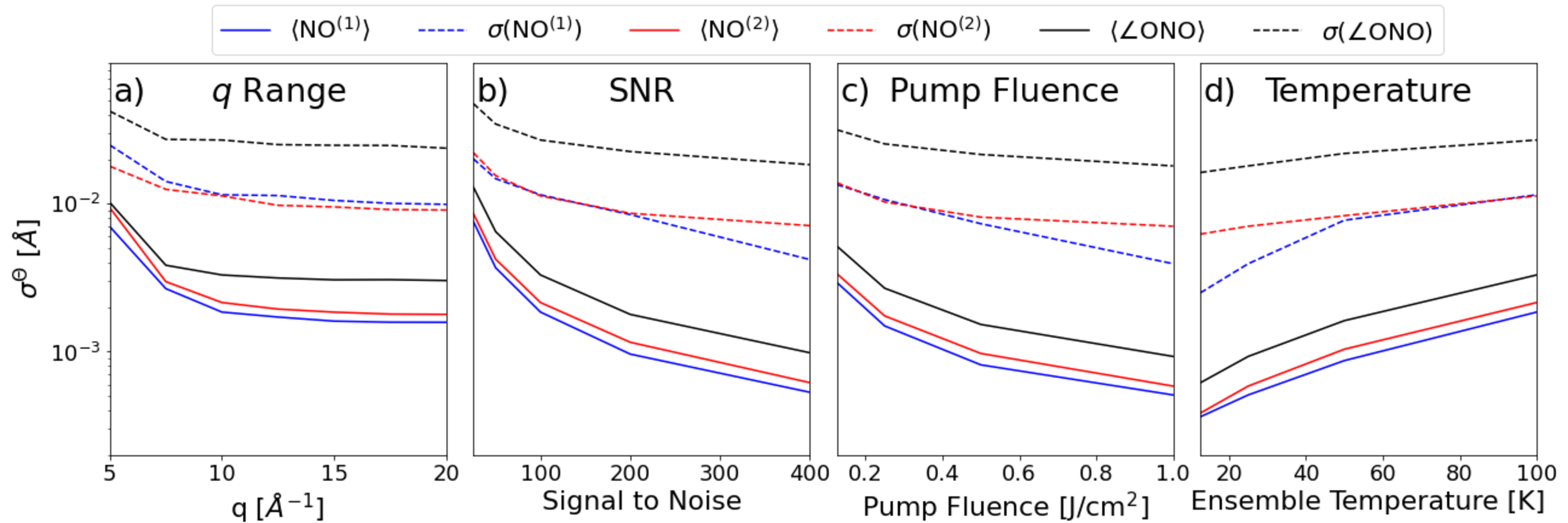}
        \caption{\textbf{The effects of various experimental parameters on \pthetag} Varying experimental parameters affects the resolution (width) of \pthetag, but our method is most sensitive to the measured signal-to-noise ratio (SNR). Panel a shows how the uncorrelated widths of \pthetag, denoted by \sth, change by increasing the $q$ range. Panel b similarly shows the dependence of \sth{} versus SNR. Panel c shows the dependence of \sth{} versus pump fluence (width of the rotational wavepacket) at 25~K. Panel d shows the dependence of \sth{} versus the molecular ensemble temperature at a constant pump fluence of 1~J/cm$^2$.}
        \label{fig:trends}
    \end{figure*}
}
    
\subsection{Solving for the high dimensional model parameters $\boldsymbol{\Theta}$}
%\paragraph*{Bayesian inference}

We retrieve \ptheta{} with the Metropolis-Hastings algorithm (MHA) from the following system of equations:
\begin{align}
    C_{lmk}(q) &= \int H_{lmk} \left( q, \boldsymbol{R} \right) \left| \Psi \left(\boldsymbol{R}\right) \right|^2 d\boldsymbol{R} \label{eq:intEqn}\\
    C_{lmk}^{(\text{calc})}(q, \boldsymbol{\Theta}) &= \int H_{lmk} \left( q, \boldsymbol{R} \right) \mmprth d\boldsymbol{R} \label{eq:intEqnSim}\\
    \begin{split}
        H_{lmk} \left( q, \boldsymbol{R} \right) &= \mathcal{I} \text{Re} \bigg\{ (-1)^{k} \frac{32 \pi^3 i^l}{2l+1} \\
        \times \sum_{\mu,\nu : \mu \neq \nu} |f_\mu(q)| &|f_\nu(q)| j_l(q\Delta R_{\mu\nu}) Y_l^{-k} \left( \theta_{\mu\nu}^{(\text{mf})}, \phi_{\mu\nu}^{(\text{mf})}\right)\bigg\}.
    \end{split} \label{eq:MHAfit}
\end{align}
We note the high dimensionality and complexity of Eq.~\ref{eq:intEqnSim}, which is a system of order 10 equations, each with order 100 terms, embedded in an order 100-dimensional space of measurements in $q$.
This must be evaluated on a $\boldsymbol{\Theta}$-dimensional space of all possible molecular structures and width parameters.
The MHA is chosen for its ability to retrieve probability distributions from high dimensional integral equations \cite{hastings.metropolis_hastings.1970, foreman.mcmc.2013} like Eq.~\ref{eq:intEqnSim}.

The MHA is designed to efficiently and preferentially sample regions of $\boldsymbol{\Theta}$-space  proportional to the agreement with data, spending the vast majority of its time sampling regions of high probability (agreement).
The MHA builds \ptheta{} by accumulating $\boldsymbol{\Theta}$ parameters based their relative posteriors
\begin{equation}
    \frac{P \left(\boldsymbol{\Theta}' | C\right)}{\mmptheta} = \frac{P \left(C | \boldsymbol{\Theta}' \right)}{P \left(C | \boldsymbol{\Theta}\right)} \label{eq:posterior_ratio}
\end{equation}
where $\boldsymbol{\Theta}$ and $\boldsymbol{\Theta}'$ are both physical, and the prior and the marginal likelihood cancel out.
We note Eq.~\ref{eq:posterior_ratio}, and hence the MHA, is theory independent and is analogous to a random walk guided by the relative agreement of neighboring $\boldsymbol{\Theta}$ parameters to the data.
For instance, if the likelihood of $\boldsymbol{\Theta}$ is 2 times larger than $\boldsymbol{\Theta}'$, the MHA will sample twice as many structures around $\boldsymbol{\Theta}$ than $\boldsymbol{\Theta}'$.
Similarly, if the likelihood for $\boldsymbol{\Theta}$ is 1000 times larger than for $\boldsymbol{\Theta}'$, then the MHA will effectively remove structures around $\boldsymbol{\Theta}'$ from the search space. 
Reference~\cite{foreman.mcmc.2013} 
The MHA python package \cite{foreman.mcmc.2013} 
used in this work and Ref.~\cite{Hegazy.dissertation.2023} give detailed descriptions of combining Bayesian Inference and the MHA.
\ref{ap:mcmc} describes our use of the MHA and Bayesian Inference in greater detail and how one can introduce physical intuition, or \textit{a priori} knowledge, into the MHA.

%To investigate these effects we use the $C_{lmk}(q)$ from Fig.~\ref{fig:Mlmk_vs_error}b with a $q$ range of [0.5, 10]~\iang and a SNR of 100 while using the ADMs in Fig.~\ref{fig:Mlmk_vs_error}c to calculate $\sigma_{lmk}(q)$.
%For all other results the the pump fluence is 1~J/cm$^2$ and the ensemble temperature is 100~K.

%So far we've discussed simulations of NO$_2$, we also performed an experiment with N$_2$O at UED in order to validate this method.
%We induced the rotational wavepack by using a train of 8 identical pulses of 800~nm light, each separated by the full quantum revival of the induced rotational wavepacket.
%We measured the first field free full quantum revival over a window of \textapprox3~ps.
%We masked out the $q$ regions of $[0,3.5]$~\iang{} due to ellipticity in the imaging of the diffraction pattern.
%Unfortunately the poor SNR precluded all contributions except $C_{200}(q)$ and signal above 7.25~\iang{}.
%Both the $B^0_2(q,t)$ and $C_{200}(q)$ coefficients are shown in Fig.~\ref{fig:analysis_steps}.
%Supplementary Section~\ref{ap:fit_error_prop} describes the data processing and retrieval of $\sigma_{lmk}(q)$.

This method ultimately yields the following three results; a distribution of $\boldsymbol{\Theta}$ parameters (the posterior \ptheta), the optimal set of model parameters (\thopt), and a parameterized probability of molecular structures \prthopt.
For each individual $\boldsymbol{\Theta}$ parameter, where the $i^\text{th}$ parameter is denoted as $\Theta_i$, we calculate its resolution as the standard deviation of the projection of \ptheta{} onto $\Theta_i$.
This resolution, \sth, is the one-dimensional standard deviation after marginalizing over all other parameters, which removes the correlations between $\boldsymbol{\Theta}$ parameters.
That is, if one randomly draws some parameters $\boldsymbol{\Theta}$ from \ptheta, the distribution of parameter $\Theta_i$ will have a width of \sth.
In this work, we focus on how Bayesian Inference and Eq.~\ref{eq:ImolGen} effectively invert data for \prthopt{} via an unambiguous and sharp \ptheta.
It is this \ptheta{} and its width (resolution) that are our figures of merit for the inversion.
The accuracy of \thopt{} depends on one's method for finding the mode, of which there are many methods.
The precision of \thopt{} is a function of its local region.
The mean and mode of said marginalized distribution will likely not correspond to \thopt, since \thopt{} is the mode of the full $\boldsymbol{\Theta}$-space distribution.
We find \thopt{} via a simple mode search algorithm described in \ref{ap:mode_search}.

The measured $q$ range, the induced rotational wavepacket, and the $\sigma_{lmk}(q)$ are vital in determining the width, shape, and parameter correlations of \ptheta.
To investigate such dependencies we first define a $|\Psi(\boldsymbol{R})|^2$ distribution for NO$_2$ to calculate $C_{lmk}(q)$.
Figure~\ref{fig:Mlmk_vs_error}a and Table~\ref{tab:mcmcValidation} show and describe this distribution, respectively. 
Measuring more diffraction patterns increases the signal-to-noise ratio (SNR) by reducing $\sigma_{lmk}(q)$ which scales as $1/\sqrt{N}$.
Here, the SNR is the geometric mean of $C_{000}(q)/\sigma_{000}(q)$ between $0.5 < q < 4$~\iang.
Figure~\ref{fig:Mlmk_vs_error}b illustrates the $C_{lmk}(q)$ coefficients used in this analysis with the following SNRs based on previous UED \cite{Wolf.CHD.2019} and x-ray \cite{Ware.TRXS.2019} diffraction experiments.
Unless otherwise stated, the standard configuration of experimental parameters for our NO$_2$ results is a $q$ range of [0.5, 10]~\iang, a SNR of 100, a pump fluence of 1~J/cm$^2$ and a 100~K ensemble temperature.

%% file: sections/04_results.tex
\ifthenelse{\equal{\natcommphys}{0}}{
    \begin{figure}[!htbp]
        \centering
        \includegraphics[scale=0.33]{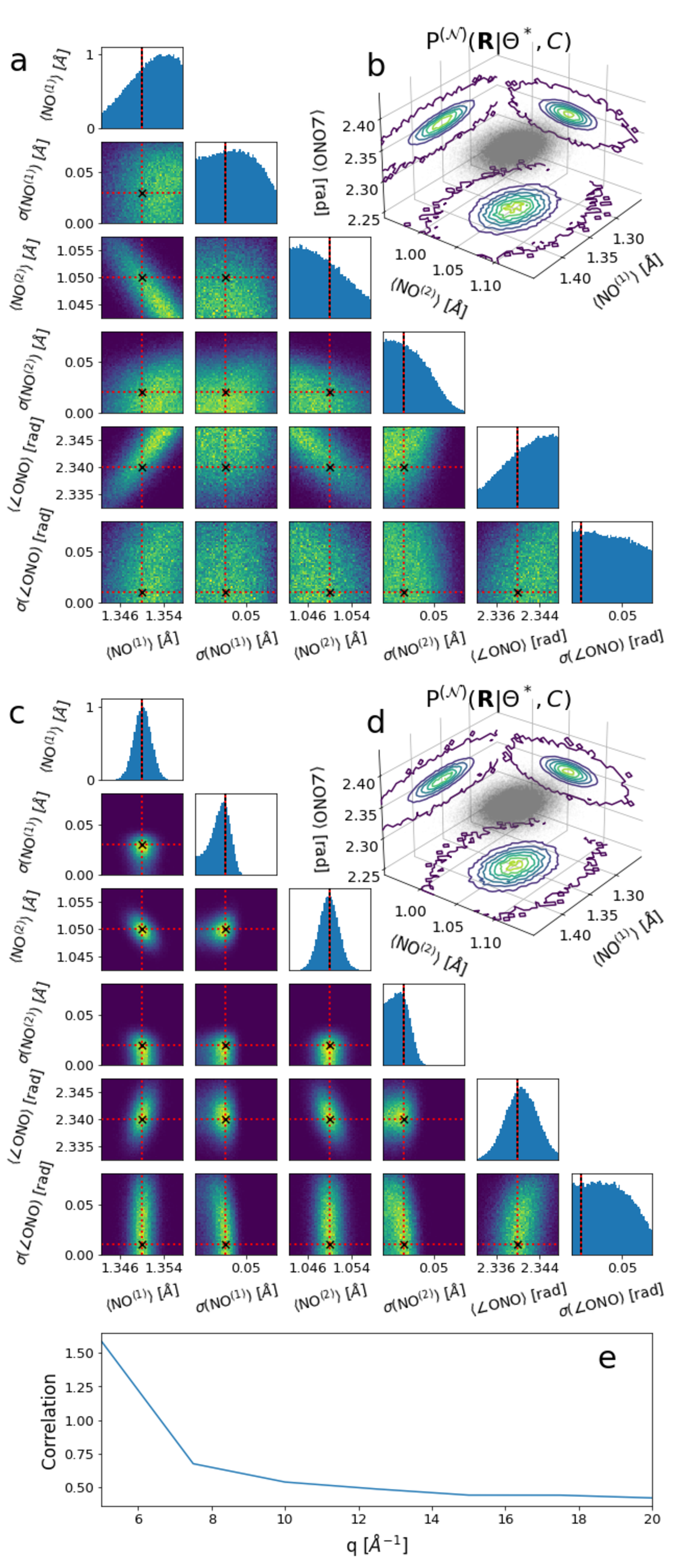} %0.36/0.31
        \caption{\textbf{Effects of varying the measured $q$ range on \pthetag} Varying the measured $q$ range affects false correlations in \pthetag{} for NO$_2$; a larger reciprocal space provides more information and dampens false correlations. Panel a shows the 1d and 2d projections of \pthetag{} for a limited $q$ range of $[0.5,5]$~\iang. The red dashed lines illustrate \thopt, while the black ``X" and solid lines indicate the ground truth values. Panel b shows the corresponding \prthoptg. Similarly, panel c shows the 1d and 2d projections of \pthetag{} for the broader $q$ range of $[0.5,20]$~\iang. Panel d shows the corresponding \prthoptg. Panel e shows the correlation between all $\boldsymbol{\Theta}$ parameters as a function of $q$ range. We note the decrease in correlations with larger $q$, where panels a and b illustrate how the width and false correlations in \pthetag{} decrease with higher $q$.}
        \label{fig:compare_qrange_correlations}
    \end{figure}
    
    \begin{figure}[!htbp]
        \centering
        %0.28
        \includegraphics[scale=0.28]{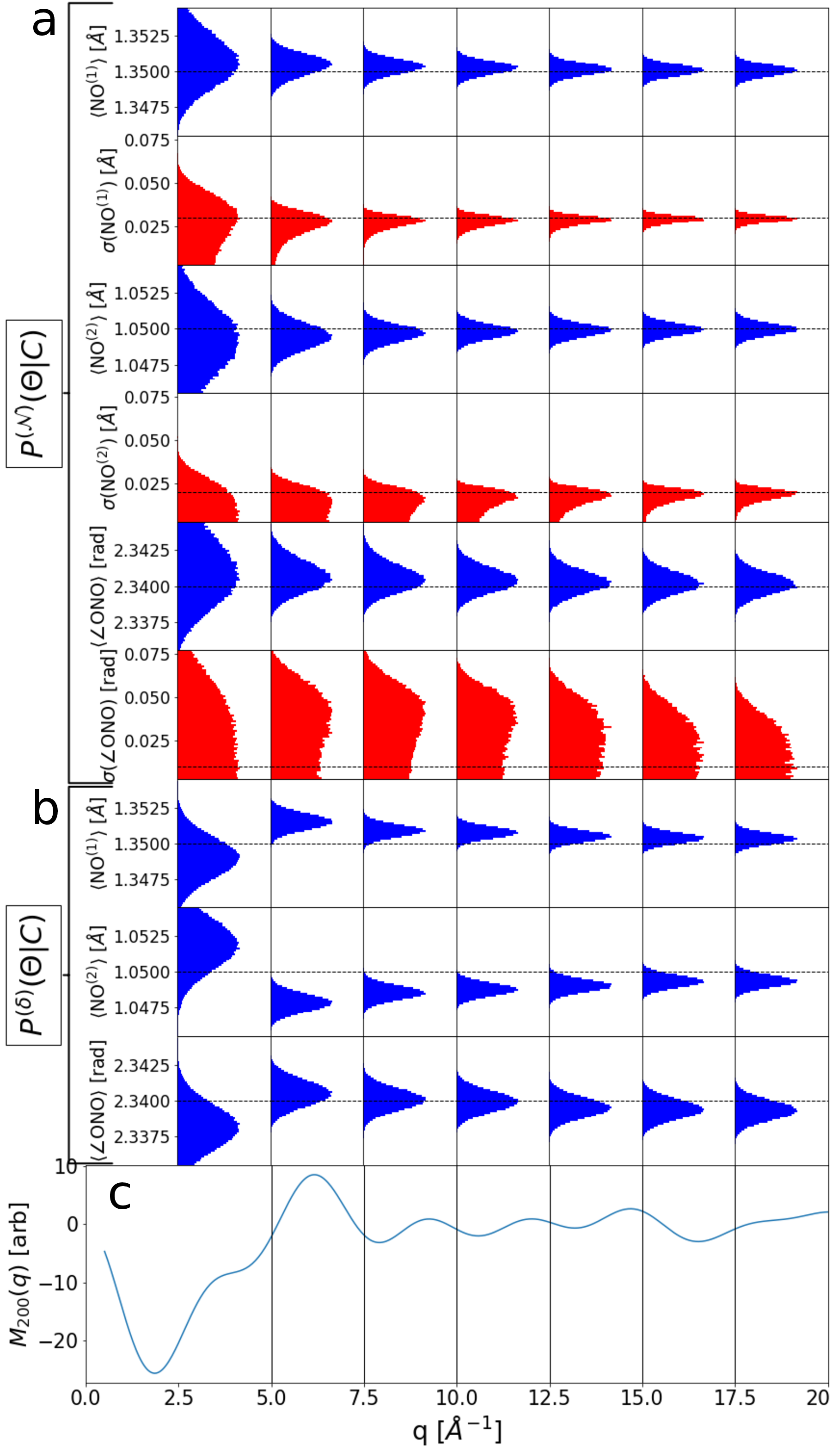}
        \caption{\textbf{Systematic errors from selecting incorrect $|\Psi(\boldsymbol{R})|^2$ distributions}The \pthetad{} distribution suffers from a $q$ dependent systematic error stemming from the false assumption that a single structure describes the results measured from an ensemble. Here we show the 1d projections of \pthetag{} (section a) and \pthetad{} (section b) as a function of the measured $q$ range (panel c) and a signal-to-noise ratio (SNR) of 400. Each column indicates a different $q$ range starting at 0.5~\iang{} with the end of said $q$ range indicated by the rightmost border of that column. The dashed lines are the ground truth values. The bottom plot in panel c is the simulated $C_{200}(q)$ coefficient used for both posteriors and is intersected by black lines that indicate the upper $q$ range of each column.}
        \label{fig:theta_peaks_q}
    \end{figure}
}

Both the simulated NO$_2$ and measured N$_2$O diffraction patterns are from the SLAC UED facility.
Elastic electron diffraction is sensitive to the nuclei and diffraction from electronic transience occurs within the removed low $q$ region.
Using the independent atom approximation we are only concerned with the nuclear structure.
Our stretched NO$_2$ molecule is simulated in the ground vibrational state due to its altered structure and we observe that 99.99\% of the N$_2$O molecules occupy the vibrational ground state (\ref{ap:n2o_thermal}).
The normal distribution, \prthg, is a good description of both our NO$_2$ and N$_2$O vibronic ground state systems as it is the ground state eigenfunction of the harmonic oscillator.
For N$_2$O, our ADM simulations account for centrifugal distortion.
In our main result, we illustrate our method's efficacy by retrieving \prthoptg{} from both simulated NO$_2$ and measured N$_2$O $C_{lmk}(q)$ coefficients.
After, we further investigate our method's behavior and sensitivity to varying experimental conditions for the simulated NO$_2$ system.
Finally, we observe how our Bayesian Inference method significantly improves real-space resolution.

\subsection{Molecular structure distribution retrieval}

To retrieve \prthoptg, we first built the posterior \pthetag, shown in Fig.~\ref{fig:response_gauss} for simulated NO$_2$ (a) and measured N$_2$O data (c).
Panels b and d show \prthoptg{} for NO$_2$ and the simulated PDF for N$_2$O, respectively.
Tables~\ref{tab:mcmcValidation} and \ref{tab:n2o_results} give the extracted \thopt{} (the most probable $\boldsymbol{\Theta}$ parameters) and \sth, respectively, for N$_2$O and NO$_2$.
For the NO$_2$ simulation, the SNR is 400.
For NO$_2$, \pthetag's resolution (\sth) for the nuclear distances and angles is \textapprox0.5~m\AA{} and fully encompasses the ground truth values.
Despite the largely flat \maono{} distribution, \thopt{} still converges on the ground truth values.
For N$_2$O data, the retrieved \pthetag{} encompasses the previously measured results of the vibronic ground state \cite{herzberg.structure_spectroscopy.1966, Teffo.n2o_structure.1989}.
The resolution of this distribution is of order 10~m\AA{} even with our limited $q$ range of $[3.5,7.25]$~\iang{} and the very poor SNR.
Moreover, the retrieved $\ev{\angle \text{N}\text{N}\text{O}}$ is $\pi$ and we resolve the \textapprox50~m\AA{} difference between the N$^\text{T}$N$^\text{C}$ and N$^\text{C}$O bond distances (Table~\ref{tab:n2o_results}).
The retrieved widths $\sigma \left( \text{N}^{\text{T}}\text{N}^{\text{C}} \right)$ and 
$\sigma \left( \angle \text{N}\text{N}\text{O}\right)$ are unphysical due to the limited $q$ range, as discussed later.
Compared to the PDF (Fig.~\ref{fig:response_gauss}d), with a \textapprox2~\AA{} Fourier resolution, this method improves resolution by a factor of 50.
In the PDF, the missing low and high $q$ components produce ringing artifacts in this inverse Fourier transform because of the incomplete Fourier space.
This confuses the PDF results as they are not positive definite and falsely indicate population at large distances.

We observe (Fig.~\ref{fig:response_gauss}a and c) that \thopt{} does not correspond to the mean or mode of most 1-dimensional projections of \pthetag.
This is due to the nonlinearity and correlations of \ptheta{} in $\boldsymbol{\Theta}$ space.
%Tracing over variables removes correlations that are important to accurately finding \thopt.
This illustrates the importance of finding \thopt{} in this correlated space since the structure parameters are indeed correlated.

\begin{table}[!htbp]
    \centering
    \begin{tabular}{|c|c|cc|cc|} \hline
         \multirow{2}{*}{$\boldsymbol{\Theta}$ Parameters} &  \multirow{2}{*}{Input} & \multicolumn{2}{c|}{\pthetag} & \multicolumn{2}{c|}{\pthetad} \\
         &  & \thopt & $\sigma^{\Theta}$ & \thopt & $\sigma^{\Theta}$ \\ \hline 
         \mrnoa [\angs] & 1.35 & 1.3500 & 0.0005 & 1.3509 & 0.0004 \\
         \srnoa [\angs] & 0.03 & 0.030 & 0.004 & -- & -- \\
         \mrnob [\angs] & 1.05 & 1.0500 & 0.0006 & 1.0485 & 0.0005\\
         \srnob [\angs] & 0.02 & 0.020 & 0.007 & -- & --\\
         \maono [rad] & 2.34 & 2.340 & 0.001 & 2.3401 & 0.0007 \\
         \saono [rad] & 0.01 & 0.01 & 0.02 & -- & -- \\
         \hline
    \end{tabular}
    \caption{\textbf{Retrieved molecular frame structure parameters for simulated NO$_2$} Our approximation of $|\Psi(\boldsymbol{R})|^2$ (\prth) is parameterized by molecular frame distances, angles, and their corresponding widths ($\boldsymbol{\Theta}$ parameters). The optimal parameters, denoted as \thopt, correspond to the mode of \ptheta. We provide the retrieved \thopt{} parameters along with their corresponding resolutions for the simulated NO$_2$. The input $\boldsymbol{\Theta}$ parameters are those used to simulate the NO$_2$ $C_{lmk}(q)$ coefficients with a signal-to-noise ratio (SNR) of 400. The retrieved \thopt{} parameters are those found when applying \prthg{} and \prthd{} to the NO$_2$ $C_{lmk}(q)$ simulated using \prthg. The \sth{} values are the resolution of \thopt{} and the uncorrelated widths of \pthetag{} and \pthetad, respectively.}
    \label{tab:mcmcValidation}
\end{table}
\begin{table}[!htbp]
    \centering
    \begin{tabular}{|c|ccc|} \hline
         & $\mmthopt_{\text{Literature}}$ & \thopt & $\sigma^{\Theta}$ \\ \hline 
         $\ev{\text{N}^{\text{T}}\text{N}^{\text{C}}}$ [\angs] & 1.128 & 1.14 & 0.04 \\
         $\sigma \left( \text{N}^{\text{T}}\text{N}^{\text{C}} \right)$ [\angs] &  & 0.08 & 0.03 \\
         $\ev{\text{N}^{\text{C}}\text{O}}$ [\angs] & 1.184 & 1.18 & 0.04 \\
         $\sigma \left(\text{N}^{\text{C}}\text{O}\right)$ [\angs] &  & 3$\times 10^{-8}$ & 0.03 \\
         $\ev{\angle \text{N}\text{N}\text{O}}$ [rad] & 3.142 & 3.14 & 0.06 \\
         $\sigma \left( \angle \text{N}\text{N}\text{O}\right)$ [rad] &  & 6$\times 10^{-12}$ & 0.06 \\
         \hline
    \end{tabular}
    \caption{\textbf{Retrieved molecular frame structure parameters for measured N$_2$O} We provide the optimal molecular frame pairwise distance and angle parameters (\thopt) for the measured N$_2$O dataset. The \thopt{} parameters correspond to the mode of \ptheta. The resolution of \thopt{} (\sth) is the standard deviation of the 1d uncorrelated projection of \pthetag. We also provide the corresponding literature values for the vibronic ground state \cite{herzberg.structure_spectroscopy.1966, Teffo.n2o_structure.1989}, denoted as $\mmthopt_{\text{Literature}}$.}
    \label{tab:n2o_results}
\end{table}

\subsection{Exploring experimental effects and systematics}

The measured $q$ range is a critical component of gas-phase ultrafast diffraction, determining the information content and the PDF's resolution.
When expanding this range, Figs.~\ref{fig:trends}a and ~\ref{fig:compare_qrange_correlations}, we observe resolution (\sth) improvements only until \textapprox8~\AA$^{-1}$, after which it plateaus.
This indicates that after a modest $q$ range our method is not very sensitive to further increases.
The false correlations between $\boldsymbol{\Theta}$ parameters (Fig.~\ref{fig:compare_qrange_correlations}e), still, continue to decline as we increase this range.
The plotted correlation in Fig.~\ref{fig:compare_qrange_correlations}e is between all 6 $\boldsymbol{\Theta}$ parameters.
The correlations seen in Figs.~\ref{fig:compare_qrange_correlations}a and c are termed false correlations since the simulated $|\Psi(\boldsymbol{R})|^2$ is a multivariate normal distribution with a diagonal covariance matrix. 
Increasing the measured reciprocal range $q$ provides more information about the system and reduces these correlations, seen in Figs.~\ref{fig:compare_qrange_correlations}a, c, and e.
%Figures~\ref{fig:compare_qrange_correlations}a and c illustrate how the false correlations diminish from a $q$ range of [0.5, 5]~\iang{} to [0.5, 20]~\iang, respectively, under conditions of SNR of 100 and a fluence of 1~J/cm$^2$.

When varying the SNR, Fig.~\ref{fig:trends}b, \sth{} rapidly decreases with increasing SNR.
Increasing SNR by an order of magnitude decreases \sth{} by an order of magnitude for pairwise distances and angles.
This strong and continuous dependence indicates that our method is sensitive to SNR due to our statistical interpretation.
Although \pthetag{} becomes more peaked, the general shape from the correlations does not change since higher SNR improves resolution but does not add more information, in terms of the $q$ range.

Increasing the induced rotational coherence and lowering the ensemble temperature rapidly improves resolution (Fig.~\ref{fig:trends}c and d) similar to increasing SNR.
In Fig.~\ref{fig:trends}c, the gas was at 25~K while varying the rotational coherence.
In Fig.~\ref{fig:trends}c, the pump fluence was 1~J/cm$^2$ while varying the ensemble temperature.
Increasing the rotational coherence and decreasing the temperature increases the magnitude and complexity of the ADMs (Fig.~\ref{fig:Mlmk_vs_error}c).
This is because higher average pump fluences induce larger rotational coherence and lowering the ensemble temperature diminishes the spread of initial rotational states that incoherently interfere.
The result is an increase in signal, a larger SNR, and consequently the similarly continuous behavior in Fig.~\ref{fig:trends}b.

Generally, when varying the $q$ range, SNR levels, pump fluence, and ensemble temperature we find the pairwise distances' \sth{} to be of order 1~m\AA; for the width parameters, \sth{} is order 10~m\AA.
Our retrieved \thopt{} values are generally within a relative error of \textapprox$10^{-7}$ and \textapprox$10^{-3}$ from the ground truth values for structural and width parameters, respectively.
This resolution is often \textapprox100 times better than PDF-based methods because our statistical treatment is highly sensitive to SNR.

Aside from experimental parameters, we investigate systematics induced by incorrectly selecting the functional form of \prth.
We assert the simulated NO$_2$ vibronic ground state $|\Psi(\boldsymbol{R})|^2$ distribution is a multivariate normal distribution (Fig.~\ref{fig:Mlmk_vs_error}a).
We evaluate both \pthetag{} and \pthetad{} on this simulation, and in Fig.~\ref{fig:theta_peaks_q} we compare their 1d projections as a function of $q$ range.
%With \prthd, we assume a single molecule response can describe a signal averaged over an ensemble of structures.
%In Fig.~\ref{fig:theta_peaks_q} we compare the 1d projections of \pthetag{} and \pthetad{} as a function of $q$ range.
The \pthetag{} distribution consistently encompasses the correct values, but the \pthetad{} distribution fails to do so for $q$ ranges of [0.5,7.5], [0.5,10], and [0.5,12.5]~\iang.
This is because \prthd{} assumes a single molecule response can describe a signal averaged over an ensemble of structures.
With increasing $q$ ranges, \pthetad{} converges in an unstable fashion on the ground truth (Fig.~\ref{fig:theta_peaks_q}b), unlike the smooth convergence in \pthetag.
We note that for NO$_2$, retrieving \pthetad{} is \textapprox100 times faster than \pthetag, which respectively take order 10~s to 1~minute and 1~hour to 1~day on 10~CPUs.
This is because \prthd{} doesn't have to sum over structures in Eq.~\ref{eq:intEqnSim}.
\ref{ap:delta_dist} and Ref.~\cite{Hegazy.dissertation.2023} provides plots and further discussion of these results.

\subsection{Effects of Bayesian Inference}

Our method retrieves the labeled pairwise distances with \textapprox100 times better resolution than the PDF.
This is due to our statistical treatment using Bayesian Inference where each $lmk$ and $q$ contribution is itself an independent probability distribution; each is an experiment of its own.
The MHA discrimination power grows exponentially with more $C_{lmk}(q)$, which increases the magnitude of the negative exponent in the relative ratio of likelihood functions $P(C|\boldsymbol{\Theta})$ (Eq.~\ref{eq:MHlikelihood}).
%That is, with more measurements Eq.~\ref{eq:MHlikelihood} becomes smaller and \ptheta{} becomes sharper.
Our method therefore heavily relies on $\sigma_{lmk}(q)$ and $C_{lmk}(q)$ (seen in Fig~\ref{fig:trends}b.
Statistical noise increases $\sigma_{lmk}(q)$, making \ptheta{} wider (Fig~\ref{fig:trends}b), while systematic errors in $C_{lmk}(q)$ shift the centriod of \ptheta{} (Fig~\ref{fig:response_gauss}c).
\ref{ap:fit_error_prop} describes our method for consistently accounting for both statistical and systematic errors.
%The PDF approach, alternatively, takes the inverse Fourier transform of the contiguous $q\,M_{000}(q)$.
%\rnc{check below}
The PDF error adds in quadrature in $\sigma_{lmk}(q)$; its scale is set by the largest error bar and disproportionately suffers from poorly measured data points.
Conversely, MHA amplifies the contribution of high precision measurements while reducing contributions from poorly measured data points by weighting each term in the likelihood by $1/\sigma_{lmk}(q)$ (Eq.~\ref{eq:MHlikelihood}).

Our Bayesian Inference approach expands the utility of gas-phase ultrafast diffraction to measure previously inaccessible variables.
Given \prth{} is a generic function parameterized by $\boldsymbol{\Theta}$, one can introduce variables through $\boldsymbol{\Theta}$ by selecting a \prth{} that depends on them.
Here, we expanded the measurable parameters of gas-phase ultrafast diffraction to include the width of $|\Psi(\boldsymbol{R})|^2$ in \prthg, shown in Fig.~\ref{fig:response_gauss} and given in Table~\ref{tab:mcmcValidation}.
Depending on one's system and desired accuracy, \textit{a priori} knowledge is needed to select the form of \prth, e.g. harmonic oscillator eigenstates for vibrational excited states.
Outside of the vibronic ground state, \prthg{} is a ``first-order" measurement of the $|\Psi(\boldsymbol{R})|^2$ width.
It also reduces the systematic effects of assuming a single structure (\pthetad) as illustrated in Fig.~\ref{fig:theta_peaks_q}.
This was the case for our measured N$_2$O data where our $q$ range of [3.5, 7.25]~\AA$^{-1}$ is insufficient to resolve the width of $|\Psi^{(\text{N}_2\text{O})}(\boldsymbol{R})|^2$.
Therefore, the widths become nuisance parameters used to avoid these systematic errors.
Still, \pthetad{} is accurate on the 10~m\AA{} scale and runs \textapprox100 times faster than \pthetag.
Therefore, \prthd{} serves as an intermediate test analysis before switching to the normal or any other distribution.
For very large molecules with many degrees of freedom, \prthd{} may be the only tractable method.

The MHA performs an unbiased search through $\boldsymbol{\Theta}$ space guided by the $C_{lmk}(q)$ coefficients and correlates each $\boldsymbol{\Theta}$ parameter.
%Selecting $\boldsymbol{\Theta}$ parameters in an unbiased manner and simultaneously evaluating each parameter further constrains \ptheta{} and correlates these $\boldsymbol{\Theta}$ parameters.
Our method is model independent and does not suffer from model bias as might be a concern for conventional methods.
Limited $q$ range artificially introduces correlations between $\boldsymbol{\Theta}$ parameters.
Since $\boldsymbol{\Theta}$ is the minimal set of parameters to define \prth, we expect the parameters to be uncorrelated.
%This is because the $\boldsymbol{\Theta}$ parameters is the minimal set 
%These correlations are due to insufficient information to correctly distinguish them, caused by our limited $q$ range.
Figure~\ref{fig:compare_qrange_correlations} shows how adding information by extending the $q$ range decreases false correlations. 
For the N$_2$O data, we observe these false correlations between $\ev{\text{N}^{\text{T}}\text{N}^{\text{C}}}$ and $\ev{\text{N}^{\text{C}}\text{O}}$ (Fig.~\ref{fig:response_gauss}c). 
%The required $q$ range to extinguish these correlations may be molecule specific, for NO$_2$ the correlations do not affect our ability to resolve the 0.3~\AA{} difference in NO distances for all $q$ ranges.
Simultaneously evaluating all $\boldsymbol{\Theta}$ parameters leverages well-resolved parameters to constrain poorly resolved parameters.
For example, the long OO bond (or $\angle \text{ONO}$) in our asymmetric NO$_2$ is the best constrained parameter as it produces the most $q$ oscillations.
The MHA removes structures where the two NO distances are inconsistent with the well-resolved OO distance.
These correlations similarly help find \thopt, as observed with N$_2$O, where the \pthetag{} uncorrelated widths do not distinguish the $\text{N}^{\text{T}}\text{N}^{\text{C}}$ and $\text{N}^{\text{C}}\text{O}$ bonds but \thopt{} does.

The width of \ptheta{} (\sth) relies heavily on SNR rather than increasing $q$ range (Fig.~\ref{fig:trends}b), which is ideal since it is generally prohibitively difficult to change the $q$ range at ultrafast diffraction facilities and easier to reduce the SNR by taking more measurements \cite{Lingyu.electron_xray.2020}.
%%%Below 8~\iang, \sth{} for the pairwise distances and angles quickly decreases as the $q$ range increases, shown in Figs.~\ref{fig:trends}a and \ref{fig:compare_qrange_correlations}.
%%%Above 8~\iang, \sth{} for the pairwise distances and angles does not change significantly.
%%%In contrast, \sth{} continuously decreases with improved SNR, shown in Fig.~\ref{fig:trends}b.
This is because smaller $\sigma_{lmk}(q)$ makes it less probable for the MHA to visit $\boldsymbol{\Theta}$ parameters with larger residuals.
%This is again due to increasing the magnitude of the exponential's argument in Eq.~\ref{eq:MHlikelihood}.
For the PDF, the resolution is $2\pi/\Delta q$, or 1.26, 0.63, and 0.31~\AA{} for $q$ ranges of 5, 10, and 20~\iang{} respectively, which is roughly 100 to 1000 times larger than our observed resolution for simulated NO$_2$ at typical to high SNR, respectively.
For the measured N$_2$O data with a very poor SNR and 0.04~\AA{} resolution, we observe a 50X improvement over the 1.7~\AA{} Fourier resolution.
This agrees with our simulated results that have more than a factor of 2 better SNR and indicates we may observe these 100--1000X improvements in future measurements.
Our method, therefore, lends itself well to high repetition-rate machines, such as the upcoming LCLS II.
We note that increasing the $q$ range above 8~\iang{} has a larger effect on the width parameters (Fig.~\ref{fig:trends}a).

%% file: sections/05_discussion_intro.tex
In the following, we provide intuition about and describe how this method is able to approximate $|\Psi(\boldsymbol{R})|^2$ while significantly improving upon real-space resolution.
We first provide intuition for how induced anisotropy accesses the molecular frame structural angles $\theta^{(\text{mf})}_{\mu\nu}$ and $\phi^{(\text{mf})}_{\mu\nu}$.
We then provide a brief intuitive discussion, that compliments the Methods section, of how our Bayesian Inference approach inverts $\langle I(\boldsymbol{q},t) \rangle$ for $\boldsymbol{\Theta}$ while improving upon resolution.
Finally, we introduce methods to evaluate excited electronic state dynamics.

%We divide the following discussion into three categories.
%Firstly, we provide intuition for the need of induced anisotropy to access the molecular frame structural angles $\theta^{(\text{mf})}_{\mu\nu}$ and $\phi^{(\text{mf})}_{\mu\nu}$.
%Secondly, we will compare conventional PDF methods to our Bayesian Inference approach.
%We also discuss how systematic errors and various experimental parameters affect the MHA results. 
%Finally, we introduce methods to evaluate excited electronic state dynamics.

%% file: sections/06_discussion_derivation.tex
To provide intuition for the distinct angular terms, we condense and label the reference frames from Eq.~\ref{eq:ImolGen}
\begin{widetext}
\begin{equation}
        \ev**{I(q)}_{\text{mol}} = \cdots \sum_{m, k} (-1)^{k} \underbrace{Y^{m}_l \left( \theta^{(\text{lf})}_q, \phi^{(\text{lf})}_q \right)}_{\text{Lab Frame}} \ev{\underbrace{D^l_{mk} \left( \phi_{\text{I}}^{(\text{lf})}, \theta_{\text{I}}^{(\text{lf})}, \chi_{\text{I}}^{(\text{lf})} \right)}_{\text{Ensemble Anisotropy}} \underbrace{j_l(q \Delta R_{\mu\nu}) Y_l^{-k} \left( \theta_{\mu\nu}^{(\text{mf})}, \phi_{\mu\nu}^{(\text{mf})}\right)}_{\text{Molecular Frame Structure}}}{\Psi(t)}.
        \label{eq:ImolDiscussion}
    \end{equation}
\end{widetext}
Equation~\ref{eq:ImolDiscussion} highlights the anisotropic contributions at each level of this method.
The molecular frame structure component separates into pairwise distance ($j_l(q \Delta R_{\mu\nu})$) and angular $(Y_l^{-k} ( \theta_{\mu\nu}^{(\text{mf})}, \phi_{\mu\nu}^{(\text{mf})}))$ terms.
The former governs the $q$ dependence and the latter is the angular decomposition of the molecular structure which acts as a scaling parameter.
The ensemble anisotropy $\left(D^l_{m k} \left( \phi_{\text{I}}^{(\text{lf})}, \theta_{\text{I}}^{(\text{lf})}, \chi_{\text{I}}^{(\text{lf})} \right)\right)$ acts as a key from the measured lab frame anisotropy ($Y^{m}_l ( \theta^{(\text{lf})}_q, \phi^{(\text{lf})}_q )$) to the molecular frame structure by coupling these two reference frames.
Similar derivations~\cite{baskin.anisotropy_calc.2006, Xiong.LFMF_calc.2022, parrish.rotation_average.2019} exist but do not stress the dependence on the 3d molecular frame coordinates; Ref.~\cite{baskin.anisotropy_calc.2006} is not treated fully quantum mechanically as done here in \ref{ap:anisotropy_derivation}.
Anisotropy is required for our method to have an explicit dependence on the pairwise angles.
Without anisotropy, $C_{000}(q)$ has no explicit angular dependence (Eq.~\ref{eq:Ccoeffs}), just like the PDF.

Stronger impulsive alignment produces a broader coherent rotational wavepacket which exhibits higher amplitude signals with more variations (Fig.~\ref{fig:Mlmk_vs_error}c).
Larger amplitude ADMs improve $C_{lmk}(q)$ SNR by lifting higher order coefficients up out of the noise, resulting in similar resolution improvements to only increasing SNR, shown in Fig.~\ref{fig:trends}c.
Increasing the number of $C_{lmk}(q)$ coefficients improves the $\theta_{\mu\nu}^{(\text{mf})}$ and $\phi_{\mu\nu}^{(\text{mf})}$ resolution since each $C_{lmk}(q)$ provides a new angular constraint via $Y_l^{-k} ( \theta_{\mu\nu}^{(\text{mf})}, \phi_{\mu\nu}^{(\text{mf})})$ (Eq.~\ref{eq:Ccoeffs}).

One can produce fast signal variations with an initially broad hot thermal ensemble.
Writing coherence onto hotter molecular ensembles produces weak but fast varying ADMs, shown in Fig.~\ref{fig:Mlmk_vs_error}c.
Figure~\ref{fig:trends}d shows how quickly the resolution worsens at higher temperatures.
When fitting the ADMs to \blm, one ideally measures particular points that include two separate regions where the ADMs have high variation and sufficiently before and after the prominent anisotropy signal where their magnitude dampens.
One need not strictly measure the entire transient rotational signal.

%Depending on the system, one must strike a balance between fast varying ADMs with a hot ensemble versus slower variations with high SNR at lower temperatures. 
%In both cases, inducing a broader rotational wavepacket will increase the SNR and make the ADMs vary more quickly.
To simulate the ADMs one will need to measure the rotational constants or calculate them from the vibronic ground state structure.
Measured constants remove structural biases potentially induced by calculating these coefficients from a simulated or presumed structure and decouple the rotational signal from the MHA sampling.
When simulating or inducing molecular tumbling is prohibitively difficult, one may use the induced anisotropy from the dipole alignment of the initial photo-excitation.
%Producing sufficiently large rotational coherences in large molecules is challenging.
%The option of using dipole alignment from the excitation pulse allows this technique to be more generally applicable to most molecular systems of interest.
This method can be made more general as our Bayesian Inference approach does not require anisotropy and is applicable to the traditionally used isotropic component.

%% file: sections/07_discussion_mcmc.tex
With Bayesian Inference, we use data to effectively invert $\langle I(\boldsymbol{q},t) \rangle$ for $\boldsymbol{\Theta}$.
%\prthopt{} is significantly more information rich than the PDF.
We use the $C_{lmk}(q)$ coefficients to independently constrain \ptheta, from which we find \thopt{} to parameterize \prthopt.
The \prthopt{} distribution, which approximates $|\Psi(\boldsymbol{R})|^2$, provides the most probable (and unique) molecular structure.
Traditionally, the PDF, being the inverse Fourier transform of $q\,M_{000}(q)$, is at best a weighted histogram of unlabeled pairwise distances from which one generally cannot obtain a unique structure.
%Traditionally, an ideal PDF derived from an infinite $q$ range, being the inverse Fourier transform of $q\,M_{000}(q)$, is at best a weighted histogram of unlabeled pairwise distances.
Since our measurements necessarily exclude $q$ all the way to 0, and the strong signal drop-off limits high $q$ measurements, our $q$ range is always limited.
These limitations obfuscate the PDF interpretations by introducing sinusoidal systematics that result in negative probabilities, e.g. in Fig~\ref{fig:response_gauss}d where we do not expect any distance above 2.3~\AA.
Therefore, we typically simulate $|\Psi(\boldsymbol{R})|^2$ with \textit{a priori} knowledge and validate simulation against the measured PDF.
%This conventional methodology acts as a check on theory.
Our method instead uncovers the globally optimal parameters (\thopt) from the data for a given \prth.
This requires only the initial vibronic ground state structure, simulations of the coherent rotational wavepacket when using $C_{lmk}(q)$ for $l>0$, and for excited state dynamics one additionally needs relevant transition dipole moments.
As made clear by comparing Figs.~\ref{fig:response_gauss}b and d, the \prthopt{} distribution is significantly more information-rich than the PDF, e.g. it provides the 3d molecular structure and width of the $|\Psi(\boldsymbol{R})|^2$.
This method thus has the potential to shift ultrafast diffraction to a discovery method applicable even to systems that extend beyond the scope of theory.

%\ifthenelse{\equal{\natcommphys}{0}}{
%    \input sections/07.5_discussion_mcmc.tex
%}

We find that building \ptheta{} to later find its mode (\thopt) and its resolution (\sth) is more informative and robust than using a gradient-based optimization routine to find \thopt{} and its precision.
In either case, an optimization routine is used to find \thopt, but given \ptheta{} our method starts near the global minima and is more robust to local minima.
If either routine finds a local minima, one can avoid reporting misleading results by citing the resolution of \ptheta{} (\sth) as its error.
Since \sth{} is the standard deviation of all $\boldsymbol{\Theta}$ parameters consistent with the data, it is a conservative estimate that very likely encompasses the global minimum.
The precision, used by an optimization routine, is determined by the loss landscape around \thopt{} and is unaware of the entire $\boldsymbol{\Theta}$ distribution.
The \ptheta{} distribution can also inform the experimentalist which values are best measured, which ones are correlated, and potentially how to improve the experimental apparatus through the false correlations and widths in Figs.~\ref{fig:response_gauss} (a and c) and \ref{fig:compare_qrange_correlations} (a and c).
One does so by varying experimental parameters, in simulation, to determine how isolated and resolved $\boldsymbol{\Theta}$ parameters become.

%% file: sections/08_discussion_excited_state.tex
Our method is broadly applicable to diffraction experiments with laser excitation, including dynamics from excited electronic states.
Laser excitation imparts one or more units of angular momentum providing at least $C_{20k}(q)$.
From low SNR N$_2$O data we see the $C_{200}(q)$ alone recovers \textapprox40~m\AA{} resolution.
The primary difficulty with extending our method to excited states dynamics lies in isolating the ADMs in rovibronically coupled systems at sufficiently long timescales.
Since the principle moments of inertia change with the structure, one must reorient the altered excited state structure by adding three molecular frame Euler angles to the $\boldsymbol{\Theta}$ parameters (\ref{ap:anisotropy_derivation}).
The generally much wider excited state $|\Psi(\boldsymbol{R}, t)|^2$ dampens $C_{lmk}(q)$ coefficients and reduces the need for extended $q$.
We discuss two variants to isolate the ADMs, a time-separable method and an isotropic method.

The time-separable method introduces a separation of time scales by assuming the ADMs are relatively stationary during the vibronic motion.
%Here we assume the ADMs are relatively stationary during the vibronic motion.
This approximation is analogous to the Born-Oppenheimer approximation.
%and allows us to separate the rotational and vibronic timescales.
For a single excitation pulse, the dipole selection rule introduces ensemble anisotropy independent of the difficulty to create a rotational wavepacket:
\begin{widetext}
    \begin{equation}
        \begin{split}
            \langle I(\boldsymbol{q}) \rangle_{\text{sep}}^{(1)} (t) \approx{} & \mathcal{I} \bigg( \sum_\mu |f_\mu(q)|^2 + \sum_{\mu,\nu : \mu \neq \nu} \text{Re} \bigg\{ f_\mu(q) f^*_\nu(q) \sum_l \frac{32\pi^3 i^l}{2l+1} \sum_{m_1, m_2} (-1)^{m_1} Y^{m_2}_l \left( \theta^{(\text{lf})}_q, \phi^{(\text{lf})}_q \right) \\
            &\times \sum_{n,n'} \tilde{\mathcal{A}}^{(1)l}_{m_2m_1} (n,n') \mel**{\psi^{n'}_{\text{el-vib}}(t)}{j_l(q \Delta R_{\mu\nu}) Y_l^{-m_1} \left( \theta_{\mu\nu}^{(\text{mf})}, \phi_{\mu\nu}^{(\text{mf})}\right)}{\psi^n_{\text{el-vib}}(t)}  \bigg\} \bigg).
        \end{split}
        \label{eq:diffSepAvg}
    \end{equation}
\end{widetext}
Here, $\tilde{\mathcal{A}}^{(1)l}_{m_2m_1} (n,n')$ are the ADMs calculated with the rovibronic ground state structure, the ground rovibronic transition dipole, and evaluated immediately after laser excitation.
%%%The dipole selection rule induces the above anisotropy, and is independent of the difficulty to induce a rotational wavepacket, particularly for large molecules.
%and for a single photon transition provides at least the $C_{20k}(q, t)$ coefficients.
%This variation is independent of one's ability to induce a rotational wavepacket, particularly for large molecules.
This requires knowledge of either the transition dipole moment or the Frank-Condon factor and the vibronic ground state dipole.

To further constrain \ptheta, one can couple to more $C_{lmk}(q)$ coefficients by introducing a precursor pulse that excites a rotational wavepacket.
This precursor pulse, assumed to be a rotational Raman impulse, is chosen to have a negligible effect on the vibronic system thus maintaining consistency with our separation of timescale approximation.
The Raman impulse first induces rotational coherence.
Following the Raman impulse, the system evolves for a rotational time $\tau$, at this point the vibronic excitation pulse arrives.
One would measure the vibronic dynamics over a small window $(t\ll \tau)$.
This is repeated for different orientations by scanning the delay $\tau$ over an appreciable portion of the rotational evolution.
This window, measured by $t$, is typically of order picosecond or less such that the ADMs do not appreciably change.
The measured diffraction images are given by Eq.~\ref{eq:IGen_BO}
\begin{widetext}
\begin{equation}
    \begin{split}
        \langle I(\boldsymbol{q}) \rangle_{\text{sep}}^{(2)} (t, \tau) \approx{} & \mathcal{I} \bigg( \sum_\mu |f_\mu(q)|^2 + \sum_{\mu,\nu : \mu \neq \nu} \text{Re} \bigg\{ f_\mu(q) f^*_\nu(q) \sum_l \frac{32\pi^3 i^l}{2l+1} \sum_{m_1, m_2} (-1)^{m_1}  Y^{m_2}_l \left( \theta^{(\text{lf})}_q, \phi^{(\text{lf})}_q \right) \\
        & \times \sum_{n,n'} \tilde{\mathcal{A}}^{(2) l}_{m_2m_1} (n,n';\tau) \mel**{\psi^{n'}_{\text{el-vib}}(t)}{j_l(q \Delta R_{\mu\nu}) Y_l^{-m_1} \left( \theta_{\mu\nu}^{(\text{mf})}, \phi_{\mu\nu}^{(\text{mf})}\right)}{\psi^n_{\text{el-vib}}(t)}  \bigg\} \bigg).
        \end{split} \label{eq:IGen_BO}
\end{equation}    
\end{widetext}
where $n$ labels the vibronic states, $\ket{\psi^n_{\text{el-vib}}(t)}$ is the vibronic wavefunction (assumed unknown), $\tilde{\mathcal{A}}^{(2)l}_{mk} (n,n'; \tau)$ are the modified ADMs, and $t$ is the arrival time of the probe after the second excitation pulse.
These modified ADMs consider the angular momentum transfer by the vibronic excitation photon and require the vibronic ground state transition dipole moment.
One then follows the above analysis procedure for each time $t$.
In such an experiment, one should measure the ensemble anisotropy without the vibronic excitation pulse to find the best-fit ADMs.
\ref{ap:anisotropy_derivation} further describes our separation of timescale approximation and provides the derivations for Eqs.~\ref{eq:diffSepAvg} and \ref{eq:IGen_BO}.

The isotropic method uses only the $C_{000}(q,t)$ term, similar to conventional analyses.
Since $\tilde{\mathcal{A}}^{(\alpha)0}_{00}(n,n`;t,\tau)$ becomes a constant absorbed by $\mathcal{I}$, this method can be applied to single (Eq.~\ref{eq:diffSepAvg}) and double pulse (Eq.~\ref{eq:IGen_BO}) experiments.
%One then performs the MHA for each $C_{000}(q,t)$.
The $C_{000}(q,t)$ term only implicitly depends on the pairwise angles through $\Delta R_{\mu\nu}$.
This is in contrast to the explicit pairwise angle dependence in the higher order $C_{lmk}(q)$ terms.
Our statistical treatment likely provides adequate pairwise angle resolution because we have more pairwise distances than are required to specify a unique structure.

For a Raman-inducing precursor pulse, one will likely use a combination of the isotropic and time-separable methods.
For fast dynamics, one would use the time-separable method for small windows shortly following the rotation time $\tau$.
Longer-lived dynamics can be retrieved by the isotropic method.
When retrieving \ptheta, in either case, one initiates the MHA with the vibronic ground state \thopt{} parameters.
For each subsequent time step one initiates MHA with the \thopt{} parameters from the previous time step.

Electronic and vibrational excited state wavepackets bifurcate into multiple states, e.g.~at conical intersections, causing \prthoptt{} to bifurcate as well.
We account for these different states by
\begin{equation}
    \mmprthoptt = \sum_{i}^{N_\text{ex}} c_i P(\boldsymbol{R}, t | \mmthopt_i, C ) 
\end{equation}
where $N_{\text{ex}}$ is the number of excited state distributions with appreciable population.
Conical intersections will induce bifurcations that spawn a new distribution that adds to $N_{\text{ex}}$.
In this way we consider this method to be fully data-driven since we can change our theoretical description ($c_i$) based on data alone.

%In the excited state, $|\Psi(\boldsymbol{R},t)|^2$ often takes on shapes that are poorly represented by normal distributions~\cite{Makhija.NO2_movie.2021}.
%Nevertheless, the normal distribution acts as a second order improvement upon fitting with a single structure~\cite{Stankus.Fits.2019, Natan.PDF_inversion.2021}.
%This improvement dampens the single structure systematics shown in Fig.~\ref{fig:theta_peaks_q} and retrieves a quantifiable measurement of $|\Psi(\boldsymbol{R},t)|^2$ widths.
%This is particularly useful for vibronic wavepackets with appreciable width.
%One can better describe amorphous $|\Psi(\boldsymbol{R},t)|^2$~\cite{Felker.vibration_flow.1984} by using more representative distributions, such as harmonic oscillator eigenstates, for \prthoptt.

Thus far we have only considered diffraction consistent with the independent atom approximation and all the equations above have been derived under this approximation.
Recently, diffraction beyond the independent atom approximation has been observed in both electron~\cite{Yang.inelastic.2020} and x-ray diffraction~\cite{Yong.lightResponse.2020}.
Under such conditions, this method must be modified by either re-deriving the above equations to consider these effects or by accounting for this signal in the $C_{lmk}(q)$ coefficients.
For MeV electron diffraction, inelastic scattering is limited to the low $q<1$~\iang{} region and can be easily removed from the $C_{lmk}(q)$ coefficients.
For x-ray diffraction beyond the independent atom approximation, contributions from excited Rydberg states create a constant offset after the initial signal turn-on that spans the entire $q$ range \cite{Yong.lightResponse.2020, Stankus.Fits.2019}.
%References~\cite{Yong.lightResponse.2020, Stankus.Fits.2019} observed a constant signal from the electron vacancy after exciting the molecule into a Rydberg state.
Due to the diffuse nature of the Rydberg state this signal does not vary appreciably in time and can be subtracted out.
%Alternatively, one can alter \prthoptt{} to include this offset by adding a corresponding parameter to $\mathbf{\Theta}$.

%% file: sections/09_conclusion.tex
We have shown that our method can approximate $|\Psi(\boldsymbol{R})|^2$ with \prthopt{} for the vibronic ground states of NO$_2$ and N$_2$O.
In simulation, we retrieve \textapprox0.5~m\AA{} resolution for  NO$_2$.
From measured N$_2$O UED data, we retrieve \textapprox40~m\AA{} resolution despite a short $q$ range of [3.5, 7.25]~\iang{} and very poor SNR.
Compared to PDF-based methods, this returns the labeled pairwise distances and angles with 50 and 100--1000 times better resolution in measurement and simulation respectively.
%Furthermore, we can retrieve the unique molecular structure more directly from the data.
%We note that NO$_2$ and N$_2$O are two of the most difficult molecules due to their high symmetry and large central angles; similar bond distances and scattering amplitudes obfuscate the labelling of pairwise variables.
In spite of similar bond distances and atomic scattering amplitudes for NO$_2$ and N$_2$O, our method distinguishes these distances.
We begin to resolve the $\langle \text{N}^{\text{T}}\text{N}^{\text{C}} \rangle$ and $\langle \text{N}^{\text{C}}O \rangle$ distances in our low SNR and narrow $q$ range UED measurement.
These results are highly encouraging and illustrate the viability of our Bayesian Inference approach.
They also inspire further expansion into excited state dynamics.
The code repository~\cite{Hegazy.github.2022} contains the algorithms used for this work and instructions on how to reproduce these results.
It also contains instructions on how to run this analysis and templates for applying this method to new molecules.

This Bayesian Inference approach is best suited for gas-phase ultrafast diffraction instruments that have high SNR such as high repetition-rate free electron facilities, e.g. LCLS-II-HE.
Resolution quickly improves with SNR considerably faster than if one increases $q$ beyond \textapprox8~\iang.
%Illustrated in our measurement of N$_2$O with a limited $q$ range below 8~\iang, [3.5, 7.25]~\iang, we demonstrate \textapprox40 times better resolution than the Fourier limit.
Nevertheless, larger $q$ ranges improve resolution for widths of $|\Psi(\boldsymbol{R})|^2$ and diminish false correlations between $\boldsymbol{\Theta}$ parameters.

Our general method has the potential to become commonplace for ultrafast gas-phase diffraction measurements due to its broad applicability and its independence from complex excited state simulations.
%To do so, we must expand this method to interpret time varying molecular structures without the need of any excited state simulations.
%In this paper we outlined two possible methods.
%With the joint probability distributions \pthetat{} we pull out \thopt$|_t$ from the data to ultimately model $|\Psi(\boldsymbol{r},t)|^2$ by \prthoptt.
%This is in contrast to traditional PDF methods that use complex simulations to calculate $|\Psi(\boldsymbol{r},t)|^2$ and validate such simulations by comparing the PDFs.
In this work, we validated its use for standard pump-probe setups.
One can extend this method to excited state dynamics either with or without anisotropy.
%When utilizing anisotropy, one can employ a single or double excitation pump setup.
Our isotropic method is well suited for current pump-probe setups that generally focus on the isotropic component.
This method greatly benefits from deterministic anisotropy that can either be induced by impulsive Raman or by the dipole moment selection from the excitation pulse.
Beyond ultrafast gas-phase diffraction, one can apply this general framework to other classes of experiments, e.g. the previously mentioned photo-electron experiments \cite{makhija.orientation.2016, Marceau.mFrame.2017, gregory.mfpad.2021,Peter.mFrame_SFI.2019,Peter.mFrame_SFI_OCS.2018}.
This is done by deriving the molecular frame response (Eq.~\ref{eq:ImolGen}) and applying this Bayesian Inference approach.

Given its broad applicability, high resolution, amenability to various measurements, and independence from complex molecular dynamic simulations, our method has the potential to effectively turn ultrafast gas-phase molecular diffraction into a discovery-oriented technique.
This method can retrieve a unique molecular structure distribution for general molecules with $\lessapprox 10$~m\AA.
%{} resolution without relying on complex molecular simulations.
Moreover, because our method is parameterized by $\boldsymbol{\Theta}$, we have the opportunity to expand the scope of ultrafast gas-phase diffraction into previously inaccessible measurements.
For instance, we demonstrated the use of this parameterization to measure the width of $|\Psi(\boldsymbol{R}, t)|^2$; this width is important in the excited state where single structures lose their meaning.
This method unlocks our ability to study larger and more complex systems that are currently too difficult to simulate.

%% file: sections/07.5_discussion_mcmc.tex
Our method retrieves the labeled pairwise distances with \textapprox100 times better resolution than the PDF.
This is due to our statistical treatment using Bayesian Inference where each $lmk$ and $q$ contribution is itself an independent probability distribution; each is an experiment of its own.
The MHA discrimination power grows exponentially with more $C_{lmk}(q)$, which increases the magnitude of the negative exponent in the relative ratio of likelihood functions $P(C|\boldsymbol{\Theta})$ (Eq.~\ref{eq:MHlikelihood}).
%That is, with more measurements Eq.~\ref{eq:MHlikelihood} becomes smaller and \ptheta{} becomes sharper.
Our method therefore heavily relies on $\sigma_{lmk}(q)$ and $C_{lmk}(q)$.
Statistical noise increases $\sigma_{lmk}(q)$, making \ptheta{} wider (\sth{} larger), while systematic errors in $C_{lmk}(q)$ shift the centriod of \ptheta, as seen in the N$_2$O data.
Supplementary Section~\ref{ap:fit_error_prop} describes our method for consistently accounting for both statistical and systematic errors.
%The PDF approach, alternatively, takes the inverse Fourier transform of the contiguous $q\,M_{000}(q)$.
%\rnc{check below}
The PDF error adds in quadrature in $\sigma_{lmk}(q)$; its scale is set by the largest error bar and disproportionately suffers from poorly measured data points.
Conversely, MHA amplifies the contribution of high precision measurements while reducing contributions from poorly measured data points by weighting each term in the likelihood by $1/\sigma_{lmk}(q)$ (Eq.~\ref{eq:MHlikelihood}).

Our Bayesian Inference approach expands the utility of gas-phase ultrafast diffraction to measure previously inaccessible variables.
Given \prth{} is a generic function parameterized by $\boldsymbol{\Theta}$, one can introduce variables through $\boldsymbol{\Theta}$ by selecting a \prth{} that depends on it.
Here, we expanded the measurable parameters of gas-phase ultrafast diffraction to include the width of $|\Psi(\boldsymbol{R})|^2$ in \prthg.
\edit{Depending on one's system and desired accuracy, \textit{a priori} knowledge is needed to select the form of \prth, e.g. harmonic oscillator eigenstates for vibrational excited states.
Outside of the vibronic ground state, \prthg{} is a ``first-order" measurement of the $|\Psi(\boldsymbol{R})|^2$ width.
It also reduces the systematic effects of assuming a single structure (\pthetad) as illustrated in Fig.~\ref{fig:theta_peaks_q}.}
This was the case for our measured N$_2$O data where our $q$ range of [3.5, 7.25]~\AA$^{-1}$ is insufficient to resolve the width of $|\Psi^{(\text{N}_2\text{O})}(\boldsymbol{R})|^2$.
Therefore, the widths become nuisance parameters used to avoid these systematic errors.
\edit{Still, \pthetad{} is accurate on the 10~m\AA{} scale and runs \textapprox100 times faster than \pthetag.
Therefore, \prthd{} serves as an intermediate test analysis before switching to the normal or any other distribution.
For very large molecules with many degrees of freedom, \prthd{} may be the only tractable method.}

The MHA performs an unbiased search through $\boldsymbol{\Theta}$ space guided by the $C_{lmk}(q)$ coefficients and correlates each $\boldsymbol{\Theta}$ parameter.
%Selecting $\boldsymbol{\Theta}$ parameters in an unbiased manner and simultaneously evaluating each parameter further constrains \ptheta{} and correlates these $\boldsymbol{\Theta}$ parameters.
Our method is model independent and does not suffer from model bias as might be a concern for conventional methods.
Limited $q$ range artificially introduces correlations between $\boldsymbol{\Theta}$ parameters.
Since $\boldsymbol{\Theta}$ is the minimal set of parameters to define \prth, we expect the parameters to be uncorrelated.
%This is because the $\boldsymbol{\Theta}$ parameters is the minimal set 
%These correlations are due to insufficient information to correctly distinguish them, caused by our limited $q$ range.
Figure~\ref{fig:compare_qrange_correlations} shows how adding information by extending the $q$ range decreases false correlations. 
For the N$_2$O data, we observe these false correlations, most notably between $\ev{\text{N}^{\text{T}}\text{N}^{\text{C}}}$ vs $\ev{\text{N}^{\text{C}}\text{O}}$ (Fig.~\ref{fig:response_gauss}c). 
%The required $q$ range to extinguish these correlations may be molecule specific, for NO$_2$ the correlations do not affect our ability to resolve the 0.3~\AA{} difference in NO distances for all $q$ ranges.
Simultaneously evaluating all $\boldsymbol{\Theta}$ parameters leverages well-resolved parameters to constrain poorly resolved parameters.
For example, the long OO bond (or $\angle \text{ONO}$) in our asymmetric NO$_2$ is the best constrained parameter as it produces the most $q$ oscillations.
The MHA removes structures where the two NO distances are inconsistent with the well-resolved OO distance.
These correlations similarly help find \thopt, as observed with N$_2$O, where the \pthetag{} uncorrelated widths do not distinguish the $\text{N}^{\text{T}}\text{N}^{\text{C}}$ and $\text{N}^{\text{C}}\text{O}$ bonds but \thopt{} does.

The width of \ptheta{} (\sth) relies heavily on SNR rather than increasing $q$ range, which is ideal since it is generally prohibitively difficult to change the $q$ range at ultrafast diffraction facilities and easier to reduce the SNR by taking more measurements.
Similarly, Ref.~\cite{Lingyu.electron_xray.2020} illustrated the importance of SNR.
Below 8~\iang, \sth{} for the pairwise distances and angles quickly decreases as the $q$ range increases, shown in Figs.~\ref{fig:trends}a and \ref{fig:compare_qrange_correlations}.
Above 8~\iang, \sth{} for the pairwise distances and angles does not change significantly.
In contrast, \sth{} continuously decreases with improved SNR, shown in Fig.~\ref{fig:trends}b.
This is because smaller $\sigma_{lmk}(q)$ makes it less probable for the MHA to visit $\boldsymbol{\Theta}$ parameters with larger residuals.
%This is again due to increasing the magnitude of the exponential's argument in Eq.~\ref{eq:MHlikelihood}.
\edit{For the PDF, the resolution is $2\pi/\Delta q$, or 1.26, 0.63, and 0.31~\AA{} for $q$ ranges of 5, 10, and 20~\iang{} respectively, which is roughly 100 to 1000 times larger than our observed resolution for simulated NO$_2$ at typical to high SNR, respectively.
For the measured N$_2$O data with a very poor SNR and 0.04~\AA{} resolution, we observe a 50X improvement over the 1.7~\AA{} Fourier resolution.
This agrees with our simulated results that have more than a factor of 2 better SNR and indicates we may observe these 100--1000X improvements in future measurements.}
Our method, therefore, lends itself well to high repetition-rate machines, such as the upcoming LCLS II.
We note that increasing the $q$ range above 8~\iang{} has a larger effect on the width parameters (Fig.~\ref{fig:trends}a).

%% file: sections/10_availability.tex
\section{Data Availability}
The UED N$_2$O data used in this analysis will be provided by the corresponding authors upon reasonable request.
The simulated NO$_2$ data, $C_{lmk}(q)$, can be calculated by the supplied analysis code in Ref.~\cite{Hegazy.github.2022}.

\section{Code Availability}
The code used in this analysis can be found in Ref.~\cite{Hegazy.github.2022}.
Here, one will find a detailed description of the code and how to run it in order to reproduce the NO$_2$ results.
This repository also includes templates for one to apply this algorithm to new molecules.

%% file: sections/11_acknowledgements.tex
Use of the Linac Coherent Light Source (LCLS), SLAC National Accelerator Laboratory, is supported by the U.S. Department of Energy, Office of Science, Office of Basic Energy Sciences under Contract No. DE-AC02-76SF00515.
The UED work was performed at SLAC MeVUED, which is supported in part by the DOE BES SUF Division Accelerator and Detector research and development program, the LCLS Facility, and SLAC under contract Nos. DE-AC02-05-CH11231 and DE-AC02-76SF00515.
Markus Ilchen acknowledges funding by the Volkswagen foundation for a Peter-Paul-Ewald Fellowship.
We thank Markus G{\"u}hr for his help in setting up the SLAC gas phase UED experiment.
We thank Theodore Vecchione for his help operating the SLAC UED facility.
We thank Gregory Stewart for creating Figs.~\ref{fig:rotations_LFMF} and \ref{fig:rotations}. 

%% file: sections/ap_ADM_calculations.tex
The axis distribution moments (ADMs) decompose the molecular ensemble anisotropy into a sparse 3d angular basis~\cite{stolow.calc_adm.2008,underwood.calc_adm.2000,gregory.mfpad.2021}.
This basis is the expansion of $|\Psi(t)|^2$ in terms of the Wigner D basis
\begin{equation}
    \mathcal{A}^l_{mk}(t) = \frac{2l+1}{8\pi^2}\expectationvalue**{D^{l}_{mk}\left(\phi, \theta, \chi \right)}{\Psi(t)}
    \label{eq:ap:ADMs}
\end{equation}
where $\phi$, $\theta$, and $\chi$ are the lab frame Euler angles that orient the molecular frame with respect to the lab frame.
The principal moments of inertia for the rovibronic  state structure define the molecular frame.
In decreasing order, the principle moments of inertia ($I_A$, $I_B$, and $I_C$) define the $\hat{\mathbf{z}}^{(\text{mf})}$, $\hat{\mathbf{x}}^{(\text{mf})}$, and $\hat{\mathbf{y}}^{(\text{mf})}$, respectively.
For a given molecular structure, or state, these principal moments of inertia also define the rotational constants $A=(2I_A)^{-1}$, $B=(2I_B)^{-1}$, and $C=(2I_C)^{-1}$ that are used to calculate the rotational kinetic energy.
The rotational Hamiltonian is given by
\begin{equation}
    H_R = AJ_A^2 + BJ_B^2 + CJ_C^2
\end{equation}
where $J_i$ is the total angular momentum operator about the $i^\text{th}$ principal moment of inertia.

\subsection{Linear Symmetric Rigid Rotors and N$_2$O}

We first address how to calculate the ADMs for a simple linear symmetric rigid rotor, like N$_2$O.
A linear rotor has two unique principal components of inertia, where the single unique moment is much much smaller than the other two equal components.
This is due to the cylindrical symmetry of the linear rotor which removes the ADMs' dependence on $\chi$.
In the rigid rotor approximation,
\begin{align}
    \bra{\theta, \phi}\ket{jm} &= Y_j^{m}(\theta, \phi) \\
    E_{jm} &= Bj(j+1) \\
    B &= \frac{\hbar^2}{2I}
\end{align}
where the Wigner D matrix is reduced to the spherical harmonics under cylindrical symmetry.
Before the alignment pulse ($t<0$) the molecular ensemble is in a thermal distribution of rotational $\ket{j,m}$ and vibrational eigenstate $\ket{\nu}$.
Here, $\nu$ labels the vibrational harmonic oscillator state.
We presume the alignment pulse intensity is not sufficient to change the thermal distribution, and the pulse width is long enough that vibrational Raman excitation is negligible.
Consequently, we separate the rotational and vibrational wavefunctions.
It is still important to consider the initial vibrational state as the moments of inertia, and therefore the rotational constants, will vary between vibrational states.
The alignment pulse launches a rotational wavepacket by introducing a rotational coherence between eigenstates
\begin{equation}
    \ket{\psi^{(i)}(t)} = \sum_{j,m} c_{j_i m_i \nu_i j m}(t) \ket{j,m}\ket{\nu_i}
    \label{ap:eq:n2o_single_wp}
\end{equation}
where $j_i$ and $m_i$ label the initial ($t<0$) rotational eigenstate for a single molecule.
This thermal ensemble is represented by the density matrix where each state is weighted by the Boltzmann distribution
\begin{align}
    \rho(t) &= \sum_{i} p_{i} \ket{\psi^{(i)}(t)} \bra{\psi^{(i)}(t)} \\
    p_{i} &= \frac{\exp{-E_{j_i m_i \nu_i}/(k_b T)}}{Z}
\end{align}
where we sum over the initial $\ket{j_i,m_i}\ket{\nu_i}$ states, $Z$ is the partition function, $k_b$ is the Boltzmann, and $T$ is the temperature.

Evaluating Eq.~\ref{eq:ap:ADMs} with respect to our density matrix representation of our thermal ensemble, we find that $\mathcal{A}^l_{m}(t)$ for a symmetric linear rigid rotor is given by
%\begin{equation}
%    P(\theta, \phi, t) = \sum_{l,m} \mathcal{A}^l_{m}(t) Y_l^m(\theta, \phi) = \left| \bra{\theta, \phi}\ket{\psi} \right|^2
%\end{equation}
\begin{equation}
    \begin{split}
    \mathcal{A}^l_{m}(t) =& \frac{2l+1}{4\pi} \Tr(\rho Y^m_l) \\
        =& \frac{2l+1}{4\pi} \sum_{i} p_{i} \sum_{j_1,m_1} \sum_{j_2,m_2} c^*_{j_i m_i \nu_i j_1 m_1}(t)c_{j_i m_i \nu_i j_2 m_2}(t) \\
        &\hspace{4mm} \times \int Y_{j_1}^{*m_1} Y_l^m Y_{j_2}^{m_2} \sin \theta d\theta d\phi \\
        =& \frac{2l+1}{4\pi}\sum_{i} p_{i} \sum_{j_1,m_1}\sum_{j_2,m_2} c^*_{j_i m_i \nu_i j_1 m_1}(t)c_{j_i m_i \nu_i j_2 m_2}(t) \\ &\hspace{4mm} \times \sqrt{\frac{(2j_1+1)(2l+1)(2j_2+1)}{4\pi}} \\
        &\hspace{4mm} \times \threeJ{j_1}{l}{j_2}{0}{0}{0}\threeJ{j_1}{l}{j_2}{-m_1}{m}{m_2}.
    \end{split}
\end{equation}

\begin{figure}[!htbp]
    \centering
    \includegraphics[scale=0.5]{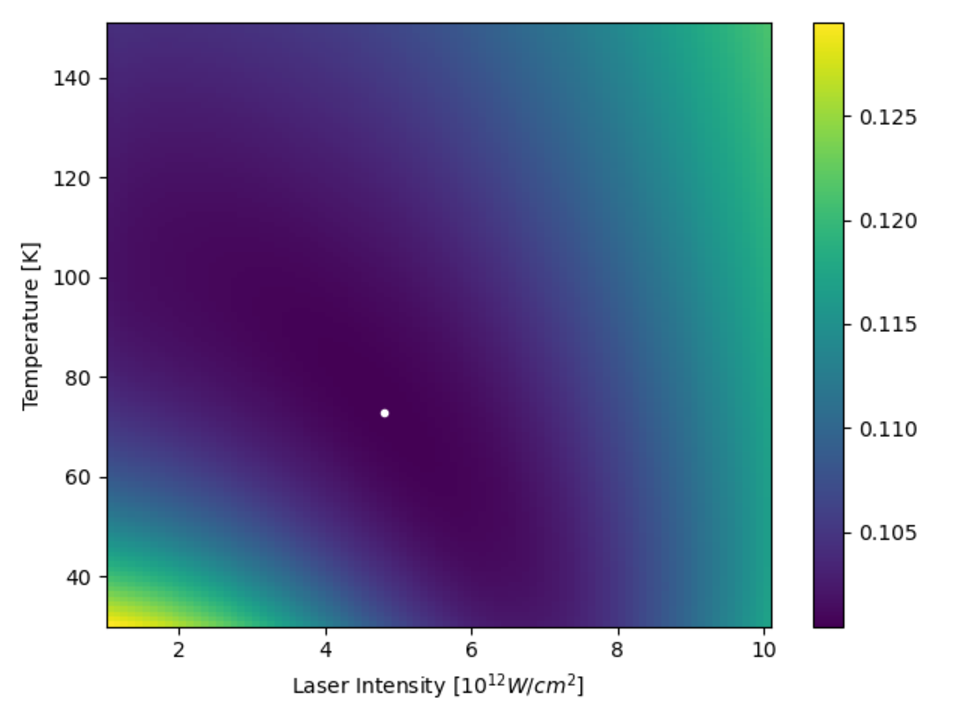}
    \caption{\textbf{Best fit ensemble temperature and Raman pump pulse intensity} We show the $\chi^2$ fit value between simulated Axis Distribution Moments (ADMs) and the N$_2$O temporal variations. We vary the ADMs by changing the molecular ensemble's temperature and the pump beam intensity. The lowest $\chi^2$ value is marked by the white dot.}
    \label{fig:ap:ADM_chiSq_fit}
\end{figure}

\begin{figure}[!htbp]
    \centering
    \includegraphics[scale=0.9]{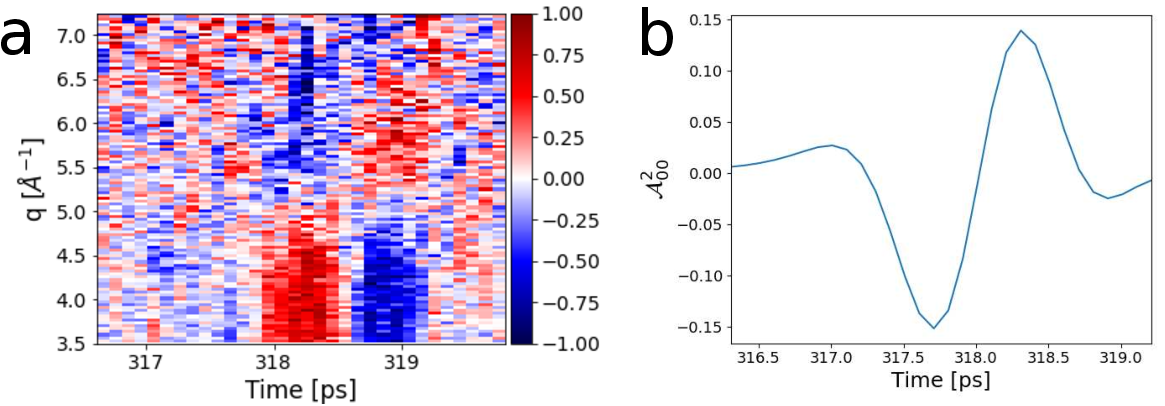}
    \caption{\textbf{Comparing the measured N$_2$O dynamics with the fitted ADMs} We show the measured time-depended anisotropy parameter $B_{200}(q,t)$ (panel a) along with the best fit Axis Distribution Moment (ADM) (panel b). The data in panel a was fitted with the simulated ADM ($\mathcal{A}^2_{0}(t)$) as a function of ensemble temperature and pump laser intensity.}
    \label{fig:ap:n2o_Blmk_A20}
\end{figure}

For N$_2$O, we simulated $\mathcal{A}^2_0(t)$ by solving the TDSE for the $c_{j_i m_i \nu_i j m}(t)$ coefficients using a split step operator.
The non-resonant excitation laser field induces the potential 
\begin{equation}
    V(t) = -\frac{1}{4}E^2_0(t)\Delta \alpha \cos^2 \theta
\end{equation}
where $E_0(t)$ is the pulse field envelope and $\Delta \alpha = \alpha_{\parallel} - \alpha_\perp$ is the molecule's differential polarizability.
The total Hamiltonian becomes
\begin{equation}
    H(t) = H_R(t) + V(t).
\end{equation}
Numerically simulating the $c_{j_i m_i \nu_i j m}(t)$ is easily done via the split step operator technique outlined in Ref.~\cite{Tannor.TD_QM.2007}.
This simulation, however, requires the alignment pulse intensity and the ensemble temperature.
To find these values, we simulated many variations of  $\mathcal{A}^2_0(t)$ and compared them to the measured $B_2^0(q,t)$ coefficients, shown in Fig.~\ref{fig:ap:n2o_Blmk_A20}.
For each $q$ bin, we fit the $\mathcal{A}^2_0(t)$ to the time dependence and calculate a $\chi^2(q)$ value.
Our aggregate $\chi^2$ value is a weighted average of these $\chi^2(q)$ weighted by the temporal variance.
We find that a temperature of 73~K and a laser intensity of $5\times10^{12}$~W/cm$^2$ provides the best fit.
Figure~\ref{fig:ap:ADM_chiSq_fit} shows this $\chi^2$ landscape and Fig.~\ref{fig:ap:n2o_Blmk_A20} shows the measured data we fit and the best fit $\mathcal{A}^2_0(t)$.

\subsection{Asymmetric Rigid Rotors}
\begin{figure}[!htbp]
    \centering
    \includegraphics[scale=0.23]{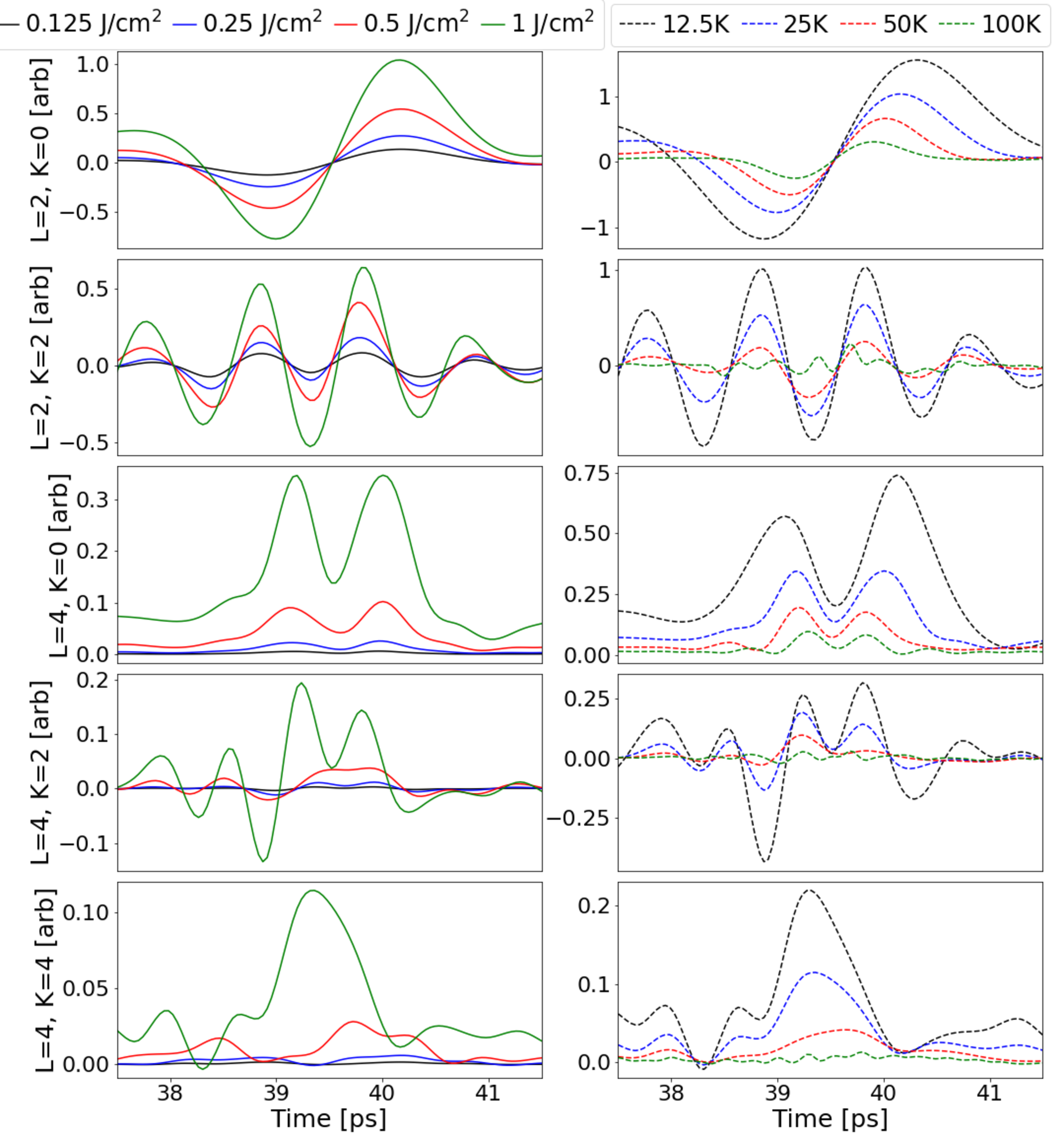}
    \caption{\textbf{Axis Distribution Moments as a function of ensemble temperature and pump laser fluence} The Axis Distribution Moments (ADMs) vary as a function of pump fluence and ensemble temperature. The left column varies the pump fluence with a constant temperature of 25~K. The right column varies the temperature with a constant fluence of 1~J/cm$^2$.}
    \label{fig:ap:ADM_fluence_temp}
\end{figure}

Asymmetric rigid rotors have three unique principal axes with $A \neq B \neq C$, such that $I_\text{A} < I_\text{B} < I_\text{C}$. As a result, they have a fundamentally different energy level structure. In general, the energy eigenvalues may be determined analytically for each $J$ using the $D_2$ symmetry group of the rigid rotor Hamiltonian. This renders the Hamiltonian matrix in the $\ket{jmk}$ symmetric top basis block diagonal~\cite{zare.angular_momentum}. Here $k$ is the angular momentum quantum number corresponding to the projection of the angular momentum on the molecular frame $\hat{\textbf{z}}$. Writing the eigenstates in this basis yields,
\begin{equation}
    \left|jm\tau\right> =\sum_k c_{jmk}\left|jmk\right>.
\end{equation}
The asymmetric top eigenstates $\ket{jm\tau}$ each correspond to an energy eigenvalue $E_{jm\tau}$, and the spacing between eigenstates determines the field-free evolution of the rotational wavepacket excited by the alignment pulse from an initial state rotational state $(i)$,
\begin{equation}
    \ket{\psi^{(i)}(t)}=\sum_{jm\tau}c_{jm\tau}\exp \bigg \{ \frac{-iE_{jm\tau}t}{\hbar}\bigg \}\ket{jm\tau}.
\end{equation}
The coefficients $c_{jm\tau}$ are determined by solving the TDSE for the asymmetric rigid rotor in a non-resonant time-dependent electric field.
The field-matter interaction is typically mediated by the molecular polarizability, resulting in a series of Raman Transitions.
Such a calculation has been detailed by several authors~\cite{poulsen2004,underwood2005,rouzee2006_1D,takemoto2008,ohshima2010,pabst2010,makhija2012,koch2019,lin2020}, so we do not discuss it here.
The density matrix $\rho_{jm\tau}^{j'm'\tau'}(t)$ can then be determined as discussed above for the linear molecule.
Finally, the ADMs can be calculated from the density matrix transformed into the $\ket{jmk}$ basis as follows,
\begin{equation}
    \begin{split}
    \mathcal{A}^K_{QS}(t) =& \frac{2K+1}{8\pi^2} \Tr(\rho(t) D^K_{QS}) \\
        =& \frac{2K+1}{8\pi^2} \sum_{j,m,k}\sum_{j',m',k'} \rho_{jmk}^{j'm'k'}(t) \\ &\hspace{4mm} \times \sqrt{(2j+1)(2j'+1)}(-1)^{m-k} \\
        &\hspace{4mm} \times \threeJ{j}{j'}{K}{-m}{m'}{Q}\threeJ{j}{j'}{K}{-k}{k'}{S}.
    \end{split}
\end{equation}
The resulting ADMs for our simulated NO$_2$ distribution as a function of ensemble temperature and the pump laser fluence is given in Fig.~\ref{fig:ap:ADM_fluence_temp}.

\iffalse
\begin{align}
    \psi(\theta, \phi) =& \bra{\psi}\ket{\theta, \phi} = \sum_{L,M} c_{L,M} Y_L^M(\theta, \phi) \\
    P(\theta, \phi, t) =& \sum_{L,M} A_{L,M}(t) Y_L^M(\theta, \phi) \\
        =& \bra{\psi}\ket{\theta,\phi} \bra{\theta, \phi}\ket{\psi} \\
    \begin{split}
    A_{L,M}(t) =& \Tr(\rho A_{L,M}) \\
        =& \sum_i p_i \expval{Y_L^M}{\psi_i(t)} \\
        =& \sum_i p_i \sum_{l_1,m_1} \sum_{l_2,m_2} c^*_{i,l_1,m_1}(t)c_{i,l_2,m_2}(t) \\
        &\hspace{4mm} \times \int Y_{l_1}^{*m_1} Y_L^M Y_{l_2}^{m_2} \sin \theta d\theta d\phi \\
        =&\sum_i p_i \sum_{l_1,m_1}\sum_{l_2,m_2} c^*_{i,l_1,m_1}(t)c_{i,l_2,m_2}(t) \\ &\hspace{4mm} \times \sqrt{\frac{(2l_1+1)(2L+1)(2l_2+1)}{4\pi}} \\
        &\hspace{4mm} \times \threeJ{l_1}{L}{l_2}{0}{0}{0}\threeJ{l_1}{L}{l_2}{-m_1}{M}{m_2}
    \end{split}
\end{align}
\fi

%% file: sections/ap_anisotropy_derivation.tex
%\paragraph*{And here's how anisotropy can reveal MF}
\begin{figure}[!htbp]
    \begin{center}
        \includegraphics[scale=0.8]{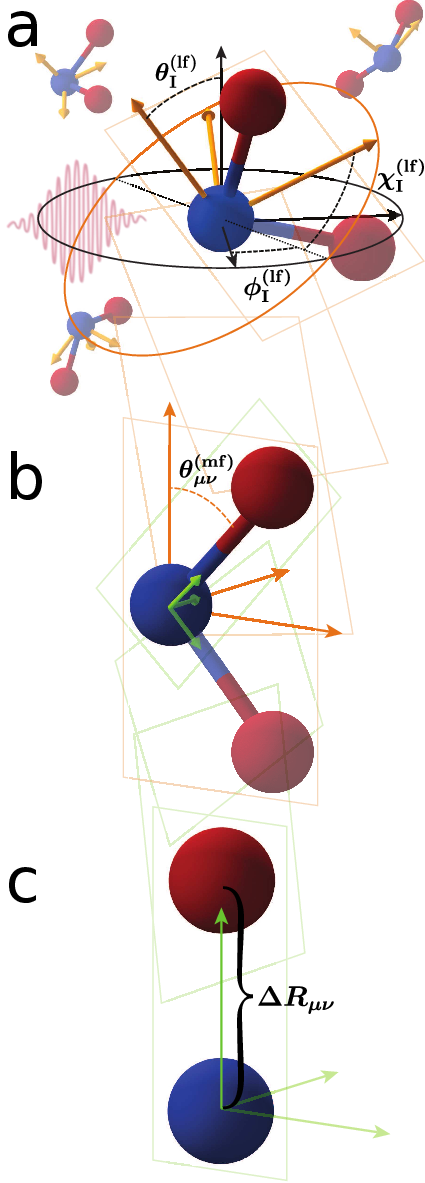}
        \caption{\label{fig:rotations}\textbf{Correspondence between the pairwise, molecular, and lab frames} Our analysis considers each pairwise distance independently and we define the origin of both the lab and molecular frames by one of the pairwise vectors. For the highlighted NO bond, the nitrogen atom (blue) defines the origin. The lab frame (panel a) is defined by the laser polarization $(\hat{\mathbf{z}})$ and propagation direction $(\hat{\mathbf{y}})$. The molecular frame (panel b) is defined by the molecule's rovibronic ground state principal moments of inertia, where the molecular A, B, and C axes define $\hat{\mathbf{z}}^{(\text{mf})}$, $\hat{\mathbf{y}}^{(\text{mf})}$, and $\hat{\mathbf{x}}^{(\text{mf})}$. The pairwise frame (panel c) is simply defined by the pairwise vector along the $(\hat{\mathbf{z}})$ axis with a pairwise length of $\Delta \boldsymbol{R}_{\mu\nu}$. To rotate the pairwise frame into the molecular frame one first rotates this vector by the polar molecular frame pairwise angle $\theta_{\mu\nu}^{(mf)}$ and then by the azimuthal $\phi_{\mu\nu}^{(mf)}$. To access the lab frame from the molecular frame, one rotates the molecule by the lab frame Euler angles $\theta^{(\text{lf})}_\text{I}$, $\phi^{(\text{lf})}_\text{I}$, and $\chi^{(\text{lf})}_\text{I}$.}
    \end{center}
\end{figure}

We now show how deterministic anisotropy allows one to access molecular frame geometric information by coupling the lab and molecular frame.
Using the Independent Atom model, the x-ray, or electron, diffraction intensity from a single molecule is given by
\begin{align}
    \begin{split}
        I(\boldsymbol{q}) &= \mathcal{I} \Big( \sum_\mu |f_\mu(q)|^2 \\
        + &\sum_{\mu,\nu : \mu \neq \nu} \text{Re} \{ f_\mu(q) f^*_\nu(q) \exp\left( i \boldsymbol{q} \cdot \left(\boldsymbol{R}_\mu - \boldsymbol{R}_\nu\right) \right) \} \Big)    \label{eq:genInts} 
    \end{split}\\
    \mathcal{I} &= \left\{ \begin{array}{cl}
         I_0 & \text{: x-ray diffraction} \\
         \frac{I_0}{R_\text{beam}^2} & \text{: electron diffraction} 
    \end{array} \right.
\end{align}
where $\mathcal{I}$ is a scaling coefficient, $I_0$ is the initial intensity of the probe, $R_\text{beam}$ is the distance between the sample and where the electron was detected, $\boldsymbol{R}_\mu$ is the position of the $\mu^{\text{th}}$ atom, and $f_\mu(q)$ is either the electron scattering amplitude or x-ray form factor of the $\mu^{\text{th}}$ atom.
Here $\boldsymbol{q}$ is the momentum transfer imparted on either the electron or x-ray after scattering from the molecule.
In the case of x-ray scattering, we assume one has already removed the anisotropic effects from Thomson Scattering.
The difference in the scattered x-ray and electron wave functions accounts for the $R_\text{beam}^{-2}$ factor in $\mathcal{I}$.
The first term $\left(\sum_\mu |f_\mu(q)|^2\right)$ is independent of the molecule's structure and is referred to as the atomic scattering contribution.
The second term depends on the pairwise distances of atoms and is known as the molecular diffraction.

Our objective is to represent the lab frame diffraction pattern, parameterized by the momentum transfer $q = \left|\boldsymbol{q}\right|$ and the detector's azimuthal angle \thd, in terms of the molecular frame pairwise distances and angles $\Delta \boldsymbol{R}_{\mu\nu}= \boldsymbol{R}_\mu - \boldsymbol{R}_\nu =\left[ \Delta R_{\mu\nu}, \theta^{(\text{mf})}_{\mu\nu}, \phi^{(\text{mf})}_{\mu\nu} \right]$.
This derivation focuses on a single $(\mu,\nu)$ pair from the molecular diffraction sum in Eq.~\ref{eq:genInts}, where the $\nu^{\text{th}}$ atom defines the origin as we rotate between the lab and various body reference frames.
Figure~\ref{fig:rotations} illustrates these various frames serving as an intuitive guide, with the $\nu^{\text{th}}$ atom translated to the origin.
Such translations are allowed since they cancel in the $\Delta \boldsymbol{R}_{\mu\nu}$ term.
For our rotations, we use the conventions in Ref.~\cite{zare.angular_momentum}.
Unless otherwise stated, $\theta$ and $\phi$ represent the polar and azimuthal angles, respectively, in a spherical coordinate system.

We define the pairwise frame (pf) such that $\hat{\mathbf{z}}^{(\text{pf})} = \Delta \hat{\boldsymbol{R}}_{\mu\nu}$, again emphasizing we translate the molecule such that the $\nu^{\text{th}}$ atom is at the origin.
% We do not define x and y, they come from rotating from mf to pf
The pairwise frame is shown in Fig.~\ref{fig:rotations}c.
The exponential term in Eq.~\ref{eq:genInts} is rewritten using the partial wave expansion 
\begin{align}
    \exp \left( i\boldsymbol{q} \cdot \Delta \boldsymbol{R}_{\mu\nu} \right) &= \sum_l i^l \left(2l+1\right) j_l(q \Delta R_{\mu\nu}) P_l \left( \cos \theta^{(\text{pf})}_q\right) \\
    &= \sum_l i^l \sqrt{4\pi \left(2l+1\right)} j_l(q \Delta R_{\mu\nu}) Y^0_l \left( \theta^{(\text{pf})}_q, \phi^{(\text{pf})}_q \right)
    \label{eq:ap:partial_wave}
\end{align}

Here, $j_l(q \Delta R_{\mu\nu})$ are the spherical Bessel functions of the first kind, $Y_l^m(\theta^{(\text{pf})}_q, \phi^{(\text{pf})}_q)$ are spherical harmonics, and $\left(\theta^{(\text{pf})}_q, \phi^{(\text{pf})}_q \right)$ are the polar and azimuthal angles that define $\boldsymbol{q}$ in the pairwise frame.
In the above equation, we determine the dependence on the labeled pairwise distance $\Delta R_{\mu\nu}$, one of our parameters of interest.

\iffalse
\begin{figure}[!htbp]
\begin{center}
\includegraphics[scale=1.0]{figs/molFrame.jpg}
\caption{\label{fig:molFrame}We show the molecular frame of CF$_2$IBr (top), defined by the molecule's principal moments of inertia, with the purple, red, gray, and white representing iodine, bromine, carbon, and fluorine, respectively. Below, in the colored boxes, we translate two pairs of atoms to the origin to show how the molecular frame pairwise angles are defined.}
\end{center}
\end{figure}
\fi

The molecular frame (mf) is defined by the molecule's principal moments of inertia, here the $\hat{\mathbf{z}}^{(\text{mf})}$, $\hat{\mathbf{y}}^{(\text{mf})}$, and $\hat{\mathbf{x}}^{(\text{mf})}$ axes correspond to the moments with increasing rotational inertia.
%Figure~\ref{fig:molFrame} shows the molecular frame for CF$_2$IBr (top), where the origin is defined by the molecule's center of mass, while emphasizing how we translate the $\nu^{\text{th}}$ atom to the origin when defining our molecular frame structure angles $(\theta_{\mu\nu}^{(mf)}, \phi_{\mu\nu}^{(mf)})$ (bottom).
Figure~\ref{fig:rotations}b shows the molecular frame for NO$_2$ with the nitrogen translated to the origin.
We rotate from the pairwise frame into the molecular frame, shown in Fig.~\ref{fig:rotations} as green and orange, respectively. 
% of \adm$.
\begin{align}
    \exp \left( i\boldsymbol{q} \cdot \Delta \boldsymbol{R}_{\mu\nu} \right) &={} \sum_l i^l \sqrt{4 \pi \left(2l+1\right)} j_l(q \Delta R_{\mu\nu}) \sum_{m_1} \left[ D^l_{m_1 0} \left( \phi_{\mu\nu}^{(\text{mf})}, \theta_{\mu\nu}^{(\text{mf})}, 0\right) \right] Y^{m_1}_l \left( \theta^{(\text{mf})}_q, \phi^{(\text{mf})}_q \right) \label{eq:ap:rotMF_init}\\
    &={}  4 \pi \sum_l i^l j_l(q \Delta R_{\mu\nu}) \sum_{m_1} (-1)^{m_1} Y_l^{-m_1} \left( \theta_{\mu\nu}^{(\text{mf})}, \phi_{\mu\nu}^{(\text{mf})} \right) Y^{m_1}_l \left( \theta^{(\text{mf})}_q, \phi^{(\text{mf})}_{q} \right)
    \label{eq:ap:rotMF}
\end{align}
The molecular frame angles $\phi_{\mu\nu}^{(\text{mf})}$ and  $\theta_{\mu\nu}^{(\text{mf})}$  define the orientation of $\Delta \hat{\textbf{r}}_{\mu\nu}$, where $\chi_{\mu\nu}^{(\text{mf})} = 0$ since $\Delta \boldsymbol{R}_{\mu\nu}$ is a vector.
We stress the importance of these molecular frame structure angles $(\theta_{\mu\nu}^{(\text{mf})}, \phi_{\mu\nu}^{(\text{mf})})$ as they are needed, along with $\Delta R_{\mu\nu}$ to define a unique molecular structure.
With PDF methods alone, one only has access to unlabeled $\Delta R_{\mu\nu}$ and generally cannot define a unique molecular structure.
These molecular frame angles are the last two geometric parameters of interest.
%One can choose a different molecular frame, but will be required to add an additional rotation from the molecular frame to principle moments of inertia.
%and are the reason one cannot uniquely retrieve the molecules structure with the pairwise distances alone.

To connect our molecular frame calculation to our measurement, we rotate into the lab frame (lf).
The lab frame $\hat{\textbf{z}}^{(\text{lf})}$ is defined as the polarization of the alignment laser $(\hat{\boldsymbol{\varepsilon}})$, and $\hat{\textbf{y}}^{(\text{lf})}$ is along the probe path and normal to the detector.
%\begin{widetext}
\begin{equation}
    \exp \left( i\boldsymbol{q} \cdot \Delta \boldsymbol{R}_{\mu\nu} \right) ={} 4\pi \sum_l i^l j_l(q \Delta R_{\mu\nu}) \sum_{m_1 m_2} (-1)^{m_1} D^l_{m_2 m_1} \left( \phi_{\text{I}}^{(\text{lf})}, \theta_{\text{I}}^{(\text{lf})}, \chi_{\text{I}}^{(\text{lf})} \right) Y_l^{-m_1} \left( \theta_{\mu\nu}^{(\text{mf})}, \phi_{\mu\nu}^{(\text{mf})}\right) Y^{m_2}_l \left( \theta^{(\text{lf})}_q, \phi^{(\text{lf})}_q \right)
    \label{eq:ap:rotExp}
\end{equation}
%\end{widetext}
Here $\phi_{\text{I}}^{(\text{lf})}$, $\theta_{\text{I}}^{(\text{lf})}$, and $\chi_{\text{I}}^{(\text{lf})}$ are the conventional Euler angles in the lab frame that describe the orientation of the molecule's principal moments of inertia with respect to the lab frame.
%We express the diffraction intensity for a single molecule in terms of the molecular frame pairwise distances and angle by inserting Eq.~\ref{eq:rotExp} into the molecular diffraction sum in Eq.~\ref{eq:genInts}
%\begin{widetext}
\begin{equation}
    \begin{split} 
        I(\boldsymbol{q}&) ={} \mathcal{I} \bigg( \sum_\mu |f_\mu(q)|^2 + \sum_{\mu,\nu : \mu \neq \nu} \text{Re} \bigg\{ f_\mu(q) f^*_\nu(q) \sum_l 4 \pi i^l j_l(q \Delta R_{\mu\nu}) \\
        &\times \sum_{m_1, m_2} (-1)^{m_1}D^l_{m_2, m_1} \left( \phi_{\text{I}}^{(\text{lf})}, \theta_{\text{I}}^{(\text{lf})}, \chi_{\text{I}}^{(\text{lf})} \right)  Y_l^{-m_1} \left( \theta_{\mu\nu}^{(\text{mf})}, \phi_{\mu\nu}^{(\text{mf})}\right) Y^{m_2}_l \left( \theta^{(\text{lf})}_q, \phi^{(\text{lf})}_q \right) \bigg\} \bigg).
    \end{split} \label{eq:ap:ImolFrame}
\end{equation}
%\end{widetext}
We have now expressed the measurable diffraction (Eq.~\ref{eq:genInts}) in terms of the pairwise molecular frame distances and angles, as well as the lab frame angles $\left(\theta^{(\text{lf})}_q, \phi^{(\text{lf})}_q \right)$ that define $\boldsymbol{q}$.

%\paragraph*{From Eq.~\ref{eq:ImolFrame} we move to the realistic situation, and show the dependencies on anisotropy}
In gas-phase diffraction experiments one measures an ensemble of molecules at different orientations, alignments, and possibly differing structures depending on the populated rovibronic states.
One samples that ensemble at a variety of times relative to the evolving ensemble anisotropy, revealing the following observable,
%We handle this ensemble smearing fully quantum mechanically, which is different from the previous semi-classical approach CITE ZEWAIL, by taking the expectation value of $D^l_{m_2 m_1} \left( \phi_{I}^{(lf)}, \theta_{I}^{(lf)}, \chi_{I}^{(lf)} \right)$ and $Y_l^{-m_1} \left( \theta_{\mu\nu}^{(mf)}, \phi_{\mu\nu}^{(mf)}\right)$
%\begin{widetext}
\begin{equation}
    \begin{aligned}
        \langle I(\boldsymbol{q}) \rangle (t) &={}\mathcal{I} \bigg( \sum_\mu |f_\mu(q)|^2 + \sum_{\mu,\nu : \mu \neq \nu} \text{Re} \bigg\{ f_\mu(q) f^*_\nu(q)  \sum_l 4 \pi i^l \sum_{m_1, m_2}(-1)^{m_1} Y^{m_2}_l \left( \theta^{(\text{lf})}_q, \phi^{(\text{lf})}_q \right) \\
        & \hspace{1cm} \times  \ev**{D^l_{m_2 m_1} \left( \phi_{\text{I}}^{(\text{lf})}, \theta_{\text{I}}^{(\text{lf})}, \chi_{\text{I}}^{(\text{lf})} \right) j_l(q \Delta R_{\mu\nu}) Y_l^{-m_1} \left( \theta_{\mu\nu}^{(\text{mf})}, \phi_{\mu\nu}^{(\text{mf})}\right)}{\Psi(t)} \bigg\} \bigg)
    \end{aligned} \label{eq:ap:ImolGen}
\end{equation}
%\end{widetext}
where $\Psi(t)$ is the molecular ensemble wavefunction that describes both the rotational and vibronic dynamics of the system.
This is the general expression for the diffraction intensity from the entire molecular ensemble.

We have derived the expected diffraction intensity in terms of the momentum transfer vector, but in an experiment we do not have direct access to $\theta^{(\text{lf})}_q$ and  $\phi^{(\text{lf})}_q$.
Instead, we measure the lab frame diffraction signal on a 2d detector, parameterized by $q=|\boldsymbol{q}|$ and \thd.
The detector lies in the x-z plane of the lab frame where \thd{} is with respect to $\hat{\mathbf{z}}^{(\text{lf})}$.
\begin{align}
    \alpha &= 2 \sin^{-1}\left(\frac{q \lambda}{4\pi}\right) + \frac{\pi}{2} \label{eq:alphad}\\
    \mmthlf &= \cos^{-1} \left(\sin\left(\alpha\right)\cos\left(\mmthd \right)\right) \label{eq:ap:thetalf}\\
    \mmphlf &= \tan^{-1}\left( \frac{\cos(\alpha)}{\sin(\alpha)\sin(\mmthd)} \right) \label{eq:ap:philf}
\end{align}
Here, $\lambda$ is either the deBroglie wavelength of the electron probe, or the x-ray wavelength, and $\alpha$ is the scattering angle rotated by $\pi/2$.
For the 3.7~MeV electron probe at the SLAC Ultrafast Electron Diffraction facility \cite{shen.UED_machine.2019} $\lambda = 3.0\times10^{-3} \text{\AA}$ and the above relations simplify to
\begin{align*}
    \alpha &\approx \frac{\pi}{2} \\
    \mmthlf &\approx \mmthd \\
    \mmphlf &\approx 0
\end{align*}
%where the anisotropy on the detector would directly match the anisotropy of the molecular ensemble as we would intuitively expect from small angle scattering.
For x-ray diffraction at \textapprox10~keV this expression does not simplify due to larger x-ray scattering angles.
%We validate Eqs.~\ref{eq:ImolGen} and \ref{eq:alphad}-\ref{eq:philf} in appendix \ref{sc:validation} by fixing the ensemble distribution $\mathcal{A}^l_{m_1 m_2}(t)$ and comparing numerically calculated diffraction patterns with those calculated using Eqs.~\ref{eq:ImolGen}, \ref{eq:alphad}-\ref{eq:philf} and find agreement up to numerical precision.
Often, one uses a linearly polarized alignment pump pulse which induces cylindrical symmetry in the ensemble rotation wave packet, which results in $m_2=0$.
Equation~\ref{eq:ap:ImolFrame} is derived for an asymmetric top, for a symmetric top there is symmetry about the molecular frame z axis, which sets $m_1=0$.

It is difficult to extract $\Delta \boldsymbol{R}_{\mu\nu}$ from Eq.~\ref{eq:ap:ImolGen} in its current form since rovibronic coupling may affect the time-dependent anisotropy.
With rovibronic coupling, to calculate the ensemble anisotropy we may be required to simulate the excited state with the complex excited state simulations we do not want to rely on.
This coupling, therefore, may render the anisotropy calculation too difficult.
Instead, we consider two methods to separate the ensemble anisotropy and the molecular frame pairwise terms by assuming the molecular structure is rigid over the measurement period.
In doing so, we aim to separate the ensemble anisotropy from the molecular frame geometry.
To do this, we decompose the ensemble anisotropy into the Axis Distribution Moments (ADMs) by projecting the ensemble of molecular frame orientations, with respect to the lab frame (Fig.~\ref{fig:rotations}a), onto the Wigner D matrices,
\begin{align}
    \mathcal{A}^l_{mk}(t) &= \frac{2l+1}{8\pi^2}\expectationvalue**{D^{l}_{mk}\left(\phi_{\text{I}}^{(\text{lf})}, \theta_{\text{I}}^{(\text{lf})}, \chi_{\text{I}}^{(\text{lf})}\right)}{\Psi(t)} \\
    \left. \mathcal{A}^l_{mk}(t)\right|_{\text{rigid}} &= \frac{2l+1}{8\pi^2} \expectationvalue**{D^{l}_{mk}\left(\phi_{\text{I}}^{(\text{lf})}, \theta_{\text{I}}^{(\text{lf})}, \chi_{\text{I}}^{(\text{lf})} \right)}{\Psi_{\text{rigid}}(t)}.
\end{align}
Simulations of the rotational wavefunction for rigid symmetric and rigid asymmetric tops \cite{Rouzee.asymm_align.2008, Rouzee.ethylene_align.2006, hamilton.symm_align.2005, holmegaard.asymm_align.2007, xiaoming.alignment.2012, Tamar.alignment.2001, stapelfeldt.aligning_molecules.2003} produce good agreement with measured alignment signatures.
To extract $\Delta \boldsymbol{R}_{\mu\nu}$ from Eq.~\ref{eq:ap:ImolGen} we consider two approximations: the typical rigid rotor approximation and a separation of time scales.
%While we are able to get the alignment wavefunction, finding the full wavefunction and accounting for how the geometric changes caused by vibrations affect the molecules' moment of inertia is often prohibitively difficult.

\subsection{Rigid Rotor Approximation}

We first consider the rigid rotor approximation, which assumes the molecular structure is constant throughout the rotational dynamics.
This allows us to take the expectation value of the molecular structure (the molecular frame terms) with respect to the ground rovibronic state structure at $t=0$.
We may also calculate the ADMs with respect to the ground rovibronic state structure, which allows us to separate the ADMs from the molecular frame terms
%\begin{widetext}
\begin{equation}
\begin{aligned}
   \langle I(\boldsymbol{q}) \rangle_{\text{rigid}} (t) ={}\mathcal{I} \bigg( \sum_\mu & |f_\mu(q)|^2 + \sum_{\mu,\nu : \mu \neq \nu} \text{Re} \bigg\{ f_\mu(q) f^*_\nu(q) \sum_l \frac{32 \pi^3 i^l}{ 2l+1} \sum_{m_1, m_2} (-1)^{m_1} Y^{m_2}_l \left( \theta^{(\text{lf})}_q, \phi^{(\text{lf})}_q \right) \\
    \times & \ev**{ j_l(q \Delta R_{\mu\nu}) Y_l^{-m_1} \left( \theta_{\mu\nu}^{(\text{mf})}, \phi_{\mu\nu}^{(\text{mf})}\right)}{\Psi(0)} \left.\mmadm{l}{m_2}{m_1}\right|_{\text{rigid}} \bigg\} \bigg).
    \label{eq:ap:diffRigAvg}
\end{aligned}
\end{equation}
%\end{widetext}
This approximation is useful when investigating the vibronic ground state structure of a molecule or when the change in the molecule's structure has a negligible impact on the moments of inertia. 

\subsection{\label{ap:excited_state}A Separation of Timescales Approximation for Excited State Dynamics}

%\paragraph*{Separation of ro and vibration -- maybe save for discussion}
The second approximation is a separation of time scales between the rotational and vibronic dynamics.
The anisotropy signature, \adm, lasts of order one to tens of picoseconds for molecules with a few to tens of atoms, respectively. %check
When the vibration or isomerization occurs on a much faster timescale than the change in anisotropy, we can calculate the rotational dynamics with respect to the known ground rovibronic state structure rather than with the unknown excited state structure.
This disparity in timescales is very common, and this approximation is analogous to the Born-Oppenheimer approximation.

We first consider the more general case of a double pump pulse experiment that first induces a rotational wavepacket and then launches a vibronic wavepacket.
The first pulse increases the ensemble anisotropy and consequently the number of $C_{lmk}(q,t)$ coefficients.
The second pulse further mixes the rotational states while exciting vibronic modes.
Let $\tau$ denote the arrival time of the second vibration-inducing pulse after the first rotation-inducing pulse, and $t$ is the elapsed time after the second pump pulse.
%For a single pump pulse experiment $\tau = 0$.
%We note that by keeping $\tau$ constant we are assuming the ensemble anisotropy does not appreciably change in the time frame we wish to measure the vibrational dynamics.
%We again separate the molecular frame and anisotropy terms in order to take the expectation of the nuclear structure with respect to the separable vibration component of the wavefunction.

In our experiment, we initially start with a thermal ensemble often dominated by the vibronic ground state.
This ensemble is made of initial rovibronic states, each indexed by ($i$), in the Born-Oppenheimer basis as
\begin{equation}
    \ket{\psi^{(i)}(0)} = \ket{J^{(i)} M^{(i)} K^{(i)}}\ket{0}
    \label{eq:init_state}
\end{equation}
prior to any pulses.
After the alignment pulse, and before the vibration-inducing pulse, our coherent rotational state evolves as
\begin{equation}
    \ket{\psi^{(i)}(\tau)} = \sum_{J,M,K}c^{(i)}_{JMK}(\tau)\ket{JMK}\ket{0}.
    \label{eq:rot_wavepacket}
\end{equation}

The vibration pump pulse induces the excited state dynamics, while the photon's angular momentum mixes the rotational states.
We project the vibronically excited state onto the Born-Oppenheimer basis,
\begin{equation}
    \ket{\Psi^{(i)}(t,\tau)} = \sum_n \sum_{J_n,M_n,K_n} \ket{J_nM_nK_n}\ket{n} \bra{J_nM_nK_nn}\ket{\Psi^{(i)}(t,\tau)}
    \label{eq:psi_BO_decomposition}
\end{equation}
where the vibronic and rotational states are mixed by the vibronic ground state dipole moment and its orientation, respectively.
To calculate the coefficients we apply time-dependent perturbation theory and assume an impulsive excitation
%\begin{widetext}
\begin{align}
    \bra{J_nM_nK_nn}\ket{\Psi^{(i,2)}(t,\tau)} =& \sum_\gamma \mel**{J_nM_nK_nn}{D^{1*}_{0\gamma} \mu^1_\gamma}{\psi^{(i)}(t,\tau)}\frac{-i}{\hbar}\int_0^\infty E_0(t')e^{-i\Delta Et'/\hbar}dt' \\
    =& \tilde{E}\sum_{J,M,K}c^{(i)}_{JMK}(\tau+t) \sum_\gamma A(J_n,J^{(i)}; K_n, K^{(i)}; M^{(i)},\gamma) \bra{n}\mu^1_\gamma \ket{0} \label{eq:overlap_2pulse} \\
    A(J_n,J; K_n, K; M, \gamma) \equiv& \bra{J_nMK_n} D^{1*}_{0\gamma}\ket{JMK} \\
    =& \sqrt{(2J+1)(2J_n+1)}(-1)^{\gamma+K-M} \threeJ{J}{1}{J_n}{-M}{0}{M} \threeJ{J}{1}{J_n}{-K}{-\gamma}{K_n}.
\end{align}
%\end{widetext}
where $\mu^1_\gamma$ is the spherical tensor of the transition dipole moment operator, $E_0(t')$ is the electric field of the vibration-inducing pulse, $\Delta E$ is the energy difference between the initial rotational state and the excited state, and
\begin{equation}
        \tilde{E} ={} \int_0^\infty E_0(t')e^{-i\Delta Et'/\hbar}dt'.
\end{equation}
Plugging Eq.~\ref{eq:overlap_2pulse} into Eq.~\ref{eq:psi_BO_decomposition} we retrieve the Born-Oppenheimer pure state immediately after the second excitation pulse
\begin{align}
        \ket{\Psi^{(i,2)}(0, \tau)} &= \tilde{E} \sum_{n,J_n,M,K_n}  \ket{nJ_nMK_n}  X^{(i,2)n}_{J_nK_n}(M;0,\tau) \label{eq:ap:psi_rhovib_t0}\\
       X^{(i,2)n}_{J_n K_n}(M;t,\tau) &= \tilde{E} \sum_{J,K} c^{(i)}_{JMK}(\tau+t) \sum_\gamma \bra{n}\mu^{1}_\gamma\ket{0}
        A(J_n,J^{(i)}; K_n, K^{(i)}; M,\gamma).
\end{align}
Since we are interested in the time dynamics of the vibronic state, we apply the time translation operator to Eq.~\ref{eq:ap:psi_rhovib_t0}.
\begin{align}
        \ket{\Psi^{(i,\alpha)}(t, \tau)} &= \sum_n \sum_{J_n,M,K_n} X^{(i,\alpha)n}_{J_nK_n}(M;t,\tau) \ket{\psi^n_\text{el-vib}(t)}\ket{J_nMK_n} \label{eq:ap:psi_rhovib}\\
    \ket{\psi^n_{\text{el-vib}}(t)} &= \hat{U}(t)\ket{n}
\end{align}

Now that we've calculated the time-dependent rovibronic state in the Born-Oppenheimer basis, we must apply it to our measurement.
We do this by taking the expectation value of our diffraction observable (Eq.~\ref{eq:ap:ImolGen}) with respect to our new rovibronic system, Eq.~\ref{eq:ap:psi_rhovib}.
Here, we only look at the molecular scattering term since it is the only term affected by $\ket{\Psi(t,\tau)}$ and pull out the scattering amplitudes $f_\mu(q)$ due to the independent atom approximation.
%\begin{widetext}
\begin{align}
    \begin{split}
    \langle I(\boldsymbol{q}) \rangle^{(2)} (t, \tau)|_{\text{mol}}={}& \mathcal{I} \sum_{\mu,\nu : \mu \neq \nu} \text{Re} \bigg\{ f_\mu(q) f^*_\nu(q) \sum_l 4\pi i^l \sum_{m_1, m_2} (-1)^{m_1}  Y^{m_2}_l \left( \theta^{(\text{lf})}_q, \phi^{(\text{lf})}_q \right) \\
    & \times \ev**{D^l_{m_2 m_1} \left( \phi_{\text{I}}^{(\text{lf})}, \theta_{\text{I}}^{(\text{lf})}, \chi_{\text{I}}^{(\text{lf})} \right) j_l(q \Delta R_{\mu\nu}) Y_l^{-m_1} \left( \theta_{\mu\nu}^{(\text{mf})}, \phi_{\mu\nu}^{(\text{mf})}\right)}{\Psi(t,\tau)} \bigg\}
    \end{split} \label{eq:ap:Imol_2} \\
    \begin{split}
    \langle I(\boldsymbol{q}) \rangle_{\text{sep}}^{(2)} (t, \tau)|_{\text{mol}} ={}&\mathcal{I} \sum_{\mu,\nu : \mu \neq \nu} \text{Re} \bigg\{ f_\mu(q) f^*_\nu(q) \sum_l \frac{32\pi^3 i^l}{2l+1} \sum_{m_1, m_2} (-1)^{m_1}  Y^{m_2}_l \left( \theta^{(\text{lf})}_q, \phi^{(\text{lf})}_q \right) \\
    & \times \sum_{n,n'} \tilde{\mathcal{A}}^{(\alpha) l}_{m_2m_1} (n,n'; t,\tau) \mel**{\psi^{n'}_{\text{el-vib}}(t)}{j_l(q \Delta R_{\mu\nu}) Y_l^{-m_1} \left( \theta_{\mu\nu}^{(\text{mf})}, \phi_{\mu\nu}^{(\text{mf})}\right)}{\psi^n_{\text{el-vib}}(t)}  \bigg\}
    \end{split} \label{eq:ap:Imol_BO} \\
    \tilde{\mathcal{A}}^{(\alpha) l}_{m_2m_1} (n,n'; t,\tau) \equiv{}& \frac{1}{Z} \sum_i p^{(i)} \\
    & \times \sum_{J_n,M,K_n}\sum_{J_{n'}',M',K_{n'}'}
    X^{(\alpha,i) n}_{J_nK_n}(M;t,\tau)X^{(\alpha,i) n'*}_{J_{n'}'K_{n'}'}(M';t,\tau) \bra{J_{n'}'M'K_{n'}'} D^l_{m_2 m_1} \ket{J_nMK_n} \label{eq:ap:ADMs_rovib}
\end{align}
%\end{widetext}
Here $Z$ is the partition function of the initial rotational states and $p^{(i)}$ is the Boltzmann weighting factor for the initial rotational states.

We have again separated the ensemble anisotropy $(\tilde{\mathcal{A}}^{(\alpha) l}_{m_1m_2} (n,n'; t,\tau))$ from the molecular frame structure term, which includes all the vibronic dynamics.
The modified ADMs, $\tilde{\mathcal{A}}^{(\alpha) l}_{mk} (n,n'; t,\tau)$, are analogous to the original ADMs, but now include the coherent rotational mixing from the vibronic inducing pulse.
That is, each vibronic state will have its own rotational coherence that must be accounted for when calculating the ensemble anisotropy.
Finally, plugging the molecular diffraction term (Eq.~\ref{eq:ap:Imol_BO}) into the full diffraction expression we get
%\begin{widetext}
\begin{equation}
\begin{aligned}
    \langle I(\boldsymbol{q}) \rangle_{\text{sep}}^{(2)} (t, \tau) ={} & \mathcal{I} \bigg( \sum_\mu |f_\mu(q)|^2 + \sum_{\mu,\nu : \mu \neq \nu} \text{Re} \bigg\{ f_\mu(q) f^*_\nu(q) \sum_l \frac{32\pi^3 i^l}{2l+1} \sum_{m_1, m_2} (-1)^{m_1} Y^{m_2}_l \left( \theta^{(\text{lf})}_q, \phi^{(\text{lf})}_q \right) \\
    &\times \sum_{n,n'} \tilde{\mathcal{A}}^{(2)l}_{m_2m_1} (n,n';t, \tau) \mel**{\psi^{n'}_{\text{el-vib}}(t)}{j_l(q \Delta R_{\mu\nu}) Y_l^{-m_1} \left( \theta_{\mu\nu}^{(\text{mf})}, \phi_{\mu\nu}^{(\text{mf})}\right)}{\psi^n_{\text{el-vib}}(t)}  \bigg\} \bigg).
\end{aligned}
\label{eq:ap:IGen_BO}
\end{equation}
%\end{widetext}

Due to the difference in timescales between the rotational and vibrational dynamics, we further simplify Eq.~\ref{eq:ap:IGen_BO}.
In its current form, Eq.~\ref{eq:ap:IGen_BO} relies on updating the ensemble anisotropy calculation as the structure changes with vibration.
This requires us to know what the structure will be at time $t$, which is what we are ultimately trying to solve for.
Instead, when the change in ensemble anisotropy is negligible with respect to the timescale of the vibration we can hold the anisotropy constant 
\begin{align}
    \tilde{\mathcal{A}}^{(2) l}_{m_2m_1} (n,n';\tau) ={}& \tilde{\mathcal{A}}^{(2) l}_{m_2m_1} (n,n';0, \tau) \approx \tilde{\mathcal{A}}^{(2) l}_{m_2m_1} (n,n';t, \tau) \label{eq:ap:rot_approx_double} \\
    \begin{split}
    \langle I(\boldsymbol{q}) \rangle_{\text{sep}}^{(2)} (t, \tau) \approx{} & \mathcal{I} \bigg( \sum_\mu |f_\mu(q)|^2 + \sum_{\mu,\nu : \mu \neq \nu} \text{Re} \bigg\{ f_\mu(q) f^*_\nu(q) \sum_l \frac{32\pi^3 i^l}{2l+1} \sum_{m_1, m_2} (-1)^{m_1}  Y^{m_2}_l \left( \theta^{(\text{lf})}_q, \phi^{(\text{lf})}_q \right) \\
    & \times \sum_{n,n'} \tilde{\mathcal{A}}^{(2) l}_{m_2m_1} (n,n';\tau) \mel**{\psi^{n'}_{\text{el-vib}}(t)}{j_l(q \Delta R_{\mu\nu}) Y_l^{-m_1} \left( \theta_{\mu\nu}^{(\text{mf})}, \phi_{\mu\nu}^{(\text{mf})}\right)}{\psi^n_{\text{el-vib}}(t)}  \bigg\} \bigg).
    \end{split}
    \label{eq:ap:IGen_BO_sep}
\end{align}
In doing so, the ensemble anisotropy and vibronic structural dependence are completely separable.
We, therefore, continue to calculate the ensemble anisotropy with respect to the ground rovibronic state structure.

In some cases, a single-pump pulse experiment is preferred over a two-pump pulse experiment when the setup is too difficult or when the anisotropy is difficult to induce or measure.
In such a case, we do not initially induce a rotational wavepacket and our initial state is given by Eq.~\ref{eq:init_state} instead of Eq.~\ref{eq:rot_wavepacket}.
Therefore, one does not sum over a coherent set of rotational states in Eq.~\ref{eq:overlap_2pulse} and 
\begin{equation}
    X^{(i,1)n}_{J_n K_n}(M;t) = \tilde{E} \sum_\gamma \bra{n}\mu^{1}_\gamma\ket{0}
        A(J_n,J^{(i)}; K_n, K^{(i)}; M,\gamma)
\end{equation}
Here, the ensemble anisotropy is imprinted immediately after the pulse by the interaction between the polarized laser and the excitation dipole.
%\begin{widetext}
\begin{align}
    \tilde{\mathcal{A}}^{(1) l}_{m_2m_1} (n,n') ={} & \tilde{\mathcal{A}}^{(1) l}_{m_2m_1} (n,n';0) \approx \tilde{\mathcal{A}}^{(1) l}_{m_2m_1} (n,n';t) \label{eq:ap:rot_approx_single} \\
    \begin{split}
    \langle I(\boldsymbol{q}) \rangle_{\text{sep}}^{(1)} (t) \approx{} & \mathcal{I} \bigg( \sum_\mu |f_\mu(q)|^2 + \sum_{\mu,\nu : \mu \neq \nu} \text{Re} \bigg\{ f_\mu(q) f^*_\nu(q) \sum_l \frac{32\pi^3 i^l}{2l+1} \sum_{m_1, m_2} (-1)^{m_1} Y^{m_2}_l \left( \theta^{(\text{lf})}_q, \phi^{(\text{lf})}_q \right) \\
    &\times \sum_{n,n'} \tilde{\mathcal{A}}^{(1)l}_{m_2m_1} (n,n') \mel**{\psi^{n'}_{\text{el-vib}}(t)}{j_l(q \Delta R_{\mu\nu}) Y_l^{-m_1} \left( \theta_{\mu\nu}^{(\text{mf})}, \phi_{\mu\nu}^{(\text{mf})}\right)}{\psi^n_{\text{el-vib}}(t)}  \bigg\} \bigg).
    \end{split}
    \label{eq:ap:IGen_BO_sep_single}
\end{align}
%\end{widetext}

Depending on the system, one may further improve this approximation by calculating the ensemble dynamics with respect to a reference structure for $t>0$.
In some cases, the vibronic transience may be on the timescale of the rotational transience.
Once Eqs.~\ref{eq:ap:rot_approx_double} or \ref{eq:ap:rot_approx_single} no longer hold at some time $t$ there are two options.
Firstly, one can use only $C_{000}(q,t)$ which does not rely on anisotropy and Eq.~\ref{eq:ap:IGen_BO} will be exact.
Secondly, one can continue calculating $\tilde{\mathcal{A}}^{(\alpha) l}_{m_2m_1} (n,n';t, \tau)$ with respect to a reference structure.
For example, if one knows an excited state structure is similar to the ground rovibronic state one can continue to use $\tilde{\mathcal{A}}^{(\alpha) l}_{m_2m_1} (n,n';t, \tau)$.
One must prove this through \textit{a priori} knowledge or through the retrieved structures at earlier times.
In the case that the dynamics do not deviate from some other known structure one may calculate the $\tilde{\mathcal{A}}^{(\alpha) l}_{m_2m_1} (n,n';t, \tau)$ with respect to this structure.

%% file: sections/ap_fitting_legendres_ADMs.tex
Our method relies heavily on two fitting procedures that will likely be the most important steps of the analysis as they define the $C_{lmk}(q)$ coefficients and $\sigma_{lmk}(q)$.
Below, we describe how one performs these fits analytically by minimizing the $\chi^2$.
These analytical methods, however, will struggle to fit the measured time dependence with ADMs if there is not enough anisotropy and/or there is poor SNR.
We highly encourage one to explore molecule-specific systematics to $C_{lmk}(q)$ by fitting simulated diffraction patterns.
One can employ L1 regularization techniques to improve these fits.
Since the derivative of $|x|$ is undefined at $x=0$ and we do not know the sign of the \blm{} and $C_{lmk}(q)$, one will need to employ coordinate or gradient descent methods when using L1 regularization.
Gradient descent will be much slower for numerous fits and should be used if the analytical approach is insufficient.
Coordinate descent is much faster than gradient descent but will likely be considerably slower as well.
%We will describe some common mistakes that may lead to incorrect results for each of the two 
fits.

Minimize the $\chi^2$ is the weighted least squares regression problem
\begin{equation}
\begin{split}
    L &= \frac{1}{2}\sum_\mu \frac{\left( \sum_\nu X_{\mu\nu} F_\nu - Y_\mu\right)^2}{\epsilon_\mu} \\
        &= \frac{1}{2}\left(\mathbf{XF} - \mathbf{Y}\right)^T \mathbf{W} \left(\mathbf{XF} - \mathbf{Y}\right).
\end{split}
\label{eq:ap:fitLoss}
\end{equation}
Here, $\mathbf{Y}$ is the data vector we wish to fit, the matrix $\mathbf{X}$ are the fit bases (features) that span the columns, $\mu$ sums over the detector pixels, and $\nu$ sums over the fit bases.
The bases are scaled by the fit coefficients $\mathbf{F}$ and each data point's contribution to the fit is weighted by $\mathbf{W}$, where
\begin{align}
    \epsilon_\mu &= \text{Var}\left(Y_\mu\right) \\
    \mathbf{W} &=
        \begin{bmatrix}
        \frac{1}{\epsilon_0} & 0 & \dots & 0 \\
        0 & \frac{1}{\epsilon_1} & & \\ 
        \vdots &  & \ddots & \\
        0 & & & \frac{1}{\epsilon_{N}}.
        \end{bmatrix} \label{eq:ap:Wgen}
\end{align}

We will discuss two common ways to solve Eq.~\ref{eq:ap:fitLoss} for the optimal fit coefficients.
The first method uses the pseudoinverse to minimize Eq.~\ref{eq:ap:fitLoss} and is commonly referred to as the normal equation.
\begin{equation}
    \mathbf{F} = \left( \mathbf{X}^T \mathbf{WX}\right)^{-1} \mathbf{X}^T \mathbf{WY}. \label{eq:ap:genNormEqn}
\end{equation}
The second method sets Eq.~\ref{eq:ap:fitLoss} to 0 and uses the QR decomposition to invert $\mathbf{X}$
\begin{align}
    \sqrt{\mathbf{W}}(\mathbf{XF} - \mathbf{Y}) &= 0 \\
    \widetilde{\mathbf{Y}} &= \sqrt{\mathbf{W}}\mathbf{Y} \\
    \widetilde{\mathbf{X}} &= \sqrt{\mathbf{W}}\mathbf{X} \\
        &= \mathbf{QR} \label{eq:ap:qr}\\
    \mathbf{F} &= \boldsymbol{R}^{-1}\mathbf{Q}^T\widetilde{\mathbf{X}} \label{eq:ap:qrfit}
\end{align}
where $\sqrt{\mathbf{W}}$ is the Cholesky decomposition and Eq~\ref{eq:ap:qr} is the QR decomposition.
The QR decomposition has a lower condition number and produces a more accurate $\mathbf{F}$.
In this work, we used the normal equation for the measured N$_2$O data and found sufficient agreement with literature values.
This may be a function of our poor SNR.
We, however, encourage the reader to use Eq.~\ref{eq:ap:qrfit} and the more accurate QR decomposition.

To retrieve the \blm{} coefficients, we fit the measured data, $\langle I(\boldsymbol{q}(\mmthd), t\rangle$, with the spherical harmonics, $Y_l^m \left( \mmthlf, \mmphlf \right)$.
Where Eqs.~\ref{eq:ap:thetalf} and \ref{eq:ap:philf} relate \thlf{} and \phlf{} in terms of \thd.
Although the spherical harmonics are orthonormal, this orthonormality is broken by the finite sampling of our detector.
To account for this now nonzero overlap between different bases, we fit the spherical harmonics to the data instead of projecting onto them.
This is most noticeable at low $q$ where one often has the best SNR and the fewest bins to resolve \thd.
We note that this can still be necessary for the isotropic component due to the Jacobian.
We use the trapezoidal rule to increase the orthonormality of our binned spherical harmonics
\begin{align}
    X_{\mu\nu} &= \frac{1}{2} \left(Y_l^\nu(\mmthlfu{\mu}, \mmphlfu{\mu}) + Y_l^\nu(\mmthlfu{\mu+1}, \mmphlfu{\mu+1})\right) \label{eq:ap:XB}\\
    Y_\mu &= \frac{1}{2}\left(\langle I(q,\mmthd_\mu,t)\rangle + \langle I(q,\mmthd_{\mu+1},t)\rangle \right) \\
    \epsilon_\mu &= \frac{1}{2} \left( \text{Var}\left(\langle I(q,\mmthd_\mu,t) \rangle \right) + \text{Var}\left(\langle I(q,\mmthd_{\mu+1},t)\rangle\right) \right) \\
    F_\nu &= B_l^\nu(q,t).
    %F(q, \mmthd, t) &= \sum_{l,m}B^l_m(q,t) Y_l^m(\mmthlf, \mmphlf) = \mathbf{X}\mathbf{B} \label{eq:FBc}\\
    %\widetilde{F}_i(q, t) &=  \frac{1}{2}\left(F(q, \mmthd_i, t) + F(q, \mmthd_{i+1}, t)\right) \\
    %\widetilde{D}_i(q,t) &= \frac{1}{2} \left(D(q,\mmthd_i,t) + D(q,\mmthd_{i+1},t)\right) \\
    %\widetilde{\epsilon}_i(q,t) &= \frac{1}{2} \left( \text{Var}\left(D(q,\mmthd_i,t)\right) + \text{Var}\left(D(q,\mmthd_{i+1},t)\right) \right) \\
    %\widetilde{\mmthlf}_i &= \frac{1}{2} \left( \mmthlf_i + \mmthlf_{i+1} \right) \\
    %\Delta \mmthlf_i &= \mmthlf_{i+1} - \mmthlf_i \label{eq:DBT}
\end{align}
%where $F$ is our spherical harmonic series we wish to fit to the diffraction data ($D$).
%The matrix $\mathbf{B}$ is a column vector of all the $B^l_m(q,t)$ coefficients, length $M$, and $\mathbf{X}$ is a matrix where each column is the corresponding $Y_l^m(\mmthlf, \mmphlf)$.
Since $Y_l^m \left( \mmthlf, \mmphlf \right) \propto P_l^m \left(\cos(\mmthlf)\right)$ we must consider the $\cos\left(\mmthlf\right)$ Jacobian when summing over $\mmthlfu{\mu}$. 
This can be resolved in two ways, by rebinning \thd{} in equally sized $\cos(\mmthlf)$ bins, or by introducing the Jacobian into $\mathbf{W}$.
Since rebinning reduces our resolution, we alter the weight matrix
\begin{align}
    %L^l_m(q,t) &= \frac{1}{2}\sum_i \frac{\left( \widetilde{F}_i(q,t ) - \widetilde{D}_i(q,t)\right)^2}{\widetilde{\epsilon}_i(q,t)} \sin\left(\widetilde{\mmthlf}_i\right) \Delta \mmthlf_i\\
     %   &= \frac{1}{2}\left(\widetilde{\mathbf{F}}\left(q, t \right) - \widetilde{\mathbf{D}}\left(q, t\right)\right)^T \mathbf{W} \left(\widetilde{\mathbf{F}}\left(q,t\right) - \widetilde{\mathbf{D}}\left(q,t\right)\right) \label{eq:fitLoss}\\
    \widetilde{\mmthlfu{\mu}} &= \frac{1}{2} \left( \mmthlfu{\mu} + \mmthlfu{\mu+1} \right) \\
    \Delta \mmthlfu{\mu} &= \mmthlfu{\mu+1} - \mmthlfu{\mu} \label{eq:DBT} \\
    \mathbf{W} &= 
        %\mathbb{1}
        %\begin{bmatrix}
        %\frac{\sin\left(\widetilde{\mmthlfu{0}}\right) \Delta %\mmthlfu{0}}{\epsilon_0} \vspace{1mm}\\
        %\frac{\sin\left(\widetilde{\mmthlfu{1}}\right) \Delta %\mmthlfu{1}}{\epsilon_1}\\
        %\vdots \\
        %\frac{\sin\left(\widetilde{\mmthlfu{N-1}}\right) \Delta %\mmthlfu{N-1}}{\epsilon_{N_{bins}-1}}
        %\end{bmatrix} 
        \begin{bmatrix}
        \frac{\sin\left(\widetilde{\mmthlfu{0}}\right) \Delta \mmthlfu{0}}{\epsilon_0} & 0 & \dots & 0 \\
        0 & \frac{\sin\left(\widetilde{\mmthlfu{1}}\right) \Delta \mmthlfu{1}}{\epsilon_1} & & \\ 
        \vdots &  & \ddots & \\
        0 & & & \frac{\sin\left(\widetilde{\mmthlfu{N-1}}\right) \Delta \mmthlfu{N-1}}{\epsilon_{N-1}} \label{eq:ap:WB}
        \end{bmatrix} 
\end{align}
where $d\left[\cos(\mmthlf)\right] \approx \sin(\mmthlf) \Delta \mmthlf$.
%Where $L^l_m(q,t)$ is our loss function and the second line is the matrix from of the above expression.
%Using Eq.~\ref{eq:fitLoss} we solve for the optimal $B^l_m(q,t)$ coefficients
%\begin{equation}
%    \mathbf{B} = \left(\widetilde{\mathbf{X}}^T\mathbf{W}\widetilde{\mathbf{X}}\right)^{-1} \widetilde{\mathbf{X}}^T\mathbf{W}\widetilde{\mathbf{D}} \label{eq:normalEqn}
%\end{equation}

Now we focus on retrieving the $C_{lmk}(q)$ coefficients by fitting the \adm{} to the \blm{} coefficients.
The \adm{} are likely not orthogonal and may vary strongly in their magnitude (L2 norm).
Consequently, the fit results from ADMs bases with larger magnitudes can easily skew the results of other bases with lower magnitudes.
These skews can completely ruin the fit for the lower magnitude bases, while not being noticeable in the fits of the larger magnitude bases.
This issue is also mitigated, or exacerbated, by increasing or decreasing the SNR, respectively.
Another way to mitigate this issue is to add regularization terms to Eq.~\ref{eq:ap:fitLoss}, which will alter Eq.~\ref{eq:ap:genNormEqn}.
One would ideally like to use L1 regularization for sparsity, but for the reasons mentioned above one would need to use the gradient descent, which is much slower.
To use Eqs.~\ref{eq:ap:fitLoss} and \ref{eq:ap:genNormEqn}, one must make the following alterations:
\begin{align}
    X_{\mu\nu} &= \frac{1}{2} \left(\mathcal{A}^l_{m_2\nu}(t_\mu) + \mathcal{A}^l_{m_2\nu}(t_{\mu+1})\right) \label{eq:ap:XC}\\
    Y_\mu &= \frac{1}{2}\left(B_l^{m_2}(q,t_\mu) + B_l^{m_2}(q,t_{\mu+1})\right) \\
    \epsilon_\mu &= \frac{1}{2} \left( \text{Var}\left(B_l^{m_2}(q,t_\mu)\right) + \text{Var}\left(B_l^{m_2}(q, t_{\mu+1})\right) \right) \\
    F_\nu &= C_{lm_2\nu}(q) \\
    \mathbf{W} &=
        \begin{bmatrix}
        \frac{1}{\epsilon_0} & 0 & \dots & 0 \\
        0 & \frac{1}{\epsilon_1} & & \\ 
        \vdots &  & \ddots & \\
        0 & & & \frac{1}{\epsilon_{N}}
        \end{bmatrix} \label{eq:ap:WC}.
    %F(q, t) &= \sum_{k}C_{lmk}(q,t) \mathcal{A}^l_{mk}(t) = \mathbf{X}\mathbf{C} \label{eq:FCc}\\
    %\widetilde{F}_i(q) &=  \frac{1}{2}\left(F(q, t_i) + F(q, t_{i+1})\right) \\
    %\widetilde{B}^l_{m,i}(q) &= \frac{1}{2} \left(B^l_m(q,t_i) + B^l_m(q,t_{i+1})\right) \\
    %\widetilde{\epsilon}_i(q) &= \frac{1}{2} \left( \text{Var}\left(B^l_m(q,t_i)\right) + \text{Var}\left(B^l_m(q,t_{i+1})\right) \right) \\
    %\mathbf{W} &=
    %    \begin{bmatrix}
    %    \frac{1}{\widetilde{\epsilon}_0(q)} & 0 & \dots & 0 \\
    %    0 & \frac{1}{\widetilde{\epsilon}_1(q)} & & \\ 
    %    \vdots &  & \ddots & \\
    %    0 & & & \frac{1}{\widetilde{\epsilon}_{N_{time}-1}(q)}
    %    \end{bmatrix} \label{eq:WC}
\end{align}

To improve the $C_{lmk}(q)$ fitting, one can increase the SNR or induce a broader rotational wavepacket.
In Fig.~\ref{fig:Mlmk_vs_error}b we illustrate how increasing the SNR improves the $C_{lmk}(q)$ resolution.
We also expect that broadening the rotational wavepacket and reducing the ensemble temperature will have a similar effect on the $C_{lmk}(q)$ to increasing the SNR as it does to \sth{} in Fig.~\ref{fig:trends}.
Again, we recommend that one runs these fit methods on the vibronic ground state structure with simulated ADMs to see which $C_{lmk}(q)$ coefficients will be retrieved with the expected anisotropy and SNR.

%% file: sections/ap_mcmc.tex
To measure $|\Psi(\boldsymbol{R})|^2$ we analytically relate the data's dependence on $|\Psi(\boldsymbol{R})|^2$ and determine a model to describe $|\Psi(\boldsymbol{R})|^2$ and its dependence on said data.
To aid the reader through this section, they may simultaneously read a simplified toy problem in Ref.~\cite{Hegazy.dissertation.2023}, which follows this discussion step by step.
Using Eq.~\ref{eq:ap:diffRigAvg} we isolate the molecular structure terms and gain access to $|\Psi(\boldsymbol{R})|^2$, as shown in Eqs.~\ref{eq:diffRigAvg}-\ref{eq:intEqn}.
\begin{align}
    C_{lmk}(q) &={} \int H_{lmk} \left( q, \boldsymbol{R} \right) \left| \Psi \left(\boldsymbol{R}\right) \right|^2 d\boldsymbol{R} \label{eq:ap:intEqn}\\
    H_{lmk} \left( q, \boldsymbol{R} \right) &={} \mathcal{I} \text{Re} \bigg\{ (-1)^{k} \frac{32 \pi^3 i^l}{2l+1} \sum_{\mu,\nu : \mu \neq \nu} |f_\mu(q)| |f_\nu(q)| j_l(q\Delta R_{\mu\nu}) Y_l^{-k} \left( \theta_{\mu\nu}^{(\text{mf})}, \phi_{\mu\nu}^{(\text{mf})}\right)\bigg\}.
    \label{eq:ap:MHAfit}
\end{align}

We approximate $|\Psi(\boldsymbol{R})|^2$ by choosing a probabilistic model that best describes our data, which we denote as \prth.
Our model \prth{} is parameterized by $\boldsymbol{\Theta}$ and dependent on the measured $C_{lmk}(q)$ coefficients, here denoted as $C$.
We now rewrite Eq.~\ref{eq:ap:intEqn} with our new model as
\begin{equation}
    C_{lmk}^{(\text{calc})}(q, \boldsymbol{\Theta}) = \int H_{lmk} \left( q, \boldsymbol{R} \right) \mmprth d\textbf{r} \label{eq:ap:intEqnSim}.
\end{equation}
Some possible forms of \prth{} include a multidimensional delta function which is analogous to a single structure, a normal distribution of structures that would describe the vibronic ground state, or harmonic oscillator eigenfunctions to describe a vibrational wavefunction.
In this work, we focus on the following \prth{} and their corresponding $\boldsymbol{\Theta}$
%\begin{widetext}
\begin{align}
    \mmprth &\approx \left| \Psi \left(\boldsymbol{R} \right) \right|^2\\
    \mmprthd &= \delta\left(\boldsymbol{\Theta}^{(\text{delta})} - \boldsymbol{R}\right) \\
    \boldsymbol{\Theta}^{(\text{delta})} &= \left[ \mmmrnoa, \mmmrnob, \mmmaono \right] \\
    \mmprthg &= \frac{1}{\sqrt{2\pi}^{N_{dof}} \prod^{i<N_{dof}}_{i=0} \boldsymbol{\Theta}^{(\text{gauss})}_{2i+1}}  \exp \Bigg \{ \frac{-1}{2}  \sum_{i=0}^{i<N_{dof}} \left( \frac{\boldsymbol{\Theta}^{(\text{gauss})}_{2i} - \boldsymbol{R}_i}{\boldsymbol{\Theta}^{(\text{gauss})}_{2i+1}} \right)^2 \Bigg \} \\
    \boldsymbol{\Theta}^{(\text{gauss})} &= \left[ \mmmrnoa, \mmsrnoa, \mmmrnob, \mmsrnob, \mmmaono, \mmsaono  \right]. \label{eq:ap:post_gauss}
\end{align}
%\end{widetext}

Given our model \prth, we use Bayesian Inference and Markov Chain Monte Carlo (MCMC) techniques to find the optimal $\boldsymbol{\Theta}$ parameters (\thopt) that best describe the observed $C_{lmk}(q)$.
Bayesian Inference encompasses methods that use Bayes' Theorem to update the hypothesis \cite{box.bayesian_inference.2011,foreman.mcmc.2013}.
The most time, and computationally, intensive step of this analysis is building the posterior \ptheta, which we define through Baye's Theorem
\begin{equation}
    P\left(\boldsymbol{\Theta} | C \right) = \frac{P\left(C| \boldsymbol{\Theta}\right) P\left( \boldsymbol{\Theta}\right)}{P\left(C\right)}\label{eq:ap:bayes_thrm}.
\end{equation}

Here, $P\left(C| \boldsymbol{\Theta}\right)$ is the likelihood function which is the probability of measuring the data $C$ given our selected model with the given $\boldsymbol{\Theta}$ parameters.
The likelihood probability plays the largest role in building the posterior and is how information from the data enters the analysis.
This can be calculated by assuming each $C_{lmk}(q)$ measurement in $q$ is its own experiment that results in a probability distribution.
That is, given many measurements $(N_{\text{images}})$ one builds a distribution of events for $C_{lmk}(q)$ which quickly becomes a normal distribution, due to the Central Limit Theorem, with a mean and standard error of the mean.
To calculate $P\left(C| \boldsymbol{\Theta}\right)$ one must multiply all of these probabilities
\begin{equation}
        P\left(C | \boldsymbol{\Theta}\right) = \left[ \prod_{lmk,q} \frac{1}{\sigma_{lmk}(q)\sqrt{2\pi}}\right] \exp{\frac{-1}{2} \sum_{lmk,q} \left(\frac{C_{lmk}(q) - C^{(\text{calc})}_{lmk}(q,\boldsymbol{\Theta})}{\sigma_{lmk}(q)} \right)^2}
    \label{eq:ap:MHlikelihood}
\end{equation}
where $\sigma_{lmk}(q)$ is the standard error of the mean of $C_{lmk}(q)$.
Since $\sigma_{lmk}(q) \propto 1/\sqrt{N_{\text{images}}}$, the summation in Eq.~\ref{eq:ap:MHlikelihood} scales as $N_{\text{images}}$.
By measuring more photons or electrons, one exponentially sharpens the probability distribution \ptheta.
As mentioned above, this assumes that each $C_{lmk}(q)$ is an independent measurement which is not the case with sufficiently large x-ray/electron beams which have widths larger than the detector pixels.
In such a scenario, one must alter Eq.~\ref{eq:ap:MHlikelihood} to account for this lack of independence.

The prior probability, $P\left(\boldsymbol{\Theta}\right)$, describes the likelihood of a given $\boldsymbol{\Theta}$.
Since $P\left(\boldsymbol{\Theta}\right)$ does not depend on data, it encapsulates our prior knowledge of the $\boldsymbol{\Theta}$ parameters.
Because we do not want to bias our search through $\boldsymbol{\Theta}$-space we define
\begin{equation}
    P\left(\boldsymbol{\Theta}\right) = e^{K\left(\boldsymbol{\Theta}\right)}
\end{equation}
where $K(\boldsymbol{\Theta})=0$ for physical values and $K(\boldsymbol{\Theta})=-\infty$ for unphysical values: $\boldsymbol{\Theta} < 0$ or $\mmmaono > \pi$.

The marginal likelihood, $P(C)$, is the probability of observing our measured data.
This probability is not something we concern ourselves with.
Since it is not dependent on $\boldsymbol{\Theta}$ it is a constant that we cancel out in our MCMC technique.

Having chosen a model to approximate $|\Psi(\boldsymbol{R})|^2$, employed Bayesian Inference to define the posterior (\ptheta) in terms of the $C_{lmk}(q)$ coefficients, we now use MCMC techniques to build for \ptheta.
%To do this, we have directly related $|\Psi(\boldsymbol{r})|^2$ and the molecular frame structure parameters to the data, modelled $|\Psi(\boldsymbol{r})|^2$ with an explicit dependence on the data (Eq.~\ref{eq:ap:bayes_thrm}), and related the posterior to to the measured data (Eq.~\ref{eq:ap:MHlikelihood}).
We ultimately aim to invert a system of integral equations, but the complexity of Eq.~\ref{eq:ap:MHAfit} greatly limits the available methods to solve for \ptheta.
For NO$_2$, we have 6 $C_{lmk}$ coefficients, each with 6 terms from summing over $\Delta\boldsymbol{R}_{\mu\nu}$ that span \textapprox100 measurement points in $q$.
When evaluating $C^{(\text{calc})}_{lmk}(q,\boldsymbol{\Theta})$, such equations are parameterized within the 6d space of $\boldsymbol{\Theta}$ parameters.
This $\boldsymbol{\Theta}$ dimensional space is where the curse of dimensionality comes in, as $\boldsymbol{\theta}$ has at least $3N_{\text{atoms}}-6$ parameters that dictates the dimensionality we must search in to build \ptheta.
To evaluate all these equations, even for a triatomic, in a random or grid-like search to find \thopt{} with femtometer resolution is computationally infeasible.
Instead, we retrieve \ptheta{} with the Metropolis-Hasting algorithm (MHA): a MCMC method developed for such high dimensional integral equations \cite{hastings.metropolis_hastings.1970}, as in Eq.~\ref{eq:ap:intEqnSim}.

The MHA is a sampling algorithm that builds the joint probability distribution \ptheta{} by randomly selecting $\boldsymbol{\Theta}$ parameters and comparing their likelihood probabilities with neighboring $\boldsymbol{\Theta}'$ parameters.
At completion, our retrieved \ptheta{} is a list of selected $\boldsymbol{\Theta}$ parameters randomly selected from the true \ptheta{} distribution.
Reference~\cite{foreman.mcmc.2013} describes the Python package used in this analysis.
To help the reader better understand our use of the MHA, we now describe one iteration.
Let $\boldsymbol{\Theta}$ be the latest addition to \ptheta.
The MHA selects a nearby $\boldsymbol{\Theta}'$ with the transition probability $Q (\boldsymbol{\Theta}, \boldsymbol{\Theta}')$.
We require $Q (\boldsymbol{\Theta}, \boldsymbol{\Theta}') = Q (\boldsymbol{\Theta}', \boldsymbol{\Theta})$ so it is equally likely to revisit every region of $\boldsymbol{\Theta}$-space.
Generally $Q (\boldsymbol{\Theta}, \boldsymbol{\Theta}')$ is uniform or Gaussian.
With $\boldsymbol{\Theta}$ and $\boldsymbol{\Theta}'$ selected, the MHA appends $\boldsymbol{\Theta}'$ to \ptheta{} with probability 
\begin{equation}
\begin{split}
    \rho \left( \boldsymbol{\Theta}, \boldsymbol{\Theta}'\right) &= \text{min}\left[1, \frac{P \left( \boldsymbol{\Theta}' | C \right) Q \left(\boldsymbol{\Theta}, \boldsymbol{\Theta}' \right)}{P\left(\boldsymbol{\Theta} | C \right) Q \left(\boldsymbol{\Theta}', \boldsymbol{\Theta} \right)} \right] \\
    = \text{min}&\left[1, \frac{P \left( C | \boldsymbol{\Theta}'\right) P \left(\boldsymbol{\Theta}'\right) Q \left(\boldsymbol{\Theta}, \boldsymbol{\Theta}' \right)}{P\left( C | \boldsymbol{\Theta}\right) P \left( \boldsymbol{\Theta}\right) Q \left(\boldsymbol{\Theta}', \boldsymbol{\Theta} \right)} \right]\label{eq:ap:MHA_selection},
\end{split}
\end{equation}
otherwise it appends $\boldsymbol{\Theta}$ again.
The ratio in Eq.~\ref{eq:ap:MHA_selection} cancels out $P(C)$, and when $P(\boldsymbol{\Theta}) = P(\boldsymbol{\Theta}')$ for all physical quantities, as it does for our case, we are only concerned with the ratio of likelihood probabilities.
The process then repeats itself by selecting a new $\boldsymbol{\Theta}'$.
Since each $\boldsymbol{\Theta}$ has either the same values or is a neighbor of the previously selected $\boldsymbol{\Theta}$ the raw \ptheta{} distribution is not an independently drawn distribution.
To remove this correlation between consecutively selected $\boldsymbol{\Theta}$ parameters, we select the $\boldsymbol{\Theta}$ parameters after every $\tau^{(\text{AC})}$.
Here $\tau^{(\text{AC})}$ is the autocorrelation time; the number of MHA steps needed to no longer by correlated with your starting position \cite{foreman.mcmc.2013}.
Thus, our retrieved \ptheta{} is a set of $\boldsymbol{\Theta}$ parameters independently drawn from the true \ptheta.
Since the early MHA selected $\boldsymbol{\Theta}$ parameters will be affected by our initial guess and the MHA requires time to equilibrate, we remove the first ~5 $\boldsymbol{\Theta}$s (after pruning by $\tau^{(\text{AC})}$).
Reference~\cite{foreman.mcmc.2013} describes in more detail how to determine when \ptheta{} has converged.

The intuition of Eq.~\ref{eq:ap:MHA_selection} is that if one cannot evaluate \ptheta{} analytically or numerically, but can calculate it up to a constant, then they can build \ptheta{} by taking the ratio of neighboring points.
The MHA uses the ratio of likelihood probabilities as a guide towards regions of higher posterior probability.
That is, the ratio of likelihood functions, where $\mmptheta \propto P( C | \boldsymbol{\Theta})$, may indicate that $\boldsymbol{\Theta}$ is twice as likely as $\boldsymbol{\Theta}'$ and consequently the MHA will visit $\boldsymbol{\Theta}$ twice as often as $\boldsymbol{\Theta}'$.
This selective sampling of $\boldsymbol{\Theta}$ parameters allows one to tackle the curse of dimensionality by efficiently sampling $\boldsymbol{\Theta}$-space while ignoring regions of low probability.
For example, if $\boldsymbol{\Theta}''$ were 100 time less likely than $\boldsymbol{\Theta}'$, and $\boldsymbol{\Theta}'$ is 50 times less likely than $\boldsymbol{\Theta}$, one would visit $\boldsymbol{\Theta}''$ once for every 50,000 visits to $\boldsymbol{\Theta}$.
This makes it very unlikely one ever visits the region near $\boldsymbol{\Theta}''$ or any region further in $\boldsymbol{\Theta}$-space that would be less likely.
This also means that one spend most of their time sampling the highly likely region around $\boldsymbol{\Theta}$ to improve resolution.
Stated more rigorously, the region of $\Delta \boldsymbol{\Theta}$ is sampled $(\int_{\Delta \boldsymbol{\Theta}} P(C|\boldsymbol{\theta}) d\boldsymbol{\theta})/(\int_{\Delta \boldsymbol{\Theta}'} P(C|\boldsymbol{\theta}) d\boldsymbol{\theta}) = (\int_{\Delta \boldsymbol{\Theta}} P(\boldsymbol{\theta}|C) d\boldsymbol{\theta})/(\int_{\Delta \boldsymbol{\Theta}'} P(\boldsymbol{\theta}|C) d\boldsymbol{\theta})$ times more than $\Delta \boldsymbol{\Theta}'$.
The MHA search is analogous to a random walk guided by the structures' relative agreement to the data, rather than a random sampling of distributions.

With the retrieved \ptheta{} is we can find \thopt, the global maximum, and evaluate the correlations between the parameters, \sth.
Since \ptheta{} is a list of $\boldsymbol{\Theta}$ parameters, we can calculate aggregate quantities.
With enough samples, one can histogram the collected $\boldsymbol{\Theta}$ parameters and/or apply a high dimensional kernel density estimator to retrieve a functional form of \ptheta{} \cite{liu.HiDimKDE.2007}.

It is important to note the MHA is theory independent when $P(\boldsymbol{\Theta})$ is constant, and that filtering the MHA results by $\tau^{(\text{AC})}$ yields independently drawn samples.
This alleviates any bias of sampling structures from physically motivated distributions that are not fully validated.
%This allows the experimentalist to interpret the data without \textit{a priori} knowledge of the system, besides the ground rovibronic state and the imparted anisotropy.
One can use $P(\boldsymbol{\Theta})$ to input chemical knowledge of the system if preferred.
Although the results will be biased by this input, one will not spend time sampling potentially erroneous $\boldsymbol{\Theta}$ parameters.

%% file: sections/ap_fitting_error_propagation.tex
The standard error of the mean of the $C_{lmk}(q)$ coefficients ($\sigma_{lmk}(q)$) contains information regarding the width and shape of \ptheta.
Similarly, the $C_{lmk}(q)$ will shift the entire distribution \ptheta {} distribution and may also change its shape.
For these reasons, it is crucial to include systematic effects in $\sigma_{lmk}(q)$ so the width of \ptheta{} will encompass the correct results even if \ptheta{} is systematically shifted.
The $\sigma_{lmk}(q)$ can be found in different ways, here we discuss three methods.
The first method is to directly measure the statistical uncertainty, as we did for N$_2$O. 
The second method is a means of estimating systematic uncertainty from experimental artifacts, also used in our N$_2$O analysis.
The third method is to analytically propagate the statistical uncertainty, which is useful for simulations.

\begin{figure}[!htbp]
\begin{center}
\includegraphics[scale=0.5]{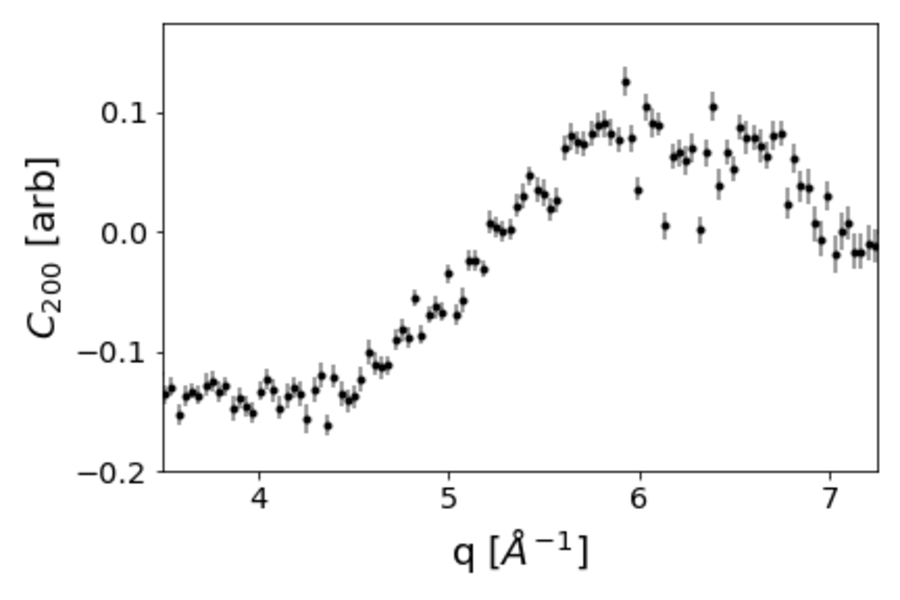}
\caption{\textbf{The raw measured $C_{200}(q)$ coefficient for N$_2$O} We show the raw $C_{200}(q)$ coefficient for the measured N$_2$O data. This is before we apply any methods to it.\label{fig:ap:raw_data}}
\end{center}
\end{figure}

The first method of directly measuring the statistical error follows standard practices.
One first fits each individual diffraction image (Supplementary Section~\ref{ap:fitting}) to retrieve the $C_{lmk}(q)$ coefficients.
One then calculates $\sigma_{lmk}(q)$ from this distribution of the $C_{lmk}(q)$ coefficients.
One can also bootstrap $\sigma_{lmk}(q)$ by fitting many different combinations of diffraction images for $C_{lmk}(q)$ and calculating the standard deviation of the resulting distribution.
In our N$_2$O analysis we fit single diffraction images for $C_{lmk}(q)$ and calculated the standard error of the mean from this distribution, shown in Fig.~\ref{fig:ap:raw_data}.

\begin{figure}[!htbp]
\begin{center}
\includegraphics[scale=0.5]{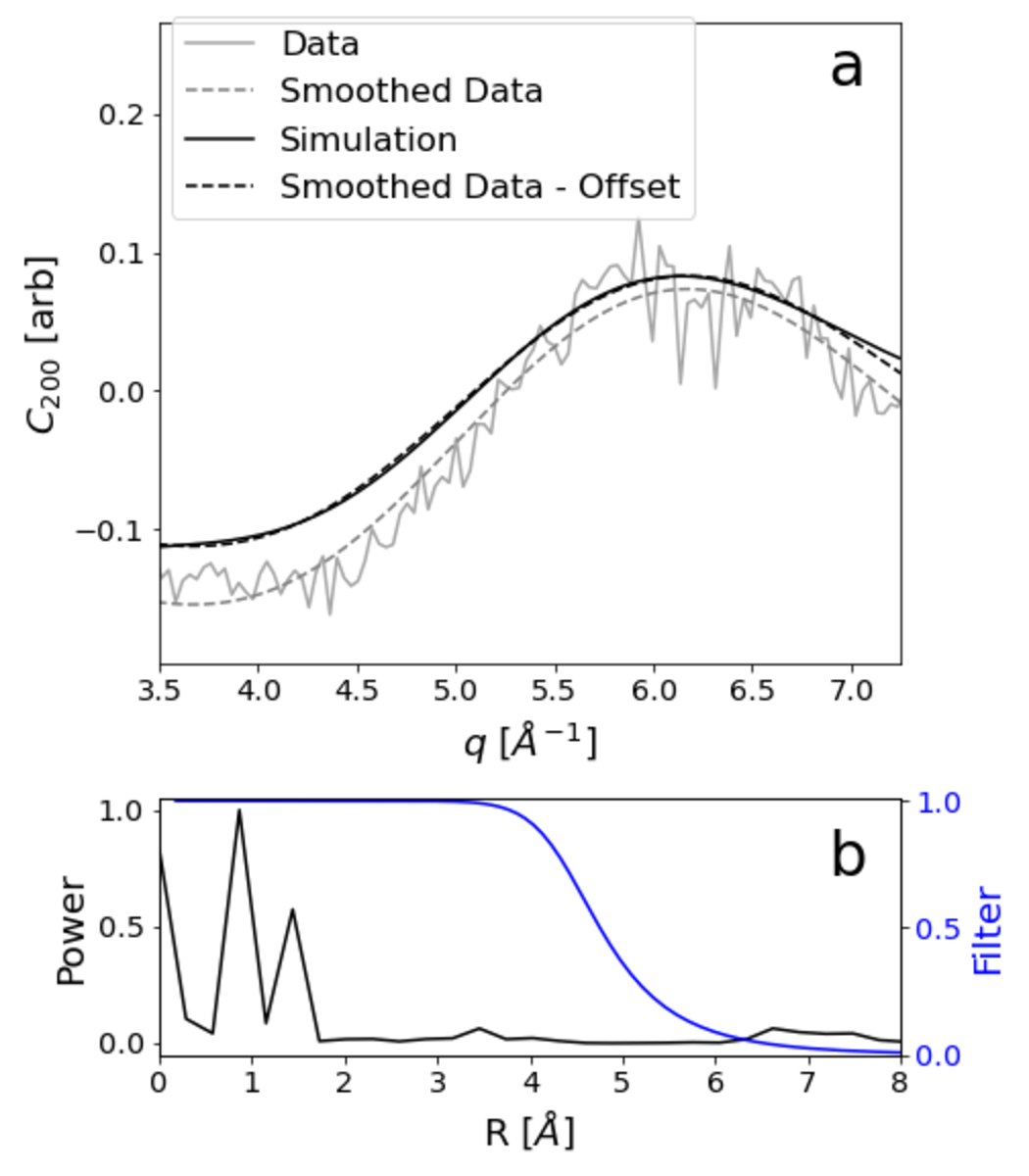}
\caption{\textbf{Removing systematic and statistical experimental contributions} Both systematic and statistical errors must be carefully addressed in measured data. Here, we show the data after accounting for noise and systematic effects. Panel a (gray) shows the original and smoothed data and its comparison to the simulated data before and after subtracting the offset. Panel b shows the Fourier power spectrum (Pair Distribution Function) of the data (black) and the applied a low-pass filter (blue).\label{fig:ap:data_filter}}
\end{center}
\end{figure}

\begin{figure}[!htbp]
\begin{center}
\includegraphics[scale=0.5]{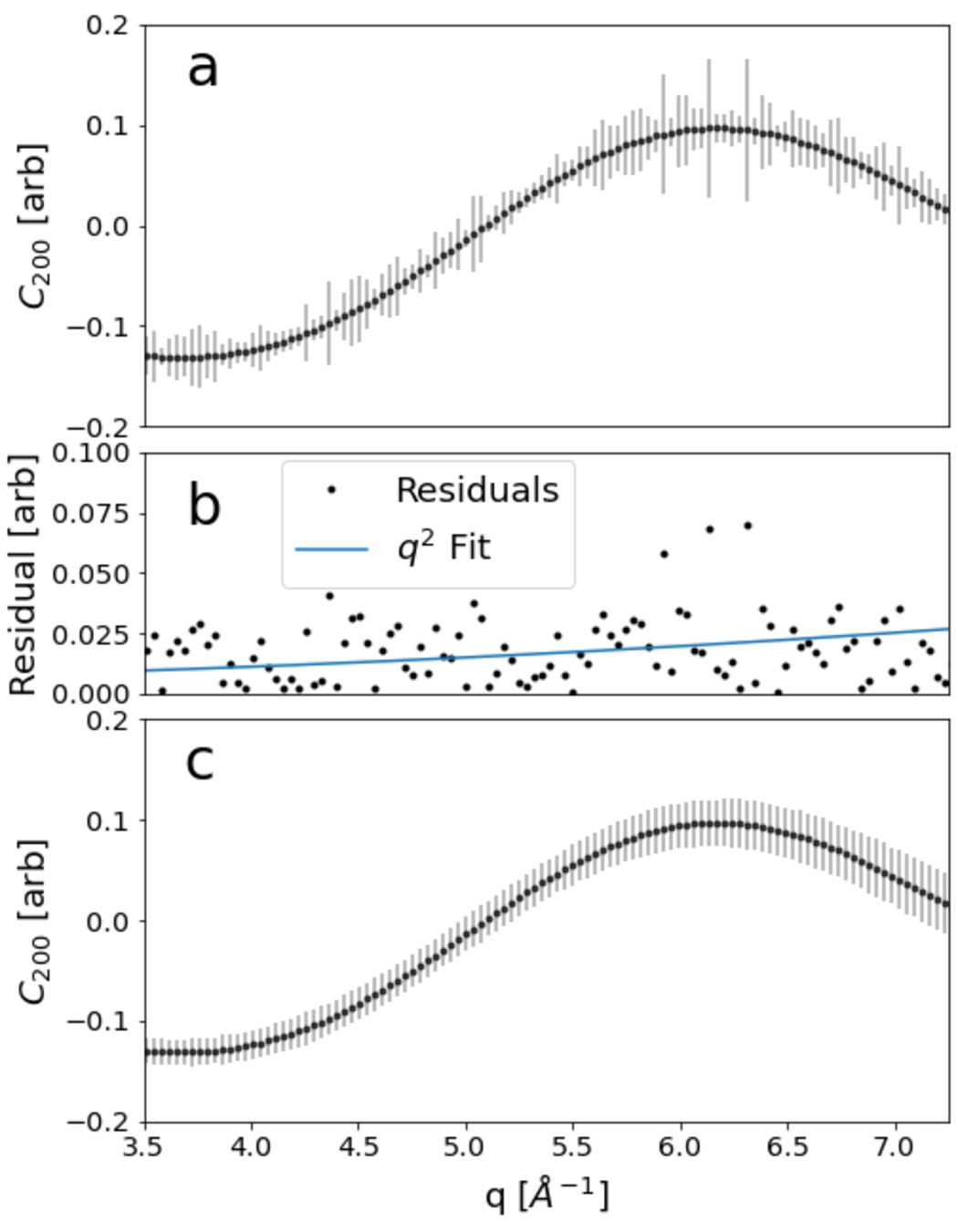}
\caption{\textbf{Defining consistent error bars after applying filters} After filtering the data, the error bars must be re-calibrated to account for the removed statistical noise. We illustrate our procedure for finding $\sigma_{200}(q)$ for the N$_2$O data. Panel a shows the low-pass filtered results where we added the residuals in quadrature with the error bars. Panel b shows these residuals between the low-pass filtered and original data, as well as a quadratic fit to them. Panel c shows the filtered data with error bars determined by the fit in panel b.\label{fig:ap:var_filter}}
\end{center}
\end{figure}

The second method addresses systematic effects from the experimental apparatus that the first method will miss.
In this dataset, the $q$ calibration changed as a function of \thd{} which washed out the signal below 3.5~\iang{} and created a time-dependent offset that varied as $\mmadm{2}{0}{0}$.
This dataset also suffered from high-frequency variations in $q$.
We removed the high frequency noise and the time-dependent offset from $C_{200}(q)$ by applying a low-pass filter and subtracting an offset, shown in Fig.~\ref{fig:ap:data_filter}.
The filter cut began around 4~\AA, far from our longest expected distance of 2.3~\AA, 
We note that
\begin{equation*}
    \mathcal{F}\left[C_{200}(q)\right] \propto \sum_{\mu\nu}\mathcal{F}\left[j_2(q\Delta R_{\mu\nu})\right]
\end{equation*}
and is not the PDF.
Using the convolution theorem, we still do not expect any signal above 2.3~\AA.
After subtracting an offset from the raw data and rescaling, we observe the dashed black line in Fig.~\ref{fig:ap:data_filter}a.

After applying the low-pass filter, we must account for the variations it removed in the error bars.
Figure~\ref{fig:ap:var_filter}a shows the filtered results with the residuals added in quadrature to the original error bars, Fig.~\ref{fig:ap:var_filter}b shows these residuals.
We fit the residuals with a quadratic since we do not expect the error to vary wildly between adjacent points after filtering.
The final error bars are shown in Fig.~\ref{fig:ap:var_filter}c.

The third method, which is only for simulation, is to propagate the error through the fitting procedure.
Since we can calculate the $C_{lmk}(q)$ coefficients, we do not need to do the fitting procedure.
However, we must calculate the error bars as though we did.
The $\text{Var}(\mathbf{F})$ is the same whether one uses the normal equation or the QR decomposition.
For the normal equation we start from Eqs.~\ref{eq:ap:genNormEqn} and \ref{eq:ap:Wgen},
\begin{align}
   \nonumber \text{Var}(\mathbf{F}) &= \left( \mathbf{X}^T \mathbf{WX}\right)^{-1} \mathbf{X}^T \mathbf{W} \text{Var}(\mathbf{Y}) \left(\left( \mathbf{X}^T \mathbf{WX}\right)^{-1} \mathbf{X}^T \mathbf{W}\right)^T \\
   \nonumber &= \left( \mathbf{X}^T \mathbf{WX}\right)^{-1} \mathbf{X}^T \mathbf{WW}^{-1} \left(\left( \mathbf{X}^T \mathbf{WX}\right)^{-1} \mathbf{X}^T \mathbf{W}\right)^T \\
    &= \left( \mathbf{X}^T \mathbf{WX}\right)^{-1}.
    \label{eq:ap:fitVar}
\end{align}
For the QR decomposition, we start from Eqs.~\ref{eq:ap:qr} and \ref{eq:ap:qrfit}
\begin{align*}
    \text{Var}(\mathbf{F}) &= \boldsymbol{R}^{-1}\mathbf{Q}^{T}\sqrt{\mathbf{W}}\text{Var}(\mathbf{Y})\sqrt{\mathbf{W}}^T\mathbf{Q}\boldsymbol{R}^{-1T} \\
    &= \boldsymbol{R}^{-1}\mathbf{Q}^{T}\sqrt{\mathbf{W}}\mathbf{W}^{-1}\sqrt{\mathbf{W}}^T\mathbf{Q}\boldsymbol{R}^{-1T} \\
    &= \widetilde{\mathbf{X}}^{-1}\widetilde{\mathbf{X}}^{-1T} \\
    &= \left( \mathbf{X}^T \mathbf{WX}\right)^{-1}.
\end{align*}

To propagate the Poissonian noise measured on the detector to the $C_{lmk}(q)$ coefficients we examine the two fitting procedures described in supplementary Section~\ref{ap:fitting}.
We first propagate through the \blm{} fit where the diffraction images are fit with spherical harmonics.
The simulated Poissonian noise on the detector is given by
\begin{equation}
    \text{Var}\left( \langle I(q,\mmthd,t)\rangle \right) = \langle I(q,\mmthd,t)\rangle
\end{equation}
where $\mathbf{X}$ and $\mathbf{W}$ are given by Eqs.~\ref{eq:ap:XB} and \ref{eq:ap:WB} respectively.
To calculate the $C_{lmk}(q)$ coefficient error bars we again use Eq.~\ref{eq:ap:fitVar}.
Instead, the $\mathbf{X}$ and $\mathbf{W}$ are given by Eqs.~\ref{eq:ap:XC} and \ref{eq:ap:WC} respectively.

\iffalse
\begin{align}
    \text{Var}\left(C_{lm_2\nu}(q)\right) &= \epsilon \left(\mathbf{X}^T\mathbf{X}\right)^{-1}_{\nu\nu}
    \mathbf{W} &=
        \begin{bmatrix}
        \frac{1}{\epsilon_0} & 0 & \dots & 0 \\
        0 & \frac{1}{\epsilon_1} & & \\ 
        \vdots &  & \ddots & \\
        0 & & & \frac{1}{\epsilon_{N}}
        \end{bmatrix} \label{eq:WC}.
\end{align}
In the case that the $\mathcal{A}^l_{m_2m_1}(t)$ are orthogonal $\text{Var}\left(C_{lm_2k}(q)\right)$ reduces to the constant error, $\epsilon$, divided by the square of the L2 norm of said $\mathcal{A}^l_{m_2m_1}(t)$.
In our case, the $\mathcal{A}^l_{m_2m_1}(t)$ are not orthogonal, so we must do the matrix inversion.
However, fitting the $C_{lm_2k}(q)$ coefficients from data will vary based upon how orthogonality of $\mathcal{A}^l_{m_2m_1}$ and consequently the fitting method employed, which may include regularization.
\fi

%% file: sections/ap_mode_search.tex
\begin{figure*}[!htbp]
    \centering
    \includegraphics[scale=0.35]{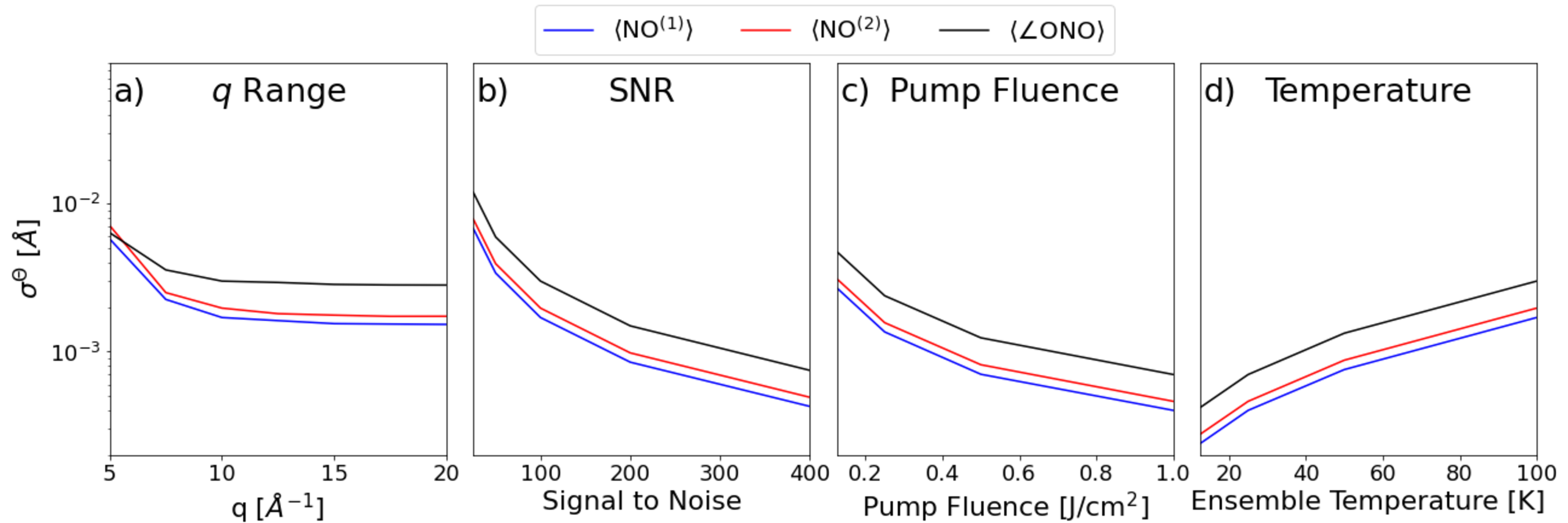}
    \caption{\textbf{The effects of various experimental parameters on \pthetad} Varying experimental parameters affects the resolution (width) of \pthetad, but our method is most sensitive to the signal-to-noise ratio (SNR). Panel a shows how the uncorrelated widths of \pthetad, denoted by \sth, change by increasing the $q$ range. Panel b similarly shows the dependence of \sth{} versus SNR. Panel c shows the dependence of \sth{} versus pump fluence (width of the rotational wavepacket) at 25~K. Panel d shows the dependence of \sth{} versus the molecular ensemble temperature at a constant pump fluence of 1~J/cm$^2$.}
    \label{fig:ap:trends_delta}
\end{figure*}

After retrieving \ptheta{} we need to find the most likely $\boldsymbol{\Theta}$ parameters (\thopt) to parameterize our probability distribution of structures \prth.
Recall, \ptheta{} is the probability distribution of parameters that parameterize our chosen probability distribution of structures (\prth), and therefore its mode corresponds to the set of parameters that best describe our measurement.
To do this, we must again address the curse of dimensionality since we are still searching within the $N_\Theta$-dimensional space, where $N_\Theta$ is the number of $\boldsymbol{\Theta}$ parameters.
%Since this high dimensionality prohibits evaluating all the parameters with a reasonable resolution, we use Bayesian Inference and the Metropolis Hastings Algorithm (MHA) to retrieve \ptheta.
We re-emphasize again that we are interested in the \thopt{} that best describes our data which is given by the mode of \ptheta, which does not necessarily correspond to the mean of \ptheta.
If one looks at a single parameter $\theta$, the mean or mode of this uncorrelated distribution may not correspond to the value that would provide the highest \ptheta{} value in the full $\boldsymbol{\Theta}$-space: illustrated in Fig.~\ref{fig:compare_qrange_correlations}.
One must therefore search the correlated $\boldsymbol{\Theta}$-space.
Once the MHA has converged, \ptheta{} may have significantly constrained $\boldsymbol{\Theta}$-space, but searching for the mode may still be infeasible for a simple grid search.
Below we describe three methods to find \thopt{} using \ptheta{} to help us overcome the curse of dimensionality.

The first and most simple way to find \thopt{} is to apply the MHA to the measured $C_{lmk}(q)$ coefficients in the same way as before, but significantly decrease $\sigma_{lmk}(q)$.
One can make \ptheta{} arbitrarily sharp, effectively zooming onto the mode, by artificially decreasing $\sigma_{lmk}(q)$.
With small enough $\sigma_{lmk}(q)$ one can zoom into \ptheta{} until it is adequately described by a quadratic, where the mean and the mode of the distribution will be the same.
The danger of using this method is that one may fall into a local maximum by decreasing $\sigma_{lmk}(q)$ too quickly without being careful.
For example, one's initial $\boldsymbol{\Theta}$ guess may be close to a local maximum and the small $\sigma_{lmk}(q)$ will force the MHA into it and not sample outside of it.
To avoid this, one must start the MHA in many different initial $\boldsymbol{\Theta}$ states and gradually decrease $\sigma_{lmk}(q)$ to find the mode and rule out any local maximum.

The second method is to interpolate \ptheta{} between the evaluated MHA points using a high dimensional Kernel Density Estimator (KDE).
We note that one can use all the MHA points rather than the points in \ptheta{} which are filtered by the auto-correlation time $\tau^{(\text{AC})}$.
This is because we are looking for the mode and not evaluating some function over the \ptheta{} distribution.
The primary difficulty with KDEs is finding the shape and width of the kernel.
Generally, KDE methods do not perform well for problems in larger than three dimensions.
More recently, there has been work to generalize KDEs to high dimensions \cite{liu.HiDimKDE.2007}.
Calculating points in \ptheta{} with a KDE will be very fast.
Such quick evaluations may allow one to find the mode through simple optimization schemes like a basic grid search or gradient descent.

The third method, used in this paper, is a mixture of simple searching methods and calculating \thopt{} by a weighted average of the most likely MHA points.
By considering only the $N_{\text{likely}}$ unique points with the highest likelihood probability we focus on the mode while disregarding tails of the \ptheta{} distribution.
Since we are only concerned with the most likely points, we look at all the points the MHA accepted.
This differs from \ptheta, which takes MHA points separated by the auto-correlation time $\tau^{\text{AC}}$.
We calculate \thopt{} by a weighted sum of the $N_{\text{likely}}$ $\boldsymbol{\Theta}$ parameters
\begin{equation}
    \begin{split}
    \mmthopt &= \frac{\sum_{n \in \{N_{\text{likely}}\}} \boldsymbol{\Theta}^{(n)} P \left(\boldsymbol{\Theta}^{(n)} | C \right) }{\sum_{n \in \{N_{\text{likely}}\}} P \left( \boldsymbol{\Theta}^{(n)} | C \right)}     \\
    &= \frac{\sum_{n \in \{N_{\text{likely}}\}} \boldsymbol{\Theta}^{(n)} P \left(C | \boldsymbol{\Theta}^{(n)} \right)}{\sum_{n \in \{N_{\text{likely}}\}} P \left(C | \boldsymbol{\Theta}^{(n)} \right)}
    \end{split}
\end{equation}
where $\{N_{\text{likely}}\}$ denotes the set of indices corresponding to the $N_{\text{likely}}$ $\boldsymbol{\Theta}$ parameters with the largest posterior, and the second equality only holds because we chose $P(\boldsymbol{\Theta}^{(n)}) = P(\boldsymbol{\Theta}^{(m)})$.
Given the most recently calculated \thopt{} value, we alternate between a grid search where points are separated by 0, $\pm1$, and $\pm1.5$ standard deviations ($\sigma^{(\text{MS})}_i$) and a random search.
Here $\sigma^{(\text{MS})}_i$ is the one dimensional standard deviation of the $i^{\text{th}}$ $\boldsymbol{\Theta}$ parameter taken over the distribution of the $N_{\text{likely}}$ $\boldsymbol{\Theta}$s.
After one iteration of the grid search, we randomly sample $\boldsymbol{\Theta}$s from a normal distribution with mean \thopt{} and standard deviation $\sigma^{(\text{MS})}$.
The point of this random sampling is to focus on the region of less than one standard deviation.
This keeps the grid search from making \thopt{} roam too far from the globally optimal parameters.
The grid and random sampling are then repeated until every parameter changes by $<3$\% for five consecutive times.
At this time we switch to a random sampling method.
We randomly sample values from a normal distribution again with mean \thopt{} and standard deviation $\sigma^{(\text{MS})}_i$.
We consider \thopt{} has converged when every parameter has changed $<0.01$\% for three consecutive random samplings, but require at least one value to change between samplings.

There are many ways to search for \thopt{} that generally trade between speed and accuracy.
Our simple search method performed well for all our experimental variations when there were sufficient samples in \ptheta, which depends on width of \ptheta.
For \ptheta{} distributions sufficiently broader than ours, one may need a more advanced method.
To calculate the precision of \thopt{} one must find the hyper curve in $\boldsymbol{\Theta}$ space with minimal precision, as outlined in Ref.~\cite{cumpson.HiDimError.1992}.

%% file: sections/ap_delta_distribution.tex
\begin{figure}[!htbp]
    \centering
    %0.44
    \includegraphics[scale=0.4]{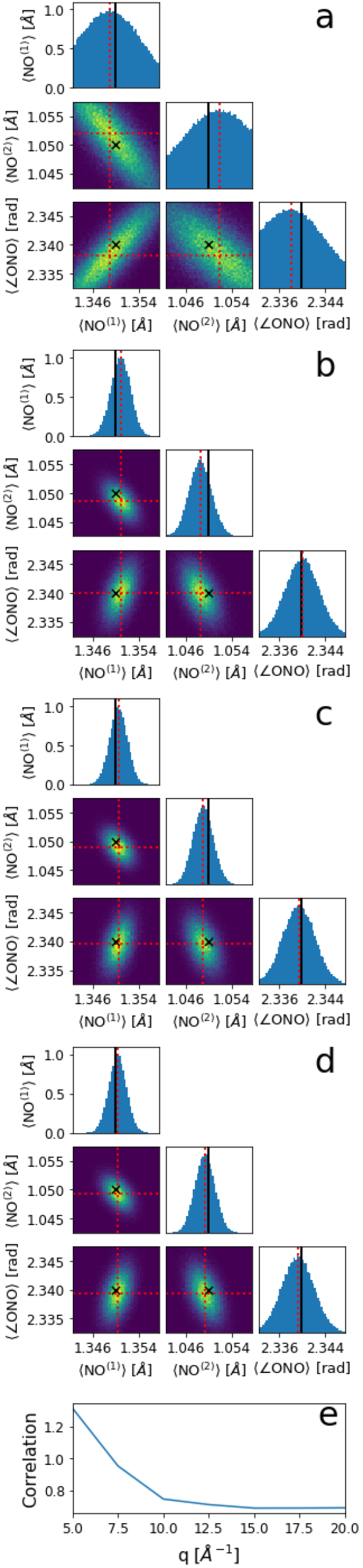}
    \caption{\textbf{Effects of varying the measured $q$ range on \pthetad} Varying the $q$ range affects false correlations in \pthetad. We show the 1d and 2d projections of the retrieved \pthetad{} distribution for varying $q$ ranges. Panel a has a $q$ range of $[0.5,5]$~\iang, b is $[0.5,10]$~\iang, c is $[0.5,15]$~\iang, and d is $[0.5,20]$~\iang. The red dashed lines illustrate \thopt, while the black ``X" and solid lines indicate the ground truth values. Panel e shows the correlation between all $\boldsymbol{\Theta}$ parameters as a function of $q$ range.}
    \label{fig:ap:correlations_delta}
\end{figure}

The delta posterior, \prthd, is quick to calculate but assumes the $C_{lmk}(q)$ coefficients calculated from a single structure and measured from an ensemble of structures are comparable.
This assumption effectively ignores the damping of the $C_{lmk}(q)$ as a function of $q$, similar to a damped oscillator, due to the width of $|\Psi(\boldsymbol{R})|^2$.
Figure~\ref{fig:theta_peaks_q} shows this $q$ dependent systematic where \pthetad{} converges on the ground truth values in an unstable fashion as $q$ increases.
The ground truth value, at times, can be considerably far from the retrieved distribution's mean, and with improved SNR may quickly be separated by $>3$ standard deviations.
The retrieved mean can also jump to either side of the ground truth values at low $q$.
This behavior, along with the systematic error, is absent in Fig.~\ref{fig:theta_peaks_q} for \pthetag.

Even with this systematic error, we find that \prthd{} follows the same trends as \prthg{} when varying experimental parameters, as shown in Fig.~\ref{fig:ap:trends_delta}.
We similarly see that for \prthd{} our method benefits more strongly from increased SNR, rather than increasing the measured $q$ range.
One will again see diminishing returns when measuring past \textapprox8~\iang.
Increasing the alignment kick strength and decreasing the ensemble temperature also have a similar effect as increasing the SNR.
In Fig.~\ref{fig:ap:correlations_delta} we also see that the correlations between $\boldsymbol{\Theta}$ parameters consistently diminishes as the $q$ range is increased.

\iffalse
The primary difference between \pthetad{} and \pthetag{} is the accuracy in retrieving the ground truth \thopt, shown in Fig.~\ref{fig:ap:trends_delta} (bottom row).
Generally, the delta distribution \thopt{} errors are roughly three orders of magnitude larger than those from the normal distribution, except for \maono{} which is comparable.
This decrease in accuracy is again from the systematic error induced by using a single structure to describe the results from an ensemble of structures.
This issue is most noticeable in Fig.~\ref{fig:ap:trends_delta}a (lower) where the \maono{} accuracy dips between 5 and 15~\iang, which corresponds to the \maono{} distribution mean alternating to either side of the ground truth value in Fig.~\ref{fig:theta_peaks_q}.
\fi

Although the delta distribution suffers from the above-mentioned systematic, it is very important when building and debugging one's analysis and is necessary for very large molecules.
Retrieving \pthetad{} is roughly 100 times faster than retrieving \pthetag{} due to dropping half the $\boldsymbol{\Theta}$ space dimensions in the retrieval of \pthetag{} and removing the integration over many structures drawn from \prthg{} when calculating $C_{lmk}^{(\text{calc})}(q)$.
We highly encourage the reader to use \pthetad{} when debugging due to its fast execution and sufficient accuracy for such intermediate evaluations.
For large molecules, the $C_{lmk}^{(\text{calc})}(q)$ integral becomes more computationally intensive as $\boldsymbol{\Theta}$-space grows.
At some point, it is computationally infeasible for the MHA to search such a large $\boldsymbol{\Theta}$-space when it must compute the $C_{lmk}^{(\text{calc})}(q)$ integral for every $\boldsymbol{\Theta}$ it randomly chooses.
For such large molecules, one will need to use the delta distribution.
To account for the delta distribution's systematic error, one can increase $\sigma_{lmk}(q)$ so \pthetad{} comfortably encompasses the ground truth values.
One can run the same simulations done in this paper on expected, or measured, structures to determine such an increase. 
By doing so, one can report results that account for the induced systematic errors from our assumption of $|\Psi(\boldsymbol{R}, t)|^2$'s shape.
\FloatBarrier

%% file: sections/ap_fittingIcoeff.tex
Both the pairwise angles and $\mathcal{I}$ act as a weighting function to the $q$ dependent Spherical Bessel functions, shown in Eq.~\ref{eq:MHAfit}.
If $\mathcal{I}$ is not correct, this may lead to a systematic offset of the molecular frame angles as the error in $\mathcal{I}$ must be absorbed by $Y_l^{-m_1} \left( \theta_{\mu\nu}^{(\text{mf})}, \phi_{\mu\nu}^{(\text{mf})}\right)$.
When fitting for $\mathcal{I}$ one will generally need to know the molecular structure, often this will be from the ground rovibronic state.
Below we describe a few methods to retrieve $\mathcal{I}$ or circumvent this issue.

Our first method cancels out the factor of $\mathcal{I}$ by using the ratio of $C_{lmk}(q)/C_{l'mk}(q)$ for the MHA.
This requires one measure multiple anisotropy components.
One can also let $l'=0$ since the isotropic component is independent of the molecular frame angles and will therefore not introduce any bias.
In this method, one does not need to use a simulated structure to fit for $\mathcal{I}$.

The second method involves having multiple datasets, or partitioning the full dataset to fit $\mathcal{I}$.
The first possible partition is in time, where one uses the $C_{lmk}(q)$ from a certain point in the alignment.
One may find it easiest to look at times before the induced rotation since one must already know the ground rovibronic state to simulate the ADMs.
The second possible partition is to use the $l=0$ signal and known $\Delta \boldsymbol{R}_{\mu\nu}$ to fit for $\mathcal{I}$.
One may also collect a second pump-off dataset to fit for $\mathcal{I}$ or randomly partition a single dataset.
Such a secondary dataset can also be used to fit the ADMs if one also induces vibrational dynamics as well.

The last method addresses the case of having few $C_{lmk}(q)$ anisotropy contributions and a small dataset.
This is the case for the N$_2$O results presented here.
One may implement a bootstrapping method that relies on fitting Eq.~\ref{eq:MHAfit} to a $C_{lmk}(q)$ for varying $q$ ranges.
One can retrieve the best fit value for $\mathcal{I}$ and its corresponding error from the resulting distribution of fits.

%% file: sections/ap_n2o_thermal_dist.tex
\begin{figure}[!htbp]
    \centering
    \includegraphics[scale=0.7]{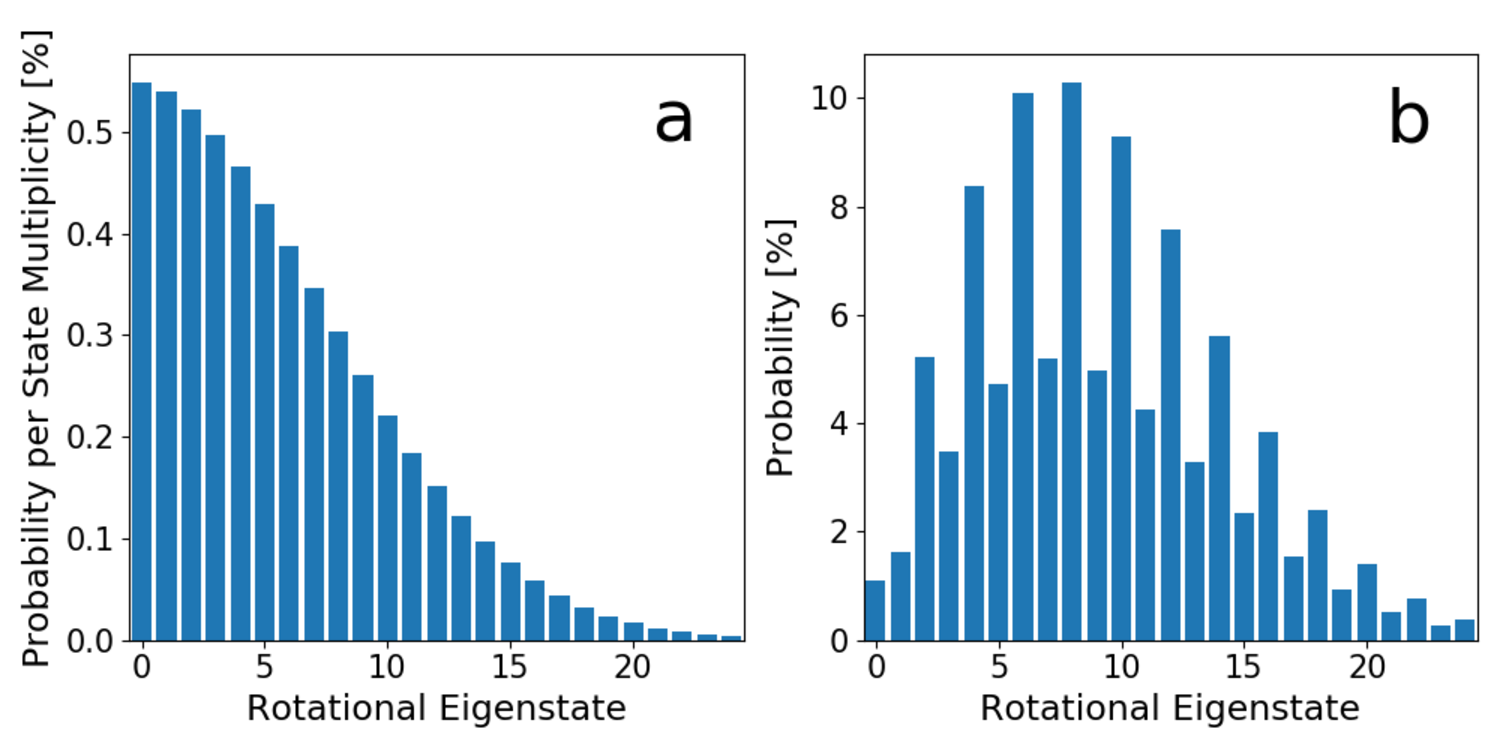}
    \caption{\textbf{Initial sample thermal distribution of rotational states} We show the thermal distribution of rotational states from the measured N$_2$O sample before impulsive Raman excitation.}
    \label{fig:ap:rot_pop}
\end{figure}

Within the sample chamber, the initial N$_2$O gas temperature was 73~K (Section~\ref{ap:ADM_calc}).
We calculated the thermal Boltzmann distribution with the lowest 22 vibrational and 100 rotational states using the measured vibrational energies and their corresponding rotational B, D, and H constants from Ref.~\cite{Toth.n2o_rovib_spect.1991}.
That is, for each vibrational state we calculated the probability of being in the lowest 100 rotational states.
For the vibrational thermal distribution, the ground vibrational state dominates with nearly 100\% population (Table~\ref{tab:ap:vib_pop}).
For the rotational thermal distribution, the distribution is shifted from 0 with a mode at the $n=8$, as shown in Fig.~\ref{fig:ap:rot_pop}.

\begin{table}[!htbp]
    \centering
    \begin{tabular}{|c|c|}
        \hline
        State & Population Percent (\%) \\ \hline
        0000 & 99.998 \\
        0110(e) & 9.124 $\times 10^{-4}$ \\
        0110(f) & 9.107 $\times 10^{-4}$ \\
        0200 & 1.000 $\times 10^{-8}$ \\
        \hline
    \end{tabular}
    \caption{\textbf{Initial sample thermal distribution of vibrational states} We show the thermal distribution for the first 4 vibrational states in the N$_2$O gas sample.}
    \label{tab:ap:vib_pop}
\end{table}

